\documentclass[aps,prd,preprint]{revtex4}

\usepackage{graphicx}
\usepackage{psfrag}

\begin{document}
\title{BCVEGPY: An Event Generator for Hadronic Production of the $B_c$ Meson}
\author{Chao-Hsi Chang$^{1,2,3}$ \footnote{email: zhangzx@itp.ac.cn},
Chafik Driouichi$^3$ \footnote{email:chafik.driouichi@hep.lu.se}, Paula
Eerola$^3$ \footnote{email:paula.eerola@hep.lu.se} and Xing-Gang
Wu$^{1}$\footnote{email: xgwu@itp.ac.cn}}
\address{$^1$Institute of Theoretical Physics, Chinese Academy of Sciences,
P.O.Box 2735, Beijing 100080, China.\\
$^2$CCAST (World Laboratory), P.O.Box 8730, Beijing 100080,
China.\footnote{Not correspondence address.}\\
$^3$Experimental High Energy Physics, Department of Physics, Lund
University, Box 118, SE-22100 Lund, Sweden.}

\begin{abstract}
We have written a Fortran programme BCVEGPY, which is an event generator for the
hadronic production of
the $B_c$ meson through the dominant hard subprocess $gg\rightarrow
B_c(B_c^*) +b+\bar{c}$.  To achieve a compact programme, we have
written the amplitude of the subprocess with the particle helicity
technique and made it as symmetric as possible, by decomposing
the gluon self couplings and then applying the symmetries. To check the
programme, various cross sections of the subprocess have been computed
numerically and compared with those in the literature. BCVEGPY is
written in a PYTHIA-compatible format, thus it is easy to implement in PYTHIA. \\
\\

\noindent {\bf PACS numbers:} 13.85.Ni, 12.38.Bx, 14.40.Nd,
14.40.Lb.

\noindent {\bf Keywords:} \(B_{c}\) meson, hadronic production,
event generator.
\end{abstract}
\maketitle

\section{Introduction}

$B_c$-physics has been attracting an increasing attention
recently, due to the experimental discovery of the $B_c$
meson\cite{CDF}, and theoretical
progress\cite{B-work,mord,qigg,prod0,prod,prod1,prod2,prod3,prod4,spec,chen,dec,dec1,life,MG,CCWZ}.
Since one can collect a high-statistics sample of $B_c$ mesons
only at high energy hadronic
colliders\cite{B-work,mord,prod0,prod,prod1,prod2,prod3}, we have
re-written a generator for the hadronic production of $B_c$. The
generator is a Fortran program package called BCVEGPY.

The hadronic production of the $B_c$ and $B_c^*$
mesons\footnote{Based on the potential model for heavy quarks, the
ground state of the $(c\bar{b})$ bound system is in $1,^1S_0$, and
according to PDG it is denoted by $B_c$. The first excited state
$1,^3S_1$ is due to the spin-splitting (spin-spin interaction) and
it is denoted by $B_c^*$ throughout the paper. Since the
spin-splitting is small, the available energy of the decays from
$B_c^*$ to $B_c$ is so small that only dipole magnetic moment
transition (M1) is allowed between them.} has been estimated by
the authors of Refs.\cite{prod,prod4} with the fragmentation
approach and by the authors of Refs.\cite{prod1,prod2,prod3} with
the so-called `complete calculation' approach i.e. to compute the
production completely at the lowest order ($\alpha_s^4$) in terms
of the dominant subprocess of perturbative QCD (pQCD) $gg\to
B_c(B_c^*)+\bar{c}+b$. Further note that since $m_b\gg m_c\gg
\Lambda_{QCD}$, apart from the fact that $c$ and $\bar b$ quarks
combine into $B_c(B_c^*)$ non-perturbatively, the rest of the
subprocess is always `hard' and can be well calculated with
pQCD\cite{prod1,prod2,prod3}. These two approaches have different
advantages but both of them are in the framework of pQCD, and
attribute the non-perturbative factor in the production to the
decay constant, $i.e.$ the non-relativistic wave function of the
$B_c(B_c^*)$ at the origin (the former through the fragmentation
function and the latter directly). Within theoretical
uncertainties, the two approaches can agree numerically,
especially when the component of gluon fragmentation is
involved\cite{prod4}. The fragmentation approach is comparatively
simple and can reach to leading logarithm order (LLO) of pQCD, but
is satisfied only in the case if one is only interested in the
produced $B_c(B_c^*)$ itself. The complete calculation approach
has the great advantage that it retains the information about the
$\bar{c}$ and $b$ quark (jets) associated with the $B_c$ meson in
the production. From the experimental point of view this is a more
relevant case. Therefore, we have written the hadronic production
programme for $B_c$ mesons based on the complete calculation
approach (full pQCD complete calculation at the lowest order
$\alpha_s^4$).

Since the non-perturbative factor in the subprocess $gg\to
B_c(B_c^*)+\cdots$, i.e. the decay constant of the $B_c(B_c^*)$
meson, can be calculated by means of the potential model for heavy
quark-antiquark systems\cite{spec}, the estimates of the
production are full theoretical predictions of pQCD without
additional experimental input. In view of future experimental
studies of the $B_c(B_c^*)$ after the CDF discovery (RUN-I, at
Tevatron), i.e., concerning the needs of experimental feasibility
studies for various topics of $B_c(B_c^*)$ meson in hadronic
collisions at Tevatron and LHC\footnote{In literature there are
calculations on the hadronic production of the $B_c$ meson with
the same approach, but the computer programs have not been
published (or not easy to find), so we would like to present ours
by re-writing the program and comparing it with others.}, we would
like to write a paper on the Fortran package BCVEGPY with detailed
explanation. In addition, we emphasize here that, as explained
later, the present generator has been formulated in a very
different way from those in Refs.\cite{prod1,prod2}, and through
carefully comparing the results obtained by BCVEGPY with those
obtained in Refs.\cite{prod1,prod2}, a solid and independent check
of BCVEGPY is well made.

At LHC the beam luminosity and the production cross section (at
such a high energy $\sqrt S=14$TeV) are so high that the rate of
producing $B_c(B_c^*)$ events can be $10^{8-9}$ per year. At
Tevatron, although the luminosity and production cross section are
lower than at LHC, the rate of producing $B_c(B_c^*)$ events still
is about $10^{4-5}$ per year, only $3\sim 4$ orders of magnitude
lower than at LHC. The $B_c$ meson has sizable and abundant weak
decay modes relating directly to $c$ or/and $\bar{b}$ flavor
respectively, so that with so high production such as at LHC, even
the detecting efficiencies being taken into account, one can reach
a statistical accuracy of $10^{-2}$ for most of the decay
modes\cite{chen,dec,dec1}. Thus there is an acute need for a
$B_c(B_c^*)$ event generator in order to be able to perform
feasibility studies, and this is why we have written the present
paper on BCVEGPY. A particularly interesting topic, which is worth
noting here, is the study of $B_s-\bar{B_s}$ mixing and $CP$
violation in $B_s$ meson decays through $B_c(B_c^*)\to B_s
\cdots$\cite{mord,qigg}. The $B_c$ meson has a very large
branching ratio to decay to a $B_s$ meson ($Br(B_c\to B_s \cdots)
\simeq$ several tens percents\cite{chen,dec,dec1}), and the $B_s$
mesons obtained by such $B_c$ decays are tagged precisely at the
$B_c$ decay vertex, if the charge of the $B_c$ meson and/or its
decay products are measured. A large-statistics sample of tagged
$B_s$ mesons at LHC, and even at Tevatron, offers a great
potential for the interesting $B_s$ physics\cite{mord}.

As pointed out in Ref.\cite{prod1}, according to pQCD there is
another production mechanism by a quark pair annihilation
subprocess $q\bar{q}\to B_c(B_c^*)+\bar{c}+b$. Nevertheless, the
`luminosity' of gluons is much higher than that of quarks in $pp$
collisions (LHC) and in $p\bar{p}$ collisions (Tevatron), and
there is a suppression factor due to the virtual gluon propagator
in the annihilation. Therefore the contribution from this
mechanism is negligible compared to the dominant one. The
calculations in Refs.\cite{prod,prod2,prod3,prod4} neglected the
contribution from the quark pair mechanism, and the BCVEGPY
package follows the same approximation.

In order to make the programme very compact, we
write BCVEGPY by applying the `helicity technique' to the
amplitude of the subprocess. The technique may be traced back to
the work in Ref.\cite{jdb}. The helicity technique has been
developed by the CALKUL collaboration\cite{cal1,cal2} for massless
gauge theories. Further development for the massive fermion case
with Abelian gauge field(s) as well as for the massless fermion
case with non-Abelian gauge field(s) was done by several
groups\cite{zdl,aes,ld,ht}. When the helicities\footnote{The
CALKUL technique is essentially based on the chirality of massless
fermions. Since in the massless case the chirality is equivalent
to the helicity with one-to-one correspondence, in the references
the authors use the term `helicity' without ambiguity. While in
this paper we adopt some techniques of CALKUL, we will frequently
use `helicity'  but not `chirality', because we consider massive
fermions here.} of all the external massless particles are fixed,
CALKUL calculates the `probability rate' in the following way.
First, Feynman diagrams with precise helicities for the external particles
are computed one by one for the concerned process with
helicity techniques, and a complex number for each Feynman diagram
is obtained. Then all the obtained complex numbers are summed up.
Finally the squared modulus of the summed result is taken,
averaging over the helicities of the particles in initial state
and summing over the helicities of the particles in final state if
an unpolarized case is considered. If a polarized process is
considered, one needs only to stop the calculation before
averaging over the helicities of the particles in initiate state
and summing over the helicities of the particles in final state.

The massless spinor technique may, in fact, be applied to the case
where massive fermion(s) and non-Abelian gauge field(s) are involved
in the concerned process,
if a suitable `generalization' could be done together with
some further rearrangements. Our present subprocess
contains non-Abelian gluons and massive fermions. Thus, in order to
make use of the massless spinor technique for obtaining a compact result,
we need a suitable generalization with appropriate rearrangements. We will
describe the procedure here below. Our strategy for the generalization
is to convert the problem into an equivalent `massless' one and to
extend the `symmetries' as much as possible. Then we try to
apply the symmetries of the converted amplitude and the helicity
technique to the present problem so as to make the programme
compact. According to pQCD, at order $\alpha_s^4$ there are 36
Feynman diagrams for the `hard' subprocess $gg\to B_c+\bar{c}+b$.
To extend the symmetries for the amplitude corresponding to the 36
diagrams, we neither consider the color factors nor distinguish
the flavors of the fermion lines at the moment. Then, these
diagrams may be grouped into only a few typical ones according to
the fermion lines and the structures of the contained
$\gamma$-matrices on the lines in the Feynman diagrams, because of
the Feynman diagram symmetries. To apply the symmetries in writing
up the program, we first focus on the numerator of the amplitude
related to each typical fermion line, and deal with the
$\gamma$-matrices precisely. We then, having suitable
four-momentum set for the numerator and denominator, respectively
implement proper numerator factors for the fermion lines: color
factors, suitable denominator and spinors (corresponding to the
external lines) $etc$. Finally we obtain an exact and full typical
fermion line, which appears in Feynman diagrams. When all kinds of
typical fermion line factors, factors for external lines of gluons
and gluon propagators are `assembled', then the full term,
corresponding to the Feynman diagram of the amplitude, is
achieved. The resulting program is indeed
very compact and potentially reduces the execution time
significantly. In general, to write an amplitude according to
helicities, each massive fermion line should be decomposed into
two light-like spinor lines, but this is not
unique\cite{cal1,cal2,zdl,aes,ld,ht}. Here we use the identity
\(\slash\!\!\!k=|k+\rangle\langle k+|+|k-\rangle\langle k-|\) ($k$
is a light-like four-momentum) to simplify the fermion (quark)
lines by  spinor products.

To verify that the program BCVEGPY is correct, we check it by
taking the same parameters as in Ref.\cite{prod2}, compute the
cross-section of the subprocess $gg\to B_c(B_c^*)+\bar{c}+b$ by
integrating out the unobserved variables numerically, and compare
the obtained results with those in Ref.\cite{prod2} carefully.
Since BCVEGPY is based on helicity techniques which is totally
different from those in Ref.\cite{prod2}, the comparison is a
very good check both for BCVEGPY and for that in Ref.\cite{prod2}.

Since the mass difference between $B_c[1,^1S_0]$ and
$B_c^*[1,^3S_1]$ is small, the $B^*_c$ decays through an
electromagnetic transition (M1) $B_c^*\to B_c+\gamma$ with an
almost 100\% branching ratio and the photon in the decay is quite
soft, so the production of $B_c^*$ can be considered as an
additional source of $B_c$ production (with an additional soft
photon). Furthermore, since $J^{P}=0^-$ for $B_c$, the
polarizations of $b$ and/or $\bar{c}$-jets will be lost during
their hadronization, so the polarization effects including those
of $b$ and/or $\bar{c}$-jets in the production of $B_c$ are
therefore not interesting, so BCVEGPY program calculates the
process for the un-polarized case only, although theoretically it
can work out certain polarization effects due to $B^*$ and/or $b$
and/or $\bar{c}$-jets.

We write the BCVEGPY package following the format of
PYTHIA\cite{pythia} so that the generator could be easily adapted
into the PYTHIA environment. In this way our generator can be used
for generating complete events. To increase the Monte Carlo
simulation efficiency for high dimensional phase space
integration, we set up a switch in BCVEGPY to choose if the VEGAS
program for obtaining the sampling importance function is used or
not. We also use several parameters, such as for the maximum
differential cross sections etc, to meet the needs for the
initializations of PYTHIA.

This paper is organized as follows. In Section II we show how to
extend the symmetries by: focussing the $\gamma$-matrix strings of
fermion lines in Feynman diagrams, disregarding color- and numeral
factors $etc$; decomposing the three- and four-gluon coupling
vertices; grouping the decomposed diagrams; establishing the
typical one in each group; applying the symmetries and helicity
techniques to the problem. In Section III we outline the structure
of BCVEGPY, explain how to use the programme and test (check) the
programme as stated in the Introduction. Section IV summarizes the
conclusions and future prospects. The details about the helicity
functions for the amplitude, polarization vector for
$B_c^*[^3S_1]$, routines and functions for the helicity amplitude
are described in the Appendices.

\section{The hard subprocess}

Based on the factorization of perturbative QCD (pQCD), the hard
subprocess play the key role of the $B_c$ generator. In hadron
collisions at high energies, the subprocess gluon-gluon fusion
$gg\rightarrow B_c(B^*_c) +b+\bar{c}$ is the dominant one in hadronic
production of $B_c(B_c^*)$ mesons\footnote{According to the
non-relativistic QCD (NRQCD), although there may be
non-color-singlet components for the physical $B_c$ meson,
with such small components one cannot find a
subprocess in an order lower in pQCD for the $B_c$ hadronic
production.
This is contrary to the case of $J/\psi$ production ($J/\psi$ has a hidden
flavor, while the $B_c$ has explicit flavors).
Therefore it is certain that the subprocess is
dominant.}.  In the following sub-sections, we will show how to
deal with this subprocess.

\subsection{The amplitude for
$gg\rightarrow B_c(B^*_c) +b+\bar{c}$}

At the lowest order $\alpha_s^4$, there are totally 36 Feynman
diagrams as shown in Figs.~1-5 for the gluon-gluon fusion process
$gg\rightarrow B_c +b+\bar{c}$, so accordingly there are 36 terms
in the production amplitude. As stated in the
Introduction, we group the Feynman diagrams into five sets
according to the structure of the fermion lines. Here there are
five subsets as indicated in Figs.~1-5. To write the amplitude
corresponding to the Feynman diagrams explicitly, let us first
focus on writing the color structures of the diagrams separately.
In fact there are five independent color factors, $C_{1ij}$,
$C_{2ij}$, $C_{3ij}$, $C_{4ij}$ and $C_{5ij}$, where $i,j=1,2,3$
are the color indices of the quarks $\bar{c}$ and $b$
respectively. They are
\begin{eqnarray}
C_{1ij}&=&(T^{c}T^{c}T^{a}T^{b})_{ij}=
\frac{N^{2}-1}{2N}(T^{a}T^{b})_{ij}\,,\nonumber\\
C_{2ij}&=&(T^{c}T^{c}T^{b}T^{a})_{ij}=
\frac{N^{2}-1}{2N}(T^{b}T^{a})_{ij}\,,\nonumber\\
C_{3ij}&=&(T^{c}T^{a}T^{c}T^{b})_{ij}=
\frac{-1}{2N}(T^{a}T^{b})_{ij}\,,\nonumber\\
C_{4ij}&=&(T^{c}T^{b}T^{c}T^{a})_{ij}=
\frac{-1}{2N}(T^{b}T^{a})_{ij}\,,\nonumber\\
C_{5ij}&=&-\frac{1}{2}\delta_{ij}Tr[T^{a}T^{b}]\,.
\end{eqnarray}
To make the final amplitude compact, we introduce two extra color
factors, $C^{\prime}_{5ij}$ and $C^{\prime}_{6ij}$, which satisfy
\begin{eqnarray}
C^{\prime}_{5ij}&=&C_{3ij}-C_{5ij}=(T^{c}T^{a}T^{b}T^{c})_{ij}=
                   \frac{1}{2}\delta_{ij}Tr[T^{a}T^{b}]-
                   \frac{1}{2N}(T^{a}T^{b})_{ij}\,,\nonumber\\
C^{\prime}_{6ij}&=&C_{4ij}-C_{6ij}=(T^{c}T^{b}T^{a}T^{c})_{ij}=
                   \frac{1}{2}\delta_{ij}Tr[T^{b}T^{a}]-
                   \frac{1}{2N}(T^{b}T^{a})_{ij}\,.
\end{eqnarray}
All the color factors in the subprocess can be expressed by the
above five independent color factors, $i.e.$ in the amplitude all
the color factors may be written in terms of these five
explicitly. To obtain the desired result, the general commutation
relation for the color factors
\begin{equation}
[T_{a},T_{b}]=i f_{abc} T_{c}
\end{equation}
has been applied to three- or four-gluon vertice of the Feynman
diagrams, where $f_{abc}$ is the antisymmetric $SU(3)$ structure
constant.

The terms of the amplitude corresponding to the grouped Feynman
diagrams are written as below.

\begin{figure}
\setlength{\unitlength}{1mm}
\begin{picture}(80,60)(30,30)
\put(-10,-25){\includegraphics{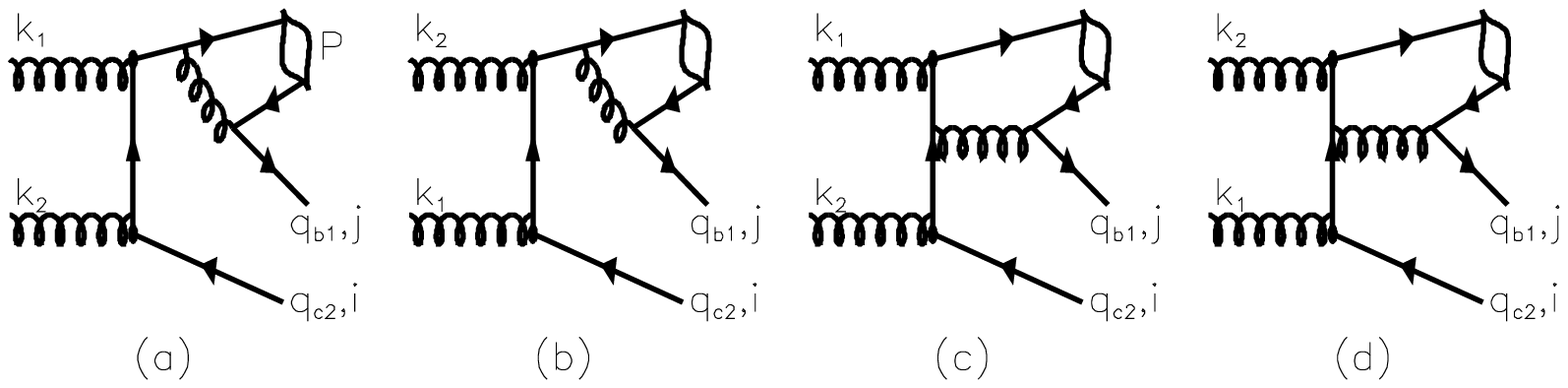}}
\put(-10,-50) {\includegraphics{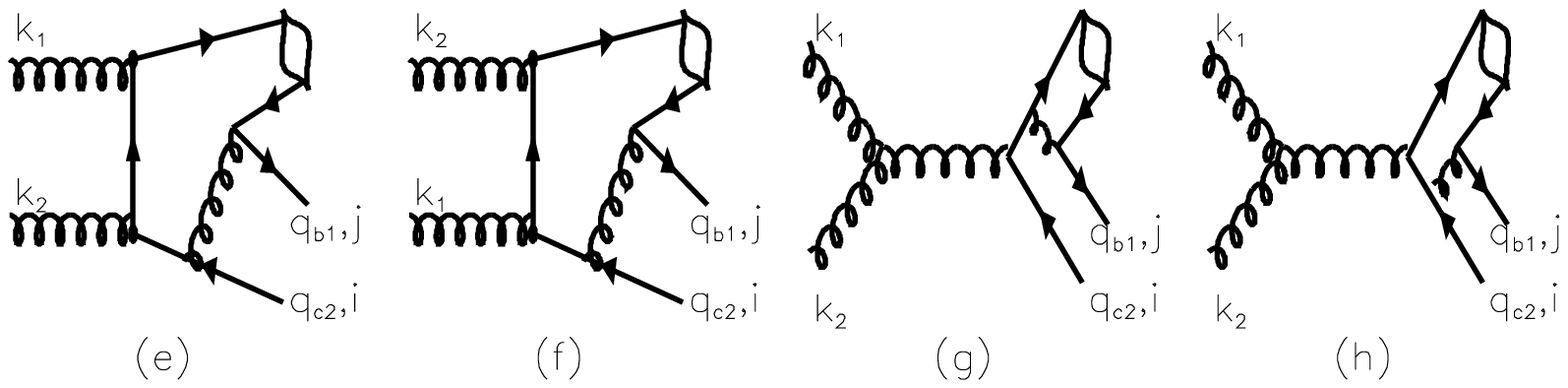}}
\end{picture}
\caption{ Feynman diagrams that can be directly grouped into the
$cc$ subset. Here $i$ and $j$ are the color indices of $\bar{c}$
and $b$ respectively.} \label{cc1}
\end{figure}

The first group,  Fig.\ref{cc1}:
\begin{eqnarray}
M_{1a}&=& C_{2ij}\bar{u}_{s}(q_{b1})i\int\frac{d^{4}q}{(2\pi)^{4}}
          \left\{ \gamma_{\delta} \frac{\bar{\chi}_{P}(q)}{(q_{b1}+q_{b2})^{2}}
           \gamma_{\delta}\frac{\slash\!\!\! k_{1}+\slash\!\!\! k_{2}-
           \slash\!\!\!q_{c2}+m_{c}}{(k_{1}+k_{2}-q_{c2})^{2}-m^{2}_{c}}
           \slash\!\!\! \epsilon_{1}^{\lambda_{1}}\cdot\right.\nonumber\\
       & & \left.\frac{\slash\!\!\! k_{2}-
           \slash\!\!\! q_{c2}+m_{c}}{(k_{2}-q_{c2})^{2}-m^{2}_{c}}
           \slash\!\!\!\epsilon_{2}^{\lambda_{2}} \right\}
           v_{s^{\prime}}(q_{c2})\;, \nonumber \\
M_{1b}&=& C_{1ij}\bar{u}_{s}(q_{b1})i\int\frac{d^{4}q}{(2\pi)^{4}}\left
           \{ \gamma_{\delta} \frac{\bar{\chi}_{P}(q)}{(q_{b1}+q_{b2})^{2}}
           \gamma_{\delta}\frac{\slash\!\!\! k_{1}+\slash\!\!\! k_{2}-
           \slash\!\!\!q_{c2}+m_{c}}{(k_{1}+k_{2}-q_{c2})^{2}-m^{2}_{c}}
           \slash\!\!\! \epsilon_{2}^{\lambda_{2}}\cdot\right.\nonumber\\
       & &\left.\frac{\slash\!\!\! k_{1}-
           \slash\!\!\! q_{c2}+m_{c}}{(k_{1}-q_{c2})^{2}-m^{2}_{c}}
           \slash\!\!\!\epsilon_{1}^{\lambda_{1}}\right\}
           v_{s^{\prime}}(q_{c2})\;,\nonumber \\
M_{1c}&=& C_{4ij}\bar{u}_{s}(q_{b1})i\int\frac{d^{4}q}{(2\pi)^{4}}\left
           \{ \gamma_{\delta} \frac{\bar{\chi}_{P}(q)}{(q_{b1}+q_{b2})^{2}}
           \slash\!\!\!\epsilon_{1}^{\lambda_{1}}\frac{\slash\!\!\! q_{c1}-
           \slash\!\!\! k_{1}+m_{c}}{(q_{c1}-k_{1})^{2}-m^{2}_{c}}
           \gamma_{\delta}\cdot\right.\nonumber\\
       & &\left. \frac{\slash\!\!\! k_{2}-\slash\!\!\! q_{c2}+m_{c}}
           {(k_{2}-q_{c2})^{2}-m^{2}_{c}}\slash\!\!\! \epsilon_{2}^{\lambda_{2}}
           \right\}v_{s^{\prime}}(q_{c2})\;,\nonumber \\
M_{1d}&=& C_{3ij}\bar{u}_{s}(q_{b1})i\int\frac{d^{4}q}{(2\pi)^{4}}\left
           \{ \gamma_{\delta} \frac{\bar{\chi}_{P}(q)}{(q_{b1}+q_{b2})^{2}}
           \slash\!\!\!\epsilon_{2}^{\lambda_{2}}\frac{\slash\!\!\! q_{c1}-
           \slash\!\!\! k_{2}+m_{c}}{(q_{c1}-k_{2})^{2}-m^{2}_{c}}
           \gamma_{\delta}\cdot\right.\nonumber\\
       & &\left.\frac{\slash\!\!\! k_{1}-\slash\!\!\! q_{c2}+m_{c}}
           {(k_{1}-q_{c2})^{2}-m^{2}_{c}}\slash\!\!\! \epsilon_{1}^{\lambda_{1}}
           \right\}v_{s^{\prime}}(q_{c2})\;,\nonumber \\
M_{1e}&=& C^{\prime}_{6ij}\bar{u}_{s}(q_{b1})i\int\frac{d^{4}q}{(2\pi)^{4}}\left
           \{ \gamma_{\delta} \frac{\bar{\chi}_{P}(q)}{(q_{b1}+q_{b2})^{2}}
           \slash\!\!\!\epsilon_{1}^{\lambda_{1}}\frac{\slash\!\!\! q_{c1}-
           \slash\!\!\! k_{1}+m_{c}}{(q_{c1}-k_{1})^{2}-m^{2}_{c}}
           \slash\!\!\! \epsilon_{2}^{\lambda_{2}}\cdot\right.\nonumber\\
      & &\left.\frac{\slash\!\!\! q_{c1}
           -\slash\!\!\! k_{1}-\slash\!\!\! k_{2}+m_{c}}
           {(q_{c1}-k_{1}-k_{2})^{2}-m^{2}_{c}}\gamma_{\delta}
           \right\}v_{s^{\prime}}(q_{c2})\;,\nonumber \\
M_{1f}&=& C^{\prime}_{5ij}\bar{u}_{s}(q_{b1})i\int\frac{d^{4}q}{(2\pi)^{4}}\left
           \{ \gamma_{\delta} \frac{\bar{\chi}_{P}(q)}{(q_{b1}+q_{b2})^{2}}
           \slash\!\!\!\epsilon_{2}^{\lambda_{2}}\frac{\slash\!\!\! q_{c1}-
           \slash\!\!\! k_{2}+m_{c}}{(q_{c1}-k_{2})^{2}-m^{2}_{c}}
           \slash\!\!\! \epsilon_{1}^{\lambda_{1}}\cdot\right.\nonumber\\
      & &\left.\frac{\slash\!\!\! q_{c1}
           -\slash\!\!\! k_{1}-\slash\!\!\! k_{2}+m_{c}}
           {(q_{c1}-k_{1}-k_{2})^{2}-m^{2}_{c}}\gamma_{\delta}
           \right\}v_{s^{\prime}}(q_{c2})\;,\nonumber \\
M_{1g}&=&(C_{2ij}-C_{1ij})\bar{u}_{s}(q_{b1})i\int\frac{d^{4}q}{(2\pi)^{4}}
           \left\{\gamma_{\delta} \frac{\bar{\chi}_{P}(q)}{(q_{b1}+q_{b2})^{2}}
           \gamma_{\delta}\frac{\slash\!\!\! k_{1}+\slash\!\!\! k_{2}-
           \slash\!\!\! q_{c2}+m_{c}}{(k_{1}+k_{2}-q_{c2})^{2}-m^{2}_{c}}
            \,\cdot \right. \nonumber \\
      & &  \frac{\gamma_{\alpha}\epsilon_{1\mu}^{\lambda_{1}}
           \epsilon_{2\nu}^{\lambda_{2}}}{(k_{1}+k_{2})^{2}}
           \Bigl((k_{1}-k_{2})_{\alpha}g_{\mu\nu}
         +(2k_{2}+k_{1})_{\mu}g_{\nu\alpha}+(-2k_{1}-k_{2})_{\nu}g_{\mu\alpha}\Bigr)
           \Biggr\}v_{s^{\prime}}(q_{c2})\;,\nonumber \\
M_{1h}&=&(C^{\prime}_{6ij}-C^{\prime}_{5ij})\bar{u}_{s}(q_{b1})i\int\frac{d^{4}q}{(2\pi)^{4}}
           \left\{\gamma_{\delta} \frac{\bar{\chi}_{P}(q)}{(q_{b1}+q_{b2})^{2}}
           \frac{\gamma_{\alpha}\epsilon_{1\mu}^{\lambda_{1}}
           \epsilon_{2\nu}^{\lambda_{2}}}{(k_{1}+k_{2})^{2}}
           \frac{\slash\!\!\! q_{c1}-\slash\!\!\! k_{1}-\slash\!\!\! k_{2}
           +m_{c}}{(k_{1}+k_{2}-q_{c1})^{2}-m^{2}_{c}} \,\cdot \right. \nonumber \\
      & &  \gamma_{\delta}\Bigl((k_{1}-k_{2})_{\alpha}g_{\mu\nu}
       +(2k_{2}+k_{1})_{\mu}g_{\nu\alpha}+(-2k_{1}-k_{2})_{\nu}g_{\mu\alpha}\Bigr)
           \Biggr\}v_{s^{\prime}}(q_{c2})\;.\nonumber \\
\label{Mocc8}
\end{eqnarray}

\begin{figure}
\setlength{\unitlength}{1mm}
\begin{picture}(80,60)(30,30)
\put(-10,-25) {\includegraphics{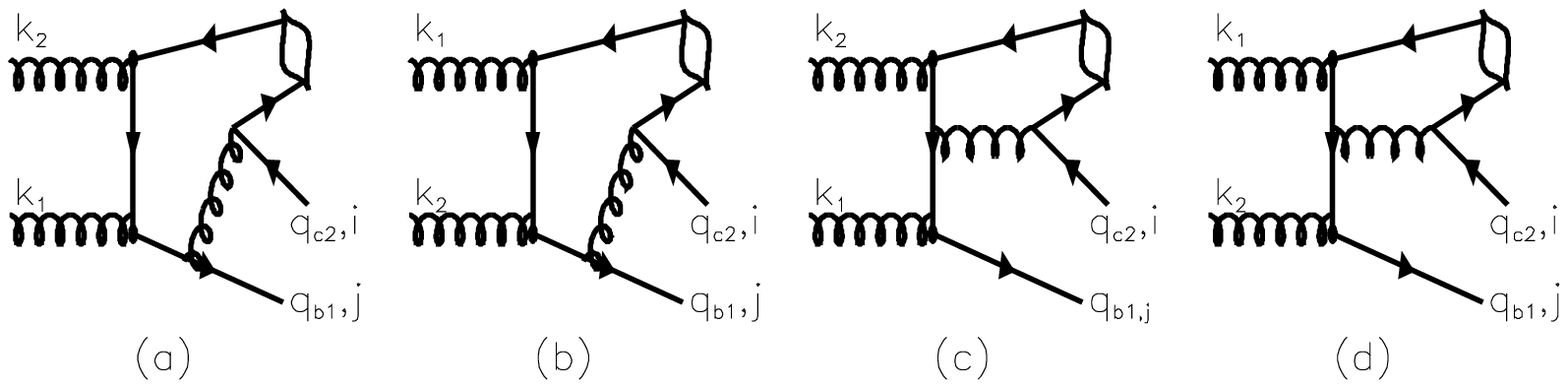}}
\put(-10,-50) {\includegraphics{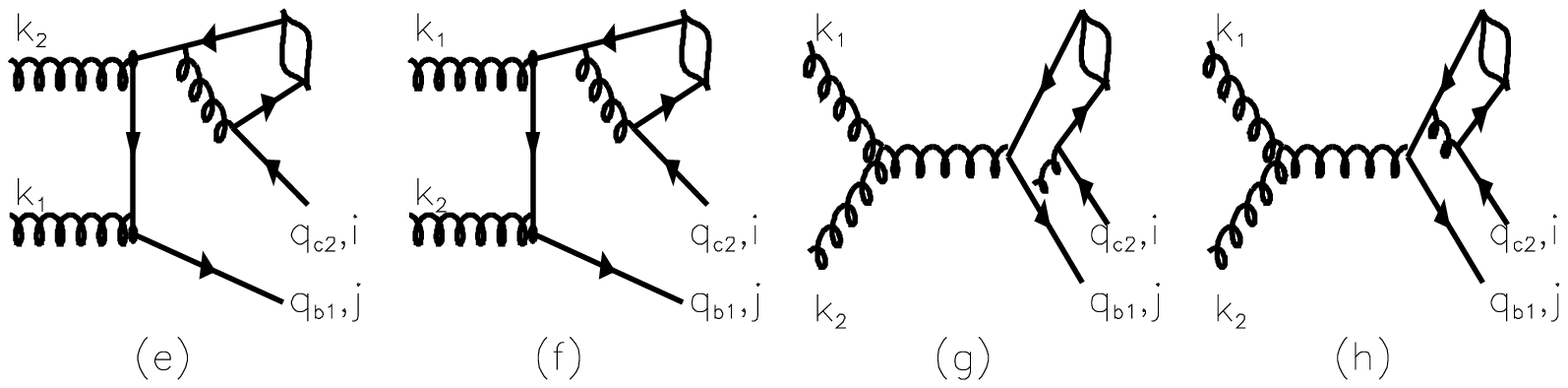}}
\end{picture}
\caption{ Feynman diagrams that can be directly grouped into the
$bb$ subset. Here $i$ and $j$ are the color indices of $\bar{c}$
and $b$ respectively.} \label{bb1}
\end{figure}

The second group, Fig.\ref{bb1}:
\begin{eqnarray}
M_{2a}&=&
(C^{\prime}_{6ij})\bar{u}_{s}(q_{b1})i\int\frac{d^{4}q}{(2\pi)^{4}}\left
           \{ \gamma_{\delta}\frac{\slash\!\!\! k_{1}+\slash\!\!\! k_{2}-
           \slash\!\!\! q_{b2}+m_{b}}{(k_{1}+k_{2}-q_{b2})^{2}-m^{2}_{b}}
           \slash\!\!\! \epsilon_{1}^{\lambda_{1}}\frac{\slash\!\!\! k_{2}-
           \slash\!\!\! q_{b2}+m_{b}}{(k_{2}-q_{b2})^{2}-m^{2}_{b}}
           \slash\!\!\!\epsilon_{2}^{\lambda_{2}}\cdot\right.\nonumber\\
       & &\left.\frac{\bar{\chi}_{P}(q)}{(q_{c1}+q_{c2})^{2}}
           \gamma_{\delta}\right\}v_{s^{\prime}}(q_{c2}) \;, \nonumber \\
M_{2b}&=& (C^{\prime}_{5ij})\bar{u}_{s}(q_{b1})i\int\frac{d^{4}q}{(2\pi)^{4}}\left
           \{ \gamma_{\delta}\frac{\slash\!\!\! k_{1}+\slash\!\!\! k_{2}-
           \slash\!\!\! q_{b2}+m_{b}}{(k_{1}+k_{2}-q_{b2})^{2}-m^{2}_{b}}
           \slash\!\!\! \epsilon_{2}^{\lambda_{2}}\frac{\slash\!\!\! k_{1}-
           \slash\!\!\! q_{b2}+m_{b}}{(k_{1}-q_{b2})^{2}-m^{2}_{b}}
           \slash\!\!\!\epsilon_{1}^{\lambda_{1}}\cdot\right.\nonumber\\
      & & \left. \frac{\bar{\chi}_{P}(q)}{(q_{c1}+q_{c2})^{2}}
           \gamma_{\delta}\right\}v_{s^{\prime}}(q_{c2}) \;, \nonumber \\
M_{2c}&=& (C_{4ij})\bar{u}_{s}(q_{b1})i\int\frac{d^{4}q}{(2\pi)^{4}}\left
           \{\slash\!\!\!\epsilon_{1}^{\lambda_{1}}\frac{\slash\!\!\! q_{b1}-
           \slash\!\!\! k_{1}+m_{b}}{(q_{b1}-k_{1})^{2}-m^{2}_{b}}
           \gamma_{\delta}\frac{\slash\!\!\! k_{2}-\slash\!\!\! q_{b2}+m_{b}}
           {(k_{2}-q_{b2})^{2}-m^{2}_{b}}\slash\!\!\! \epsilon_{2}^{\lambda_{2}}
           \cdot\right.\nonumber\\
      & &\left.\frac{\bar{\chi}_{P}(q)}{(q_{c1}+q_{c2})^{2}} \gamma_{\delta}
           \right\}v_{s^{\prime}}(q_{c2}) \;, \nonumber \\
M_{2d}&=& (C_{3ij})\bar{u}_{s}(q_{b1})i\int\frac{d^{4}q}{(2\pi)^{4}}\left
           \{\slash\!\!\!\epsilon_{2}^{\lambda_{2}}\frac{\slash\!\!\! q_{b1}-
           \slash\!\!\! k_{2}+m_{b}}{(q_{b1}-k_{2})^{2}-m^{2}_{b}}
           \gamma_{\delta}\frac{\slash\!\!\! k_{1}-\slash\!\!\! q_{b2}+m_{b}}
           {(k_{1}-q_{b2})^{2}-m^{2}_{b}}\slash\!\!\! \epsilon_{1}^{\lambda_{1}}
           \cdot\right.\nonumber\\
      & &  \left.\frac{\bar{\chi}_{P}(q)}{(q_{c1}+q_{c2})^{2}} \gamma_{\delta}
           \right\}v_{s^{\prime}}(q_{c2}) \;, \nonumber \\
M_{2e}&=& (C_{2ij})\bar{u}_{s}(q_{b1})i\int\frac{d^{4}q}{(2\pi)^{4}}\left
           \{\slash\!\!\!\epsilon_{1}^{\lambda_{1}}\frac{\slash\!\!\! q_{b1}-
           \slash\!\!\! k_{1}+m_{b}}{(q_{b1}-k_{1})^{2}-m^{2}_{b}}
           \slash\!\!\! \epsilon_{2}^{\lambda_{2}}\frac{\slash\!\!\! q_{b1}
           -\slash\!\!\! k_{1}-\slash\!\!\! k_{2}+m_{b}}
           {(q_{b1}-k_{1}-k_{2})^{2}-m^{2}_{b}}\gamma_{\delta}\cdot\right.\nonumber\\
      & &\left.\frac{\bar{\chi}_{P}(q)}{(q_{c1}+q_{c2})^{2}}\gamma_{\delta}
           \right\}v_{s^{\prime}}(q_{c2}) \;, \nonumber \\
M_{2f}&=& (C_{1ij})\bar{u}_{s}(q_{b1})i\int\frac{d^{4}q}{(2\pi)^{4}}\left
           \{\slash\!\!\!\epsilon_{2}^{\lambda_{2}}\frac{\slash\!\!\! q_{b1}-
           \slash\!\!\! k_{2}+m_{b}}{(q_{b1}-k_{2})^{2}-m^{2}_{b}}
           \slash\!\!\! \epsilon_{1}^{\lambda_{1}}\frac{\slash\!\!\! q_{b1}
           -\slash\!\!\! k_{1}-\slash\!\!\! k_{2}+m_{b}}
           {(q_{b1}-k_{1}-k_{2})^{2}-m^{2}_{b}}\gamma_{\delta}\cdot\right.\nonumber\\
      & &\left.\frac{\bar{\chi}_{P}(q)}{(q_{c1}+q_{c2})^{2}}\gamma_{\delta}
           \right\}v_{s^{\prime}}(q_{c2}) \;, \nonumber \\
M_{2g}&=&(C^{\prime}_{6ij}-C^{\prime}_{5ij})\bar{u}_{s}(q_{b1})i\int\frac{d^{4}q}{(2\pi)^{4}}
           \Biggl\{\gamma_{\delta} \frac{\slash\!\!\! k_{1}+\slash\!\!\! k_{2}-
           \slash\!\!\! q_{b2}+m_{b}}{(k_{1}+k_{2}-q_{b2})^{2}-m^{2}_{b}}
           \frac{\gamma_{\alpha}\epsilon_{1\mu}^{\lambda_{1}}
           \epsilon_{2\nu}^{\lambda_{2}}}{(k_{1}+k_{2})^{2}}\,\cdot \nonumber \\
         & &  \Bigl((k_{1}-k_{2})_{\alpha}g_{\mu\nu}+(2k_{2}+k_{1})_{\mu}g_{\nu\alpha}
        +(-2k_{1}-k_{2})_{\nu}g_{\mu\alpha}\Bigr)
           \frac{\bar{\chi}_{P}(q)}{(q_{c1}+q_{c2})^{2}}
           \gamma_{\delta}\Biggr\}v_{s^{\prime}}(q_{c2}) \;, \nonumber \\
M_{2h}&=&(C_{2ij}-C_{1ij})\bar{u}_{s}(q_{b1})i\int\frac{d^{4}q}{(2\pi)^{4}}
           \Biggl\{\frac{\gamma_{\alpha}\epsilon_{1\mu}^{\lambda_{1}}
           \epsilon_{2\nu}^{\lambda_{2}}}{(k_{1}+k_{2})^{2}}
           \frac{\slash\!\!\! q_{b1}-\slash\!\!\! k_{1}-\slash\!\!\! k_{2}+m_{b}}
           {(k_{1}+k_{2}-q_{b1})^{2}-m^{2}_{b}}\gamma_{\delta}\,\cdot \nonumber \\
          & &   \frac{\bar{\chi}_{P}(q)}
           {(q_{c1}+q_{c2})^{2}}\gamma_{\delta} \Bigl((k_{1}-k_{2})_{\alpha}g_{\mu\nu}
      +(2k_{2}+k_{1})_{\mu}g_{\nu\alpha}
           +(-2k_{1}-k_{2})g_{\mu\alpha}\Bigr)\Biggr\}v_{s^{\prime}}(q_{c2})\,.
\label{Mobb8}
\end{eqnarray}

\begin{figure}
\setlength{\unitlength}{1mm}
\begin{picture}(80,60)(30,30)
\put(-10,-25) {\includegraphics{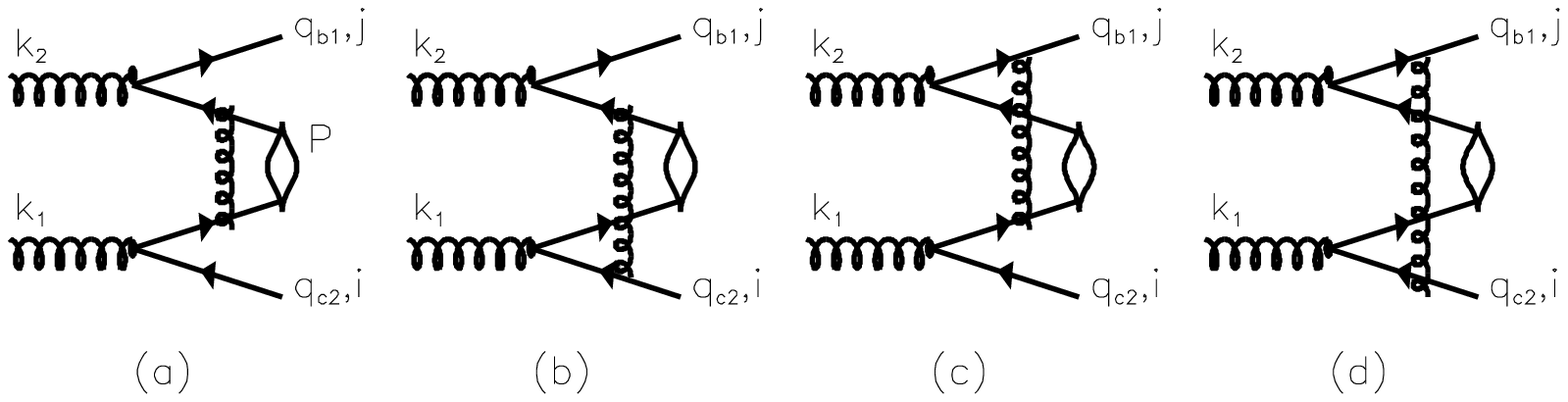}}
\put(-10,-50) {\includegraphics{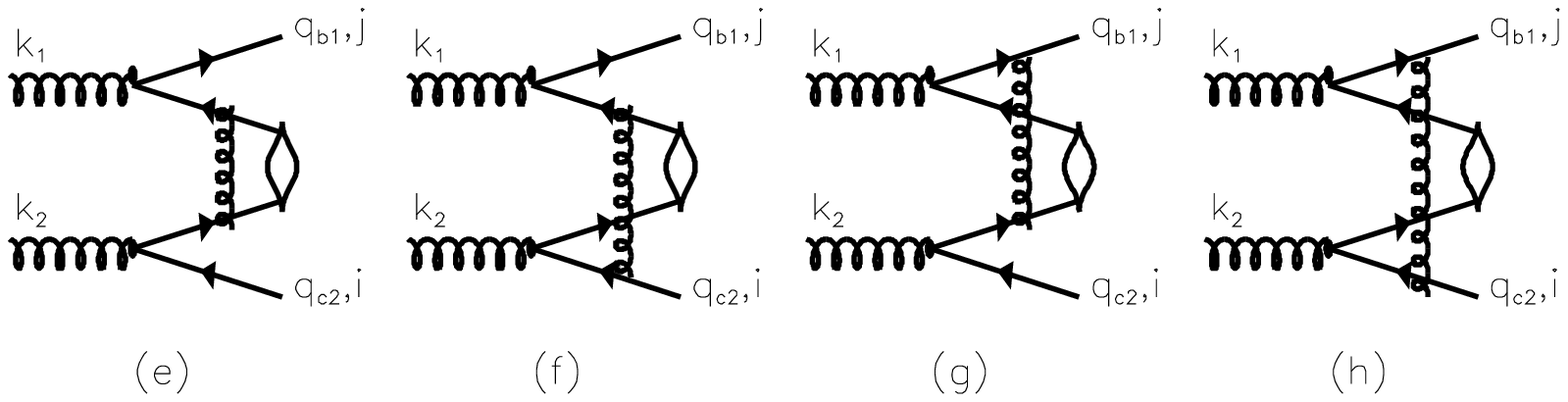}}
\end{picture}
\caption{ Feynman diagrams that can be directly grouped into the
$cb$ or $bc$ subsets, where the first four diagrams belong to the
cb subset, while the last four belong to the bc subset. Here $i$
and $j$ are the color indices of $\bar{c}$ and $b$ respectively.}
\label{bc1}
\end{figure}

The third group, Fig.\ref{bc1}:
\begin{eqnarray}
M_{3a}&=&
C_{1ij}\bar{u}_{s}(q_{b1})\slash\!\!\!\epsilon_{2}^{\lambda_{2}}
           \frac{\slash\!\!\! q_{b1}-\slash\!\!\!k_{2}+m_{b}}
           {(q_{b1}-k_{2})^{2}-m^{2}_{b}}i\int\frac{d^{4}q}{(2\pi)^{4}}
           \gamma_{\delta}\frac{\bar{\chi}_{P}(q)}{(k_{1}-q_{c1}-q_{c2})^{2}}
           \gamma_{\delta}\frac{\slash\!\!\! k_{1}-\slash\!\!\! q_{c2}+m_{c}}
           {(k_{1}-q_{c2})^{2}-m^{2}_{c}}\cdot\nonumber\\
      & &  \slash\!\!\!\epsilon_{1}^{\lambda_{1}}
           v_{s^{\prime}}(q_{c2}) \;, \nonumber \\
M_{3b}&=& C_{3ij}\bar{u}_{s}(q_{b1})\slash\!\!\!\epsilon_{2}^{\lambda_{2}}
           \frac{\slash\!\!\! q_{b1}-\slash\!\!\!k_{2}+m_{b}}
           {(q_{b1}-k_{2})^{2}-m^{2}_{b}}i\int\frac{d^{4}q}{(2\pi)^{4}}
           \gamma_{\delta}\frac{\bar{\chi}_{P}(q)}{(k_{1}-q_{c1}-q_{c2})^{2}}
           \slash\!\!\!\epsilon_{1}^{\lambda_{1}} \frac{\slash\!\!\! q_{c1}-
           \slash\!\!\! k_{1}+m_{c}}{(q_{c1}-k_{1})^{2}-m^{2}_{c}}\cdot\nonumber\\
      & &   \gamma_{\delta}v_{s^{\prime}}(q_{c2}) \;, \nonumber\\
M_{3c}&=&
C_{3ij}\bar{u}_{s}(q_{b1})i\int\frac{d^{4}q}{(2\pi)^{4}}\gamma_{\delta}
           \frac{\slash\!\!\!k_{2}-\slash\!\!\! q_{b2}+m_{b}}
           {(k_{2}-q_{b2})^{2}-m^{2}_{b}}\slash\!\!\!\epsilon_{2}^{\lambda_{2}}
           \frac{\bar{\chi}_{P}(q)}{(k_{1}-q_{c1}-q_{c2})^{2}}
           \gamma_{\delta}\frac{\slash\!\!\! k_{1}-\slash\!\!\! q_{c2}+m_{c}}
           {(k_{1}-q_{c2})^{2}-m^{2}_{c}}\cdot\nonumber\\
      & &     \slash\!\!\!\epsilon_{1}^{\lambda_{1}}
           v_{s^{\prime}}(q_{c2}) \;, \nonumber\\
M_{3d}&=& C^{\prime}_{5ij}\bar{u}_{s}(q_{b1})i\int\frac{d^{4}q}{(2\pi)^{4}}\gamma_{\delta}
           \frac{\slash\!\!\!k_{2}-\slash\!\!\! q_{b2}+m_{b}}
           {(k_{2}-q_{b2})^{2}-m^{2}_{b}}\slash\!\!\!\epsilon_{2}^{\lambda_{2}}
           \frac{\bar{\chi}_{P}(q)}{(k_{1}-q_{c1}-q_{c2})^{2}}
           \slash\!\!\!\epsilon_{1}^{\lambda_{1}} \frac{\slash\!\!\! q_{c1}-
           \slash\!\!\! k_{1}+m_{c}}{(q_{c1}-k_{1})^{2}-m^{2}_{c}}\cdot\nonumber\\
      & &  \gamma_{\delta}v_{s^{\prime}}(q_{c2}) \;, \nonumber\\
M_{3e}&=&
C_{2ij}\bar{u}_{s}(q_{b1})\slash\!\!\!\epsilon_{1}^{\lambda_{1}}
           \frac{\slash\!\!\! q_{b1}-\slash\!\!\!k_{1}+m_{b}}
           {(q_{b1}-k_{1})^{2}-m^{2}_{b}}i\int\frac{d^{4}q}{(2\pi)^{4}}
           \gamma_{\delta}\frac{\bar{\chi}_{P}(q)}{(k_{2}-q_{c1}-q_{c2})^{2}}
           \gamma_{\delta}\frac{\slash\!\!\! k_{2}-\slash\!\!\! q_{c2}+m_{c}}
           {(k_{2}-q_{c2})^{2}-m^{2}_{c}}\cdot\nonumber\\
      & &  \slash\!\!\!\epsilon_{2}^{\lambda_{2}}
           v_{s^{\prime}}(q_{c2}) \;, \nonumber \\
M_{3f}&=& C_{4ij}\bar{u}_{s}(q_{b1})\slash\!\!\!\epsilon_{1}^{\lambda_{1}}
           \frac{\slash\!\!\! q_{b1}-\slash\!\!\!k_{1}+m_{b}}
           {(q_{b1}-k_{1})^{2}-m^{2}_{b}}i\int\frac{d^{4}q}{(2\pi)^{4}}
           \gamma_{\delta}\frac{\bar{\chi}_{P}(q)}{(k_{2}-q_{c1}-q_{c2})^{2}}
           \slash\!\!\!\epsilon_{2}^{\lambda_{2}}\frac{\slash\!\!\! q_{c1}-
           \slash\!\!\! k_{2}+m_{c}}{(q_{c1}-k_{2})^{2}-m^{2}_{c}}\cdot\nonumber\\
      & &  \gamma_{\delta}v_{s^{\prime}}(q_{c2}) \;, \nonumber \\
M_{3g}&=& C_{4ij}\bar{u}_{s}(q_{b1})i\int\frac{d^{4}q}{(2\pi)^{4}}\gamma_{\delta}
           \frac{\slash\!\!\!k_{1}-\slash\!\!\! q_{b2}+m_{b}}
           {(k_{1}-q_{b2})^{2}-m^{2}_{b}}\slash\!\!\!\epsilon_{1}^{\lambda_{1}}
           \frac{\bar{\chi}_{P}(q)}{(k_{2}-q_{c1}-q_{c2})^{2}}
           \gamma_{\delta}\frac{\slash\!\!\! k_{2}-\slash\!\!\! q_{c2}+m_{c}}
           {(k_{2}-q_{c2})^{2}-m^{2}_{c}}\cdot\nonumber\\
      & &  \slash\!\!\!\epsilon_{2}^{\lambda_{2}}
           v_{s^{\prime}}(q_{c2}) \;, \nonumber \\
M_{3h}&=& C^{\prime}_{6ij}\bar{u}_{s}(q_{b1})i\int\frac{d^{4}q}{(2\pi)^{4}}\gamma_{\delta}
           \frac{\slash\!\!\!k_{1}-\slash\!\!\! q_{b2}+m_{b}}
           {(k_{1}-q_{b2})^{2}-m^{2}_{b}}\slash\!\!\!\epsilon_{1}^{\lambda_{1}}
           \frac{\bar{\chi}_{P}(q)}{(k_{2}-q_{c1}-q_{c2})^{2}}
           \slash\!\!\!\epsilon_{2}^{\lambda_{2}} \frac{\slash\!\!\! q_{c1}-
           \slash\!\!\! k_{2}+m_{c}}{(q_{c1}-k_{2})^{2}-m^{2}_{c}}\cdot \nonumber \\
      & &  \gamma_{\delta}v_{s^{\prime}}(q_{c2}) \;.
\label{Mobc4}
\end{eqnarray}

\begin{figure}
\setlength{\unitlength}{1mm}
\begin{picture}(80,60)(30,30)
\put(-10,-25) {\includegraphics{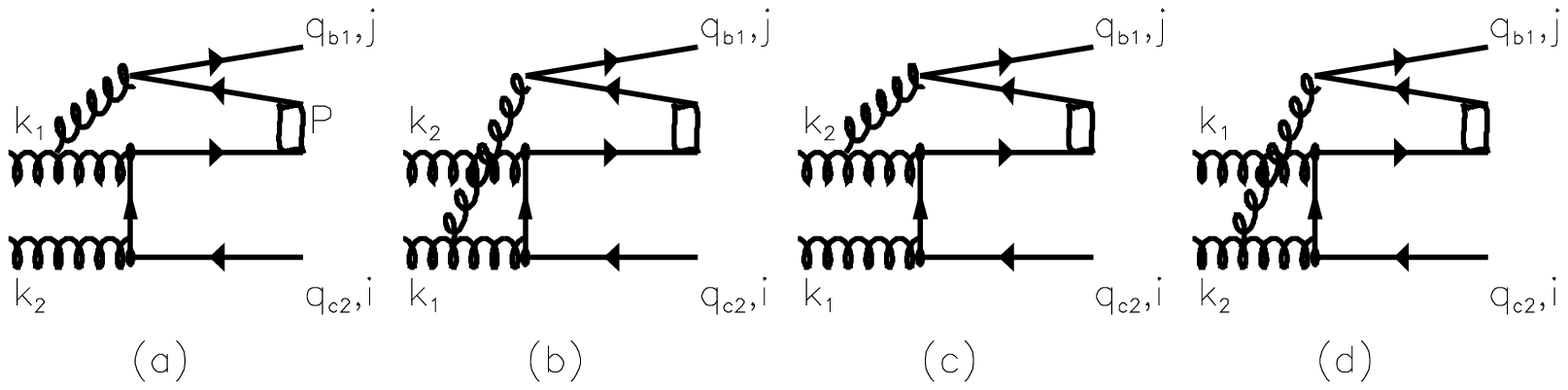}}
\put(-10,-50) {\includegraphics{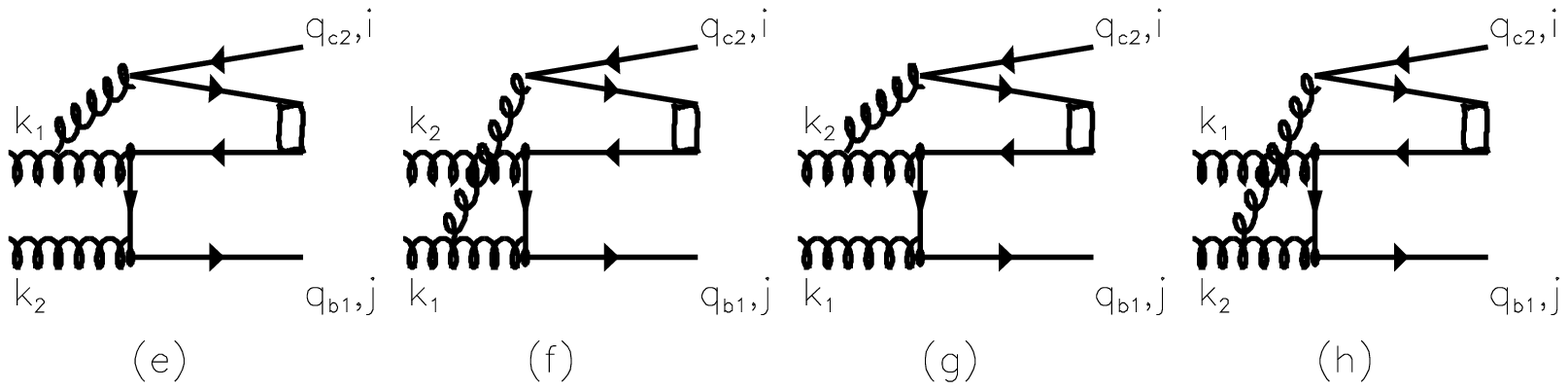}}
\end{picture}
\caption{ Feynman diagrams with only one three-gluon vertex, which
can not be directly grouped into the $cc$, $bb$, $cb$ and $bc$
subsets. Here $i$ and $j$ are the color indices of $\bar{c}$ and
$b$ respectively.} \label{obc}
\end{figure}

The fourth group, Fig.\ref{obc}:
\begin{eqnarray}
M_{4a}&=&(C_{4ij}-C_{2ij})\bar{u}_{s}(q_{b1})i\int\frac{d^{4}q}{(2\pi)^{4}}
          \gamma_{\delta}\frac{\bar{\chi}_{P}(q)}{(q_{b1}+q_{b2})^{2}}
           \frac{\gamma_{\alpha}\epsilon_{1\mu}^{\lambda_{1}}}{(k_{1}
           -q_{b1}-q_{b2})^{2}}
           \Bigl((k_{1}+q_{b1}+q_{b2})_{\alpha}g_{\mu\delta}+ \nonumber \\
       & & (k_{1}-2q_{b1}-2q_{b2})_{\mu}g_{\alpha\delta}
           +(q_{b1}+q_{b2}-2k_{1})_{\delta}g_{\mu\alpha}\Bigr)
           \frac{\slash\!\!\!k_{2}-\slash\!\!\! q_{c2}+m_{c}}
           {(k_{2}-q_{c2})^{2}-m^{2}_{c}}\slash\!\!\! \epsilon_{2}^{\lambda_{2}}
           v_{s^{\prime}}(q_{c2}) \;, \nonumber\\
M_{4b}&=&(-C_{5ij})\bar{u}_{s}(q_{b1})i\int\frac{d^{4}q}{(2\pi)^{4}}
         \gamma_{\delta}\frac{\bar{\chi}_{P}(q)}{(q_{b1}+q_{b2})^{2}}
           \slash\!\!\! \epsilon_{2}^{\lambda_{2}}
           \frac{\slash\!\!\! q_{c1}-\slash\!\!\!k_{2}+m_{c}}
           {(q_{c1}-k_{2})^{2}-m^{2}_{c}}
           \frac{\gamma_{\alpha}\epsilon_{1\mu}^{\lambda_{1}}}
           {(k_{1}-q_{b1}-q_{b2})^{2}} \cdot \nonumber \\
       & &  \Bigl((k_{1}+q_{b1}+q_{b2})_{\alpha}g_{\mu\delta}+
          (k_{1}-2q_{b1}-2q_{b2})_{\mu}g_{\alpha\delta}
           +(q_{b1}+q_{b2}-2k_{1})_{\delta}g_{\mu\alpha}\Bigr)
           v_{s^{\prime}}(q_{c2}) \;, \nonumber\\
M_{4c}&=&(C_{3ij}-C_{1ij})\bar{u}_{s}(q_{b1})i\int\frac{d^{4}q}{(2\pi)^{4}}\gamma_{\delta}
           \frac{\bar{\chi}_{P}(q)}{(q_{b1}+q_{b2})^{2}}
           \frac{\gamma_{\alpha}\epsilon_{2\mu}^{\lambda_{2}}}{(k_{2}-q_{b1}-q_{b2})^{2}}
           \Bigl((k_{2}+q_{b1}+q_{b2})_{\alpha}g_{\mu\delta}+ \nonumber\\
       & & (k_{2}-2q_{b1}-2q_{b2})_{\mu}g_{\alpha\delta}
           +(q_{b1}+q_{b2}-2k_{2})_{\delta}g_{\mu\alpha}\Bigr)
           \frac{\slash\!\!\!k_{1}-\slash\!\!\! q_{c2}+m_{c}}
           {(k_{1}-q_{c2})^{2}-m^{2}_{c}}\slash\!\!\! \epsilon_{1}^{\lambda_{1}}
           v_{s^{\prime}}(q_{c2}) \;, \nonumber\\
M_{4d}&=&(-C_{5ij})\bar{u}_{s}(q_{b1})i\int\frac{d^{4}q}{(2\pi)^{4}}\gamma_{\delta}
           \frac{\bar{\chi}_{P}(q)}{(q_{b1}+q_{b2})^{2}}
           \slash\!\!\! \epsilon_{1}^{\lambda_{1}}
           \frac{\slash\!\!\! q_{c1}-\slash\!\!\!k_{1}+m_{c}}
           {(q_{c1}-k_{1})^{2}-m^{2}_{c}}
           \frac{\gamma_{\alpha}\epsilon_{2\mu}^{\lambda_{2}}}
           {(k_{2}-q_{b1}-q_{b2})^{2}}\cdot \nonumber \\
       & &  \Bigl((k_{2}+q_{b1}+q_{b2})_{\alpha}g_{\mu\delta}
           +(k_{2}-2q_{b1}-2q_{b2})_{\mu}g_{\alpha\delta}
           +(q_{b1}+q_{b2}-2k_{2})_{\delta}g_{\mu\alpha}\Bigr)
           v_{s^{\prime}}(q_{c2})\;, \nonumber\\
M_{4e}&=&(C_{1ij}-C_{3ij})\bar{u}_{s}(q_{b1})i\int\frac{d^{4}q}{(2\pi)^{4}}
           \slash\!\!\! \epsilon_{2}^{\lambda_{2}}
           \frac{\slash\!\!\! q_{b1}-\slash\!\!\!k_{2}+m_{b}}
           {(q_{b1}-k_{2})^{2}-m^{2}_{b}}
           \frac{\gamma_{\alpha}\epsilon_{1\mu}^{\lambda_{1}}}{(k_{1}-q_{c1}-q_{c2})^{2}}
           \frac{\bar{\chi}_{P}(q)}{(q_{c1}+q_{c2})^{2}}\gamma_{\delta}\cdot \nonumber \\
         & &  \Bigl((k_{1}+q_{c1}+q_{c2})_{\alpha}g_{\mu\delta}
        +(k_{1}-2q_{c1}-2q_{c2})_{\mu}g_{\alpha\delta}
           +(q_{c1}+q_{c2}-2k_{1})_{\delta}g_{\mu\alpha}\Bigr)
           v_{s^{\prime}}(q_{c2}) \;, \nonumber\\
M_{4f}&=&(C_{5ij})\bar{u}_{s}(q_{b1})i\int\frac{d^{4}q}{(2\pi)^{4}}
           \frac{\gamma_{\alpha}\epsilon_{1\mu}^{\lambda_{1}}}{(k_{1}-q_{c1}-q_{c2})^{2}}
           \frac{\slash\!\!\!k_{2}-\slash\!\!\! q_{b2}+m_{b}}{(k_{2}-q_{b2})^{2}-m^{2}_{b}}
           \slash\!\!\! \epsilon_{2}^{\lambda_{2}}
           \frac{\bar{\chi}_{P}(q)}{(q_{c1}+q_{c2})^{2}}\gamma_{\delta}\cdot \nonumber \\
         & &  \Bigl((k_{1}+q_{c1}+q_{c2})_{\alpha}g_{\mu\delta}
        +(k_{1}-2q_{c1}-2q_{c2})_{\mu}g_{\alpha\delta}
           +(q_{c1}+q_{c2}-2k_{1})_{\delta}g_{\mu\alpha}\Bigr)
           v_{s^{\prime}}(q_{c2}) \;, \nonumber\\
M_{4g}&=&(C_{2ij}-C_{4ij})\bar{u}_{s}(q_{b1})i\int\frac{d^{4}q}{(2\pi)^{4}}
           \slash\!\!\! \epsilon_{1}^{\lambda_{1}}
           \frac{\slash\!\!\! q_{b1}-\slash\!\!\!k_{1}+m_{b}}
           {(q_{b1}-k_{1})^{2}-m^{2}_{b}}
           \frac{\gamma_{\alpha}\epsilon_{2\mu}^{\lambda_{2}}}{(k_{2}-q_{c1}-q_{c2})^{2}}
           \frac{\bar{\chi}_{P}(q)}{(q_{c1}+q_{c2})^{2}}\gamma_{\delta}\cdot \nonumber \\
          & &   \Bigl((k_{2}+q_{c1}+q_{c2})_{\alpha}g_{\mu\delta}
       +(k_{2}-2q_{c1}-2q_{c2})_{\mu}g_{\alpha\delta}
           +(q_{c1}+q_{c2}-2k_{2})_{\delta}g_{\mu\alpha}\Bigr)
           v_{s^{\prime}}(q_{c2})\;, \nonumber\\
M_{4h}&=&(C_{5ij})\bar{u}_{s}(q_{b1})i\int\frac{d^{4}q}{(2\pi)^{4}}
           \frac{\gamma_{\alpha}\epsilon_{2\mu}^{\lambda_{2}}}{(k_{2}-q_{c1}-q_{c2})^{2}}
           \frac{\slash\!\!\!k_{1}-\slash\!\!\! q_{b2}+m_{b}}{(k_{1}-q_{b2})^{2}-m^{2}_{b}}
           \slash\!\!\! \epsilon_{1}^{\lambda_{1}}
           \frac{\bar{\chi}_{P}(q)}{(q_{c1}+q_{c2})^{2}}\gamma_{\delta}\cdot \nonumber \\
        & &\Bigl((k_{2}+q_{c1}+q_{c2})_{\alpha}g_{\mu\delta}
        +(k_{2}-2q_{c1}-2q_{c2})_{\mu}g_{\alpha\delta}
           +(q_{c1}+q_{c2}-2k_{2})_{\delta}g_{\mu\alpha}\Bigr)
           v_{s^{\prime}}(q_{c2})\;.
\label{Mobo2}
\end{eqnarray}

\begin{figure}
\setlength{\unitlength}{1mm}
\begin{picture}(80,60)(30,30)
\put(-10,-45) {\includegraphics{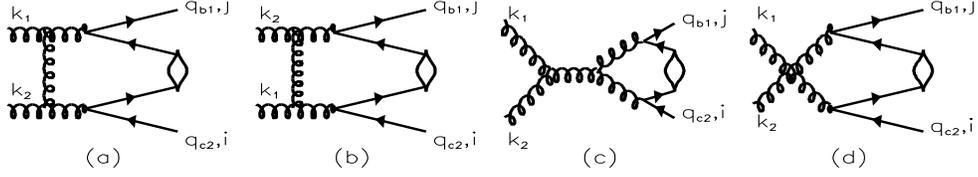}}
\end{picture}
\caption{Feynman diagrams with two three-gluon vertices or with a
four-gluon vertex, which can not be directly grouped into the
$cc$, $bb$, $cb$ and $bc$ subsets. Here $i$ and $j$ are the color
indices of $\bar{c}$ and $b$ respectively.} \label{oo1}
\end{figure}

The fifth group Fig.\ref{oo1}:
\begin{eqnarray}
M_{5a}&=&(C_{2ij}-C_{4ij}-C_{5ij})\bar{u}_{s}(q_{b1})i\int\frac{d^{4}q}{(2\pi)^{4}}
           \frac{\gamma_{\alpha}\epsilon_{1\mu}^{\lambda_{1}}}{(q_{b1}+q_{b2})^{2}}
           \frac{\bar{\chi}_{P}(q)}{(k_{2}-q_{c1}-q_{c2})^{2}}
           \frac{\gamma_{\beta}\epsilon_{2\nu}^{\lambda_{2}}}{(q_{c1}+q_{c2})^{2}}\cdot \nonumber \\
         & &  \Bigl((2k_{1}-q_{b1}-q_{b2})_{\alpha}g_{\mu\delta}
        +(2q_{b1}+2q_{b2}-k_{1})_{\mu}g_{\alpha\delta}+(-q_{b1}-q_{b2}-
           k_{1})_{\delta}g_{\mu\alpha}\Bigr)\cdot \nonumber \\
        & &\Bigl((k_{2}+q_{c1}+q_{c2})_{\delta}g_{\nu\beta}
           +(k_{2}-2q_{c1}-2q_{c2})_{\nu}g_{\beta\delta}
        +(-2k_{2}+q_{c1}+q_{c2})_{\beta}g_{\delta\nu}\Bigr)v_{s^{\prime}}(q_{c2})\;, \nonumber\\
M_{5b}&=&(C_{1ij}-C_{3ij}-C_{5ij})\bar{u}_{s}(q_{b1})i\int\frac{d^{4}q}{(2\pi)^{4}}
           \frac{\gamma_{\alpha}\epsilon_{2\mu}^{\lambda_{2}}}{(q_{b1}+q_{b2})^{2}}
           \frac{\bar{\chi}_{P}(q)}{(k_{1}-q_{c1}-q_{c2})^{2}}
           \frac{\gamma_{\beta}\epsilon_{1\nu}^{\lambda_{1}}}{(q_{c1}+q_{c2})^{2}}\cdot \nonumber\\
        & &\Bigl((2k_{2}-q_{b1}-q_{b2})_{\alpha}g_{\mu\delta}
        +(2q_{b1}+2q_{b2}-k_{2})_{\mu}g_{\alpha\delta}+(-q_{b1}-q_{b2}-
        k_{2})_{\delta}g_{\mu\alpha}\Bigr)\cdot \nonumber \\
        & &\Bigl((k_{1}+q_{c1}+q_{c2})_{\delta}g_{\nu\beta}
           +(k_{1}-2q_{c1}-2q_{c2})_{\nu}g_{\beta\delta}
        +(-2k_{1}+q_{c1}+q_{c2})_{\beta}g_{\delta\nu}\Bigr)v_{s^{\prime}}(q_{c2}) \;, \nonumber\\
M_{5c}&=& (C_{1ij}+C_{4ij}-C_{3ij}-C_{2ij})\bar{u}_{s}(q_{b1})i\int\frac{d^{4}q}{(2\pi)^{4}}
           \frac{\gamma_{\alpha}\epsilon_{1\mu}^{\lambda_{1}}}{(q_{b1}+q_{b2})^{2}}
           \frac{\bar{\chi}_{P}(q)}{(k_{1}+k_{2})^{2}}
           \frac{\gamma_{\beta}\epsilon_{2\nu}^{\lambda_{2}}}{(q_{c1}+q_{c2})^{2}}\cdot \nonumber \\
      & &  \Bigl((k_{1}+k_{2}+q_{b1}+q_{b2})_{\beta}g_{\alpha\delta}
       +(k_{1}+k_{2}-2q_{b1}-2q_{b2})_{\delta}g_{\alpha\beta}+(q_{b1}+q_{b2}-
           2k_{1}-2k_{2})_{\alpha}g_{\beta\delta}\Bigr)\cdot \nonumber \\
     & &\Bigl((k_{1}-k_{2})_{\delta}g_{\nu\mu}
           +(2k_{2}+k_{1})_{\mu}g_{\nu\delta}
       +(-2k_{1}-k_{2})_{\nu}g_{\delta\mu}\Bigr)v_{s^{\prime}}(q_{c2}) \;, \nonumber\\
M_{5d}&=& \bar{u}_{s}(q_{b1})i\int\frac{d^{4}q}{(2\pi)^{4}}
           \frac{\gamma_{\alpha}\epsilon_{1\mu}^{\lambda_{1}}}{(q_{b1}+q_{b2})^{2}}
           \bar{\chi}_{P}(q)
           \frac{\gamma_{\beta}\epsilon_{2\nu}^{\lambda_{2}}}{(q_{c1}+q_{c2})^{2}}\cdot \nonumber\\
      & &\Bigl((C_{2ij}+C_{3ij}-C_{1ij}-C_{4ij})
      (g_{\mu\beta}g_{\nu\alpha}-g_{\mu\alpha}g_{\nu\beta})+
      (C_{5ij}-C_{1ij}+C_{3ij})\cdot \nonumber \\
      & &(g_{\mu\nu}g_{\beta\alpha}-g_{\mu\alpha}g_{\nu\beta})+
      (C_{5ij}-C_{2ij}+C_{4ij})(g_{\mu\nu}g_{\alpha\beta}-g_{\mu\beta}g_{\nu\alpha})\Bigr)
      v_{s^{\prime}}(q_{c2})\;.
\label{Moo4}
\end{eqnarray}
In Eqs.~(4-8), $k_{1}$, $k_{2}$ and $\epsilon_{1}$, $\epsilon_{2}$
are the momenta and the polarization vectors of the gluons;
$q_{c1}$, $q_{b1}$ are the momenta of $c$ and $b$ quarks and
$q_{c2}$, $q_{b2}$ are the momenta of $\bar{c}$ and $\bar{b}$
anti-quarks, respectively. Note that in all the above equations
we have omitted the factor $g_s^4$ (the fourth power of the QCD
coupling constant), so we should consider it when
evaluating the final result. For convenience, we group the terms
(Feynman diagrams) according to the character of the gluon
attachment to the fermion lines. The details will be explained in
the next sections.

Under the non-relativistic approximation, for the weak binding
system of ($c\bar{b}$) we have
\begin{equation}
  M_{B_c}=M_{B_c^{*}}=M\simeq m_{b}+m_{c}\,,\;\;\;
  q_{c1}=\frac{m_{c}}{M}P\,,\;\;\;\, q_{b2}=\frac{m_{b}}{M}P\,,
   \label{eq:momentum}
\end{equation}
and the wave function $ \bar{\chi}_P(q) $ can be written as
\begin{equation}
\bar{\chi}_P(q)=\phi(q)\frac{1}{2\sqrt{M}}(\alpha\gamma_5 +
\beta\slash\!\!\!\epsilon(s_{z}))(\slash\!\!\!P+M)\,, \label{BSw}
\end{equation}
where $\alpha=1$, $\beta=0$ for the pseudoscalar $B_c
([1^{1}S_{0}])$ and $\alpha=0$, $\beta=1$ for the vector $B^*_c
([1^{3}S_{1}])$. The radial part of the momentum space wave function $\phi(q)$
is related to the space-time wave function at
origin by the integration:
$$ i\int\frac{d^{4}q}{(2\pi)^{4}} \phi(q) =\psi(0)\,.$$

\subsection{Motivation and basic formulae for decomposing
the gluon self-coupling vertices}

In Ref.\cite{zdl}, a method is proposed for treating the amplitude
of the processes, which contain massless fermions and non-abelian
gauge boson(s), in order to make the result compact and to avoid
numerical cancellations between very large numbers. The authors of
Ref.\cite{zdl} group the Feynman diagrams of the concerned process
into gauge-invariant subsets according to how the lines of the
gauge bosons attach to the fermion lines. They then choose
convenient gauges for the subsets independently of each other (not
a unique gauge for the whole process), and finally they obtain a
compact result, when all terms of the amplitude are written
according to the helicities of the fermions. The key point of the
approach (helicity technique) is that when all the massless
fermions are written on helicity state basis, then it is
straightforward to choose convenient gauges for all subsets. If
massive fermions are involved in the concerned process and one
wishes to use the same technique, then one needs to generalize it.
Some rearrangements, such as replacing one massive fermion by two
massless fermions $etc$, should be made, and the gauge choice for
each subset is more complicated than in the massless cases. In the
present case for the subprocess $gg\to B_c+b+\bar{c}$, a unique
gauge for the whole amplitude is more practical. Apart from the
gauge choice, we apply the techniques of Ref.~\cite{zdl} as much
as possible to make the program compact. The polarization vectors
of the gluons are replaced by $\gamma$-matrix elements of massless
fermion helicity states and the color factors are dealt with
independently from the Dirac $\gamma$-matrix strings. Furthermore,
we try to make the amplitude more symmetric by applying a
decomposition of the terms when writing the program. This is
achieved by decomposing first the terms, which contain three-
or/and four-gluon vertices, into terms without self-interactions
of gluons. Then, according to the structure of the contained
fermion lines, some of the decomposed terms are chosen as
`coordinators', which would be called as the typical ones, so that
all the other terms, referring to the Feynman diagrams, may be
`expressed' by the `typical ones'. Therefore when writing the
program, only different 'typical Feynman diagrams' need to be
written precisely, while the non-typical ones may be generated by
means of the typical ones according to the relationship
(expression). In this section, we show how to decompose the terms
in the Feynman diagrams containing three- or four-gluon
vertex(ices) into terms similar to the typical ones, disregarding
the difference in color factors, and list the results for the
typical Feynman diagrams.

First of all, let us introduce a massless fermion with an
arbitrary light-like reference momentum $q\; (q^2=0)$ and its
relevant helicity spinors ($|q_\pm\rangle =\frac{1\pm
\gamma_5}{2}|q \rangle$ and $\slash\!\!\! q|q\rangle=0$), and then
construct the requested massive fermion four-spinors $u(p)$ and
$v(p)$ with momentum $p\; (p^2=m^2)$ in terms of $|q_\pm\rangle$
as follows:
\begin{eqnarray}
u_{s}(p)&=& \frac{1}{\sqrt{2 p\cdot q}}
(\slash\!\!\!p+m) |q_{h}\rangle \,,\nonumber\\
v_{s}(p)&=& \frac{1}{\sqrt{2 p\cdot q}}
(\slash\!\!\!p-m)|q_{-h}\rangle\,, \label{massive}
\end{eqnarray}
where $s=\pm\frac{1}{2}$ is the spin of the massive spinors,
while $h=\pm $ is the helicity of the massless spinor. For
convenience we adopt the explicit form for the polarization
vectors $\epsilon^\pm$ of the gluon with momentum $k$ as in
CALKUL. When the helicity states $|k_{\pm}\rangle$ and
$|q'_{\pm}\rangle$ of two massless fermions are defined as
$|q_\pm\rangle$, but the fermions have a light-like momentum $k\; (k^2=0)$ and a
referred one $q^{\prime}\; (q'^2=0)$ respectively, the
polarization vectors $\epsilon^\pm$ of the gluon with momentum $k$
may be represented as follows:
\begin{eqnarray}
\epsilon^{+}_{\mu}(k,q^{\prime}) &=& \frac{\langle q^{\prime}_{-}|
\gamma_{\mu}|k_{-}\rangle} {\sqrt{2}\langle q^{\prime}\cdot k\rangle}\,,
\nonumber \\
\epsilon^{-}_{\mu}(k,q^{\prime}) &=& \frac{\langle
q^{\prime}_{+}|\gamma_{\mu}
|k_{+}\rangle} {\sqrt{2}\langle q^{\prime}\cdot k\rangle^{*}}\,,
\nonumber\\
\slash\!\!\!\epsilon^{+}(k,q^{\prime})&=&\frac{\sqrt{2}}{\langle
q^{\prime}\cdot k \rangle}(|k_{-}\rangle \langle
q^{\prime}_{-}|+|q^{\prime}_{+}\rangle \langle
k_{+}|)\,,\nonumber \\
\slash\!\!\!\epsilon^{-}(k,q^{\prime})&=&\frac{\sqrt{2}}{\langle
q^{\prime}\cdot k \rangle^{*}}(|k_{+}\rangle \langle
q^{\prime}_{+}|+|q^{\prime}_{-}\rangle \langle k_{-}|)\,.
\end{eqnarray}
Throughout the paper $\langle q'\cdot k \rangle$ and  $\langle
q'\cdot k \rangle^*$ denotes $\langle q'_{-}|k_{+} \rangle$ and
its complex conjugation respectively.

Because the light-like momenta $q$ and $q^{\prime}$ can be chosen
arbitrarily, we choose them to be the same as a light-like
momentum $q_{0}$, which is the reference momentum for all massive
spinors and gluon polarization vectors. We take one gauge for the
whole set of 36 Feynman diagrams, that is convenient here and
different from CALKUL where different gauges are taken for
different gauge-invariant subsets.

The gluon self-coupling vertices do not adapt well for being simplified
by using the polarization vector in calculating the
amplitude. Hence an effort to replace the part with the gluon
self-coupling of the diagrams by the so-called QED-like ones
(without gluon self-coupling) is described here. This approach is
not straightforward, due to the fact that the
gluon self-coupling diagrams with massive quarks
can not be mapped completely into the QED-like diagrams
as in the case of the massless quark condition\cite{zdl}. In the
massive quark case, some additional functions need to be
introduced. Fortunately, for the `simple' subprocess, these
`extra' functions for the three-gluon vertices are just parts
of the diagrams with four-gluon vertices. Thus the
diagrams involving four-gluon coupling
vertices just need to be decomposed according to their
color factors and there is no need to reduce their $\gamma$ matrix and
spinor factors any further.

To decompose Feynman diagrams with self-interactions into those
without any, we need to deal with some of the so-called `basic
structures'. The number of the `basic structures'
required for a specific process increases with the number of the
gluons involved in the concerned process. For the subprocess,
$gg\rightarrow B_c(B^{*}_c) +b+\bar{c}$, to convert the necessary
parts to the `basic structures', only the parts containing a
three-gluon coupling vertex have to be decomposed. Thus let us
outline the decomposition for the concerned subprocess.

\begin{figure}
\centering
\includegraphics[width=0.8\textwidth]{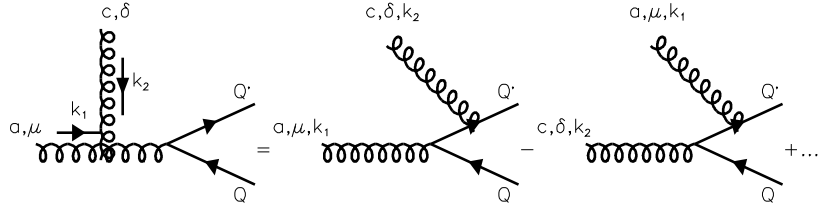}
\vspace{-2.5in} \caption{The three-gluon coupling vertex is
decomposed as in Eq.(\ref{vertf}): the first two terms are the
`basic QED-like' terms and the `remaining' terms are expressed by
several extra basic functions.} \label{vertex}
\end{figure}

The decomposition of a three-gluon coupling vertex is shown in
Fig.\ref{vertex} (the first structure): a three-gluon vertex
$f^{abc}T_{\mu\nu\delta}(k_1,K,k_2)$ through a gluon propagator
$-ig_{v v^{\prime}}/K^2$ couples to a quark line with a
quark-gluon-quark vertex $ T_{b}\gamma_{\nu^{\prime}}$, where
$K=-(k_{1}+k_{2})=-(Q+Q^{\prime})$, and $f^{abc}$, $T_b$ are color
factors at the two vertices. It is
\begin{equation}
M^{ac}_{\mu\delta}(k_{1},k_{2},Q,Q^{\prime})=-g^2_{s}f^{abc}T_{b}
T_{\mu\nu\delta}(k_1,K,k_2)\frac{-ig_{s}g^{v
v^{\prime}}}{K^{2}}\gamma_{\nu^{\prime}}\,,
\end{equation}
where $Q$ and $Q^{\prime}$ are the momenta of the fermion legs at the
vertex. Since
\begin{equation}
T_{\mu\nu\delta}(k_1,K,k_2)=(k_{1}-K)_{\delta}g_{\mu\nu}
+(K-k_{2})_{\mu}g_{\nu\delta}+(k_{2}-k_{1})_{\nu}g_{\mu\delta}\,,
\end{equation}
so the precise expression is
\begin{eqnarray}
M^{ac}_{\mu\delta}(k_{1},k_{2},Q,Q^{\prime})&=& g_{s}^2
f^{abc}T_{b}\frac{i}{K^{2}}\bigg((k_{1}+Q+Q^{\prime})_{\delta}
\gamma_{\mu}+k_{1\mu}\gamma_{\delta}-
\slash\!\!\!Q^{\prime}\gamma_{\mu}\gamma_{\delta}-
\gamma_{\mu}\slash\!\!\!Q^{\prime} \gamma_{\delta}\nonumber\\
& & -\gamma_{\delta}\slash\!\!\!Q\gamma_{\mu}- \gamma_{\delta}
\gamma_{\mu}\slash\!\!\!Q+
(\slash\!\!\!Q+\slash\!\!\!Q^{\prime}-2\slash\!\!\!
k_{1})g_{\mu\delta} \bigg)\nonumber\\
&=& g_{s}^2 f^{abc}T_{b}\frac{i}
{K^{2}}\bigg(\gamma_{\delta}(\slash\!\!\!k_{1}- \slash\!\!\!Q+m)
\gamma_{\mu}-\gamma_{\mu} (\slash\!\!\!k_{2}-\slash\!\!\!Q+m)
\gamma_{\delta}+\nonumber\\
& & (\slash\!\!\!Q^{\prime}-m)
(\gamma_{\delta}\gamma_{\mu}-g_{\mu\delta}) +\nonumber \\
& &(\gamma_{\mu}\gamma_{\delta}-g_{\mu\delta}) (\slash\!\!\!Q+m)+
k_{2\delta} \gamma_{\mu}-k_{1\mu}\gamma_{\delta}\bigg)\,.
\end{eqnarray}

If the factor $g_{s}^2$, the propagator scalar
factor and the color factor are disregarded,
and the symbol \(\simeq\) is used for an
equality modulo these factors, we have
\begin{eqnarray}
M^{ac}_{\mu\delta}&\simeq&\gamma_{\delta}(\slash\!\!\!k_{1}-
\slash\!\!\!Q+m) \gamma_{\mu}-\gamma_{\mu}
(\slash\!\!\!k_{2}-\slash\!\!\!Q+m)
\gamma_{\delta}+(\slash\!\!\!Q^{\prime}-m)
(\gamma_{\delta}\gamma_{\mu}-g_{\mu\delta})\nonumber\\
& & +(\gamma_{\mu}\gamma_{\delta}-g_{\mu\delta}) (\slash\!\!\!Q+m)
+ k_{2\delta} \gamma_{\mu}-k_{1\mu}\gamma_{\delta}+\cdots\,,
\label{vertf}
\end{eqnarray}
where $m$ is the mass of the fermion. Here the gluons and
fermions may not be on-shell. If the first and the second terms on
the right-hand side are embedded in the diagrams, the basic
QED-like diagrams may be obtained, while the rest terms $\cdots$
on the right-hand side can be absorbed into several extra
functions.

\begin{figure}
\centering
\includegraphics[width=0.8\textwidth]{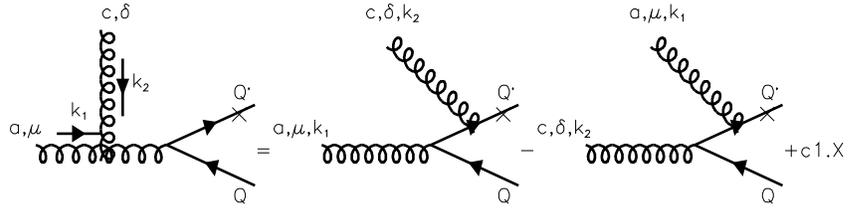}
\vspace{-2.5in} \caption{Reduction of the basic structure with a
three gluon vertex: the first two terms correspond to the basic
QED-like diagrams and the symbol `$\times$' means a quark gluon
vertex $T_{d}\gamma_{\alpha}$.} \label{basic1}
\end{figure}

\begin{figure}
\centering
\includegraphics[width=0.8\textwidth]{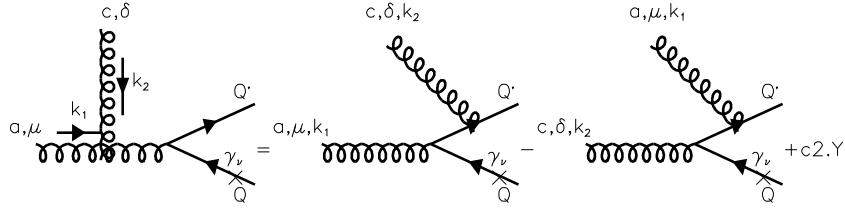}
\vspace{-2.5in} \caption{ Dividing the basic topology including
the three gluon vertex, where the first two terms correspond to
the `basic QED-like' diagrams and the symbol `$\times$' means a
quark gluon vertex $T_{d}\gamma_{\alpha}$.} \label{basic2}
\end{figure}

To relate to the `basic structures', two further decompositions
are shown in Fig.\ref{basic1} (the second structure) and
Fig.\ref{basic2} (the third structure) respectively: a three-gluon
vertex $f_{abc}T_{\mu\nu\delta} (k_1,k_1+k_2,k_2)$ is coupled to a
quark line, which has two vertices of quark-gluon-quark
$T_{b}\gamma_{\nu}$ and $T_{d}\gamma_{\alpha}$. To be precise, we
label the latter vertex  with the symbol `$\times$' on the quark
line. In the present case the two external quark lines are both
on-shell and then these two structures can be simplified by using
the on-shell conditions
\begin{equation}
\bar{u}(Q^{\prime})(\slash\!\!\!Q^{\prime}-m)=0\,, \;\;\;
(\slash\!\!\!Q+m)v(Q)=0 \,. \label{onshell}
\end{equation}

The contribution from the corresponding part in Fig.\ref{basic1}
is:
\begin{eqnarray}
M^{acd}_{\mu\delta\alpha}(k_{1},k_{2},Q,Q^{\prime})&=&-g_{s}f^{abc}T_{b}T_{d}
\bar{u}(Q^{\prime})\gamma_{\alpha}\frac{i(\slash\!\!\!k_{1}+\slash\!\!\!k_{2}
-\slash\!\!\!Q+m)}{(k_{1}+k_{2}-Q)^2-m^2}\cdot \nonumber \\
&&T_{\mu\nu\delta}(k_1,K,k_2) \frac{-ig_{s}g^{v
v^{\prime}}}{K^{2}}\gamma_{\nu^{\prime}}v(Q)\,.
\end{eqnarray}

With Eq.(\ref{vertf}) and the on-shell condition
Eq.(\ref{onshell}) for the two external fermion legs, we obtain
\begin{eqnarray}
M^{acd}_{\mu\delta\alpha}(k_{1},k_{2},Q,Q^{\prime})
&\simeq&\bar{u}(Q^{\prime})\gamma_{\alpha}(\slash\!\!\!k_{1}
+\slash\!\!\!k_{2}-\slash\!\!\!Q+m)
\Bigl(\gamma_{\delta}(\slash\!\!\!k_{1}- \slash\!\!\!Q+m)
\gamma_{\mu}-\gamma_{\mu} (\slash\!\!\!k_{2}-\slash\!\!\!Q+m)
\gamma_{\delta}\Bigr)v(Q)\nonumber\\
& & +\Bigl(m^{2}_{1}+m^{2}_{2}+ 2k_{1}\cdot k_{2}-2Q\cdot
k_{1}-2Q\cdot k_{2}\Bigr)\bar{u}(Q^{\prime})
\gamma_{\alpha}(\gamma_{\delta}\gamma_{\mu}-
g_{\mu\delta})v(Q)+\nonumber\\
& & \bar{u}(Q^{\prime})\gamma_{\alpha}\Bigl(k_{2\delta}
(\slash\!\!\!k_{1}+ \slash\!\!\!k_{2})\gamma_{\mu}-
k_{1\mu}(\slash\!\!\!k_{1}+ \slash\!\!\!k_{2})
\gamma_{\delta}-2k_{2\delta}Q_{\mu}+ 2k_{1\mu}Q_{\delta}
\Bigr)v(Q)\nonumber\\
&=&\bar{u}(Q^{\prime})\gamma_{\alpha}(\slash\!\!\!k_{1}
+\slash\!\!\!k_{2}-\slash\!\!\!Q+m)
\Bigl(\gamma_{\delta}(\slash\!\!\!k_{1}- \slash\!\!\!Q+m)
\gamma_{\mu}-\gamma_{\mu} (\slash\!\!\!k_{2}-\slash\!\!\!Q+m)
\gamma_{\delta}\Bigr)v(Q)\nonumber\\
&& +(c1\cdot X)+\cdots\;, \label{eq:basic1}
\end{eqnarray}
where `$\simeq$' means that the factor $g_{s}^2$, the propagator scalar
factor and the color factor have been omitted. The first term is
for the basic QED-like diagrams. The second term $(c1\cdot
X)$ where
$$
c1=m^{2}_{1}+m^{2}_{2}+ 2k_{1}\cdot k_{2}-2Q\cdot k_{1}-2Q\cdot
k_{2}
$$
and
$$ X=\bar{u}(Q^{\prime})\gamma_{\alpha}\{\gamma_{\delta}\gamma_{\mu}-
g_{\mu\delta}\} v(Q)$$ will be treated below. The third term
`$\cdots$', in fact, does not contribute to the amplitude, no
matter whether the gluons are virtual or real. The proof is that
when both the concerned gluons are real, it is easy to show that
the remaining terms give zero contribution by using the relation
$\epsilon(k_{l})\cdot k_{l}=0\;\;(l=1,2)$. When one of the
concerned gluons is virtual, the gluon with momentum $k_{2}$ for
example, $k_{2\delta}$ will always couple to a `simple' quark line
as $\bar{u}(R)\gamma_{\delta}v(R^{\prime})$ at the lowest order
for the amplitude of the considered process $gg\to
B_c(B_c^*)+\bar{c}+b$, where $u$ and $v$ are the quark and
anti-quark spinors, and $k_{2}=R+R^{\prime}$ (the momenta $R$,
$R^{\prime}$ satisfy the on-shell condition: $R^{2}=R^{\prime
2}=m^2$, where $m$ is the mass of the quark). It is easy to show
that the contribution of the remaining terms is zero by using a
similar on-shell condition as Eq.(\ref{onshell}). Therefore, in
Fig.\ref{basic1} and in the following Fig.\ref{basic2}, the
remaining terms have not been shown explicitly.

In the same way as above, the contribution from the corresponding
part in Fig.\ref{basic2} is:
\begin{eqnarray}
M^{acd}_{\mu\delta\alpha}(k_{1},k_{2},Q,Q^{\prime}) &\simeq&
\bar{u}(Q^{\prime})\Bigl( \gamma_{\delta}
(\slash\!\!\!Q^{\prime}-\slash\!\!\!k_{2}+M)\gamma_{\mu}-
\gamma_{\mu}(\slash\!\!\!Q^{\prime}-\slash\!\!\!k_{1}+M)\gamma_{\delta}
\Bigr)(\slash\!\!\!Q^{\prime}-\slash\!\!\!k_{1}-
\slash\!\!\!k_{2}+M)\cdot\nonumber\\
& & \gamma_{\alpha}v(Q)+\Bigl(m^{2}_{1}+m^{2}_{2}+ 2k_{1}\cdot
k_{2}-2Q^{\prime}\cdot k_{1}- 2Q^{\prime}\cdot
k_{2}\Bigr)\bar{u}(Q^{\prime})(\gamma_{\delta}\gamma_{\mu}-
g_{\mu\delta})\cdot\nonumber\\
& &\gamma_{\alpha}v(Q)+\bar{u}(Q^{\prime})
\Bigl(-k_{2\delta}\gamma_{\mu}(\slash\!\!\!k_{1}+
\slash\!\!\!k_{2})+k_{1\mu}\gamma_{\delta}(\slash\!\!\!k_{1}+
\slash\!\!\!k_{2})+2k_{2\delta}Q^{\prime}_{\mu}+\nonumber\\
& & 2k_{1\mu}Q^{\prime}_{\delta} \Bigr)
\gamma_{\alpha}v(Q)+(c2\cdot Y)+\cdots \;, \label{eq:basic2}
\end{eqnarray}
where again the factor $g_{s}^2$, the scalar factor of the
propagators and the color factor have been omitted. The
first two terms correspond to the basic QED-like diagrams, and the
term which is expressed by \((c2\cdot Y)\) is defined as
\begin{displaymath}
c2=m^{2}_{1}+m^{2}_{2}+ 2k_{1}\cdot k_{2}-2Q^{\prime}\cdot
k_{1}-2Q^{\prime}\cdot k_{2}\,,
\end{displaymath}
where $Y$ is a new extra function $$
Y=\bar{u}(Q^{\prime})\{\gamma_{\delta}\gamma_{\mu}-g_{\mu\delta}\}
\gamma_{\alpha}v(Q).$$ Similarly as for the second structure in
Eq.(19), the remaining terms in the present structure contribute
nothing, thus in Fig.\ref{basic2}, they have not been shown
explicitly.

Having made all the preparations above, and performing all the possible
interchanges, such as gluon exchange, quark and anti-quark
exchange, we relate each of the terms in the diagrams to a
combination of those with the basic QED-like diagrams
(`coordinators'), including the introduced two `extra' functions as
well.

\subsection{The decomposition}

The $B_c$ meson is a double-heavy weak-binding state. According to
pQCD each term of the amplitude for the subprocess may be
factorized into two factors: that of perturbative $gg\rightarrow
b+\bar{b}+c+\bar{c}$ (all the quarks are on shell) and that of
non-perturbative $c+\bar{b} \rightarrow B_{c}$. The binding wave
function in the Bethe-Salpeter framework may be used to dictate
the non-perturbative one, and approximately written as
Eq.(\ref{BSw}). To carry out the factorization of the amplitude,
one may apply Eq.(\ref{BSw}) and the two equations:
\begin{eqnarray}
\slash\!\!\!q_{c1}+m_{c}&=& \sum_{s}u(q_{c1},s)\bar{u}(q_{c1},s)
\simeq\alpha_1(\slash\!\!\!\!P+M)\,,\nonumber \\
\slash\!\!\!q_{b2}-m_{b}&=& \sum_{s}v(q_{b2},s)\bar{v}(q_{b2},s)
\simeq -\alpha_2(\slash\!\!\!\!P+M)
\end{eqnarray}
to each term in the amplitude in Eqs.(\ref{Mocc8}-\ref{Moo4}) with the
corresponding factors $\bar\chi_P(q)$.

Then the general structure of the amplitude in `explicit helicity'
form turns to
\begin{eqnarray}
M_{i}^{(\lambda_{1}, \lambda_{4}, \lambda_{5},
\lambda_{6})}(q_{b1},q_{b2},q_{c1},q_{c2},k_{1},k_{2})
&=&\sum_{\lambda_{2},\lambda_{3}}C_{i}X_{i}D_{1}
B_{Fi}^{(\lambda_{1},\lambda_{2},
\lambda_{3},\lambda_{4},\lambda_{5},\lambda_{6})}
(q_{b1},q_{b2},q_{c1},q_{c2},k_{1},k_{2})
\cdot\nonumber\\
& &
D_{2}B_{B_c(B_c^*)}^{(\lambda_{2},\lambda_{3})}(q_{b2},q_{c1}),
\label{matrixd}
\end{eqnarray}
where $i=1,\cdots,36$ (or labelled as Feynman diagrams: $1a, 1b,
\cdots, 5d$, respectively), $\lambda_{j}\;(j=1,\cdots,6)$ denote
the helicities (spins) of the quarks and gluons respectively
appearing in the two factorized `processes' $gg\rightarrow
b+\bar{b}+c+\bar{c}$ and $c+\bar{b} \rightarrow B_{c}$. Note that
from now on, we change the notation of the helicities of the
particles in the processes as: $\lambda_{1}$ denotes the helicity
of $b$, $\lambda_{2}$ that of $\bar{b}$, $\lambda_{3}$ that of
$c$, $\lambda_{4}$ that of $\bar{c}$; whereas $\lambda_{5}$,
instead of $\lambda_1$ in Eqs.(4-8), denotes that of gluon-1 and
$\lambda_{6}$, instead of $\lambda_2$ in Eqs.(4-8), denotes that
of gluon-2. Here $C_i$, $X_i$ denote the color factor and the
scalar factor from all the propagators as a whole for the
$i$th-diagram, respectively. $B_{Fi}^{(\lambda_{1},\lambda_{2},
\lambda_{3},\lambda_{4},\lambda_{5},\lambda_{6})}
(q_{b1},q_{b2},q_{c1},q_{c2},k_{1},k_{2})$ and
$B_{B_c(B_c^*)}^{(\lambda_{2},\lambda_{3})}(q_{b2},q_{c1})$ are
the amplitudes corresponding to the `free quark part'
$g(k_1,\lambda_5)g(k_2,\lambda_6)\rightarrow
b(q_{b1},\lambda_1)+\bar{b}(q_{b2},\lambda_2)+c(q_{c1},\lambda_3)
+\bar{c}(q_{c2},\lambda_4)$ (all the quarks are on-shell) and the
`bound state part' $c(q_{c1},\lambda_3)+\bar{b}(q_{b2},\lambda_2)
\rightarrow B_{c}(B_c^*)$, respectively. Substituting
Eq.(\ref{massive}), we have
\begin{eqnarray}
B_{Fi}^{(\lambda_{1},\lambda_{2},\lambda_{3},\lambda_{4},\lambda_{5},
\lambda_{6})}(q_{b1},q_{b2},q_{c1},q_{c2},k_{1},k_{2})&=&\langle
q_{0\lambda_{1}}|(\slash\!\!\!q_{b1}+ m_{b})\Gamma_{1i}
(\slash\!\!\!q_{b2}-m_{b})| q_{0\lambda_{2}} \rangle\cdot  \nonumber\\
&&\langle q_{0\lambda_{3}}| (\slash\!\!\!q_{c1}+m_{c})\Gamma_{2i}
(\slash\!\!\!q_{c2}-m_{c})|q_{0\lambda_{4}}\rangle\,, \\
B_{B_c(B_c^*)}^{(\lambda_{2},\lambda_{3})}(q_{b2},q_{c1})=\langle
q_{0\lambda_{2}}| (\alpha\gamma_{5}&+&\beta
\slash\!\!\!\epsilon(s_{z})) \frac{(\slash\!\!\!\!P+M)} {2
\sqrt{M}}\psi(0)|q_{0\lambda_{3}} \rangle \,,\nonumber \\
(\alpha=1, \;\;\beta=0\,, \;\; for\;\; B_c[1^1S_0];&& \alpha=0,
\;\;\beta=1\,, \;\; for\;\; B_c^*[1^3S_1])
\end{eqnarray}
where $|q_0\lambda_r\rangle$ and $\langle q_0\lambda_r|$
($\lambda_r=\pm,\;r=1,\cdots,4$) are the introduced helicity
states of the massless fermion $q_0$ which specifically relate to
those of massive fermions with the momentum $p$ and mass $m$ as in
Eq.(\ref{massive}), \(\Gamma_{1i,2i}\) are the explicit strings of
Dirac \(\gamma\) matrices corresponding to $i$-th Feynman diagram
which contain the gluon helicities $\lambda_{5}$ and
$\lambda_{6}$. $D_{1}=\frac{1}{\sqrt{2 q_{b1}\cdot q_{0}}}
\frac{1}{\sqrt{2 q_{b2}\cdot q_{0}}} \frac{1}{\sqrt{2 q_{c1}\cdot
q_{0}}} \frac{1}{\sqrt{2 q_{c2}\cdot q_{0}}}$ and
$D_{2}=\frac{1}{\sqrt{2 q_{c1}\cdot q_{0}}} \frac{1}{\sqrt{2
q_{b2}\cdot q_{0}}}$ are the two common normalization factors.

The function
$D_{2}B_{B_c(B_c^*)}^{(\lambda_{2},\lambda_{3})}(q_{b2},q_{c1})$,
which contains the bound state wave function, is:
\begin{equation}
D_{2}B_{B_c}^{(\lambda_{2},\lambda_{3})}(q_{b2},q_{c1})=\frac{\psi(0)\sqrt{M}}
{2\sqrt{m_{b}m_{c}}} \delta_{\lambda_{2},\lambda_{3}}
(\delta_{\lambda_{2}-}- \delta_{\lambda_{2}+})
\end{equation}
for $B_{c}[^1S_1]$, and
\begin{eqnarray}
D_{2}B_{B_c^*}^{(\lambda_{2},\lambda_{3})}(q_{b2},q_{c1})&=&
\frac{\psi(0)\sqrt{M}} {2\sqrt{m_{b}m_{c}}}
\delta_{\lambda_{2},\lambda_{3}} (\delta_{\lambda_{2}+}+
\delta_{\lambda_{2}-}) \left(\frac{M\epsilon(s_{z})\cdot q_{0}}
{P\cdot q_{0}}\right)+\frac{\psi(0)\sqrt{M}}
{2\sqrt{m_{b}m_{c}}}\left(\frac{1}{2P\cdot q_{0}}\right)\cdot
\nonumber \\
&&\langle q_{0\lambda_{2}}| \slash\!\!\!\epsilon(s_{z})
\slash\!\!\!\!P|q_{0\lambda_{3}} \rangle
\end{eqnarray}
for $B^{*}_{c}[^3S_1]$.

These 36 functions, $B_{Fi}^{(\lambda_{1}, \lambda_{2},
\lambda_{3}, \lambda_{4}, \lambda_{5}, \lambda_{6})}(q_{b1},
q_{b2}, q_{c1}, q_{c2}, k_{1}, k_{2})$, can be constructed by nine
`basic ones' which correspond to Fig.\ref{bc1}a, Fig.\ref{bc1}b,
Fig.\ref{bc1}c, Fig.\ref{bc1}d, Fig.\ref{oo1}d (two basic
functions), Fig.\ref{cc1}a, Fig.\ref{cc1}c and Fig.\ref{cc1}e, and
by performing all the possible interchanges of the initial gluons
and the two quark lines. In fact, the functions which correspond
to Fig.\ref{bc1}c and Fig.\ref{bc1}d can be obtained from those in
Fig.\ref{bc1}a and Fig.\ref{bc1}b by interchanging the initial
gluons and the quark lines in the diagrams, and in the following
section we will show that the functions corresponding to
Fig.\ref{cc1}a and Fig.\ref{cc1}c can also be expressed by other
seven `basic functions', although here we still treat those four
functions as `basic ones', in order to treat them on an equal
footing.

We use $E_{m,j,k}(q_{b1}, q_{b2}, q_{c1}, q_{c2}, k_{1},
k_{2})\,,\; (m=1,2,\cdots,9;\; j=1,\cdots,4;\; k=1,2,\cdots 64)$
to denote the `basic functions', where $k$ denotes 64 possible
helicities (spins) corresponding to possible `values' of
$(\lambda_{1},\lambda_{2}, \lambda_{3}, \lambda_{4},
\lambda_{5},\lambda_{6})$ as shown in Table~\ref{helicity}, and
the correspondences of the functions to the Feynman diagrams are
shown in Table~\ref{tab1} and Table~\ref{tab2}. Here $j$ means the
four possible interchanges: $1$ means identical (without any
interchange), $2$ means interchange of the gluons, $3$ means
interchange of the quark and the anti-quark, and $4$ means
interchange of the gluons and the quark and anti-quark. When
applying the interchange among the particles, the symmetries of
the decomposed amplitude guarantee that every term may be
expressed by the basic functions with helicities and momenta of
the particles in the process. For all the functions $E_{m,j,k}$,
with $m,j$ being fixed, they are related to each other by proper
complex conjugation with or without changing the whole sign. To
write the program, and to apply these relations of the element
functions among different helicity states conveniently, we take
the correspondence between $k$ and $(\lambda_{1},\lambda_{2},
\lambda_{3}, \lambda_{4}, \lambda_{5},\lambda_{6})$ as described
in Table~\ref{helicity}.
\begin{table}
\begin{center}
\caption{The correspondence between $k=1, \cdots, 64$ and
$\lambda_{1}=\pm, \lambda_{2}=\pm, \lambda_{3}=\pm,
\lambda_{4}=\pm, \lambda_{5}=\pm, \lambda_{6}=\pm$, which stand
for the helicities of the particles in the process.}
\begin{tabular}{|c||c|c|c|c|c|c||c||c|c|c|c|c|c||c||c|c|c|c|c|c||c||c|c|c|c|c|c|}
\hline
$k$&$\lambda_{1}$&$\lambda_{2}$&$\lambda_{3}$&$\lambda_{4}$&$\lambda_{5}$&$\lambda_{6}$&
$k$&$\lambda_{1}$&$\lambda_{2}$&$\lambda_{3}$&$\lambda_{4}$&$\lambda_{5}$&$\lambda_{6}$&
$k$&$\lambda_{1}$&$\lambda_{2}$&$\lambda_{3}$&$\lambda_{4}$&$\lambda_{5}$&$\lambda_{6}$&
$k$&$\lambda_{1}$&$\lambda_{2}$&$\lambda_{3}$&$\lambda_{4}$&$\lambda_{5}$&$\lambda_{6}$\\
\hline\hline
1&+&+&+&+&+&+&17&$-$&$-$&$-$&$-$&$-$&$-$&33&+&+&$-$&+&+&+&49&$-$&$-$&+&$-$&$-$&$-$\\
\hline
2&+&+&+&+&+&$-$&18&$-$&$-$&$-$&$-$&$-$&+&34&+&+&$-$&+&+&$-$&50&$-$&$-$&+&$-$&$-$&+\\
\hline
3&+&+&+&+&$-$&+&19&$-$&$-$&$-$&$-$&+&$-$&35&+&+&$-$&+&$-$&+&51&$-$&$-$&+&$-$&+&$-$\\
\hline
4&+&+&+&+&$-$&$-$&20&$-$&$-$&$-$&$-$&+&+&36&+&+&$-$&+&$-$&$-$&52&$-$&$-$&+&$-$&+&+\\
\hline
5&+&+&+&$-$&+&+&21&$-$&$-$&$-$&+&$-$&$-$&37&+&+&$-$&$-$&+&+&53&$-$&$-$&+&+&$-$&$-$\\
\hline
6&+&+&+&$-$&+&$-$&22&$-$&$-$&$-$&+&$-$&+&38&+&+&$-$&$-$&+&$-$&54&$-$&$-$&+&+&$-$&+\\
\hline
7&+&+&+&$-$&$-$&+&23&$-$&$-$&$-$&+&+&$-$&39&+&+&$-$&$-$&$-$&+&55&$-$&$-$&+&+&+&$-$\\
\hline
8&+&+&+&$-$&$-$&$-$&24&$-$&$-$&$-$&+&+&+&40&+&+&$-$&$-$&$-$&$-$&56&$-$&$-$&+&+&+&+\\
\hline
9&+&$-$&$-$&+&+&+&25&$-$&+&+&$-$&$-$&$-$&41&+&$-$&+&+&+&+&57&$-$&+&$-$&$-$&$-$&$-$\\
\hline
10&+&$-$&$-$&+&+&$-$&26&$-$&+&+&$-$&$-$&+&42&+&$-$&+&+&+&$-$&58&$-$&+&$-$&$-$&$-$&+\\
\hline
11&+&$-$&$-$&+&$-$&+&27&$-$&+&+&$-$&+&$-$&43&+&$-$&+&+&$-$&+&59&$-$&+&$-$&$-$&+&$-$\\
\hline
12&+&$-$&$-$&+&$-$&$-$&28&$-$&+&+&$-$&+&+&44&+&$-$&+&+&$-$&$-$&60&$-$&+&$-$&$-$&+&+\\
\hline
13&+&$-$&$-$&$-$&+&+&29&$-$&+&+&+&$-$&$-$&45&+&$-$&+&$-$&+&+&61&$-$&+&$-$&+&$-$&$-$\\
\hline
14&+&$-$&$-$&$-$&+&$-$&30&$-$&+&+&+&$-$&+&46&+&$-$&+&$-$&+&$-$&62&$-$&+&$-$&+&$-$&+\\
\hline
15&+&$-$&$-$&$-$&$-$&+&31&$-$&+&+&+&+&$-$&47&+&$-$&+&$-$&$-$&+&63&$-$&+&$-$&+&+&$-$\\
\hline
16&+&$-$&$-$&$-$&$-$&$-$&32&$-$&+&+&+&+&+&48&+&$-$&+&$-$&$-$&$-$&64&$-$&+&$-$&+&+&+\\
\hline
\end{tabular}
\label{helicity}
\end{center}
\end{table}
Thus we have now
\begin{eqnarray}
B_{Fi}^{(\lambda_{1},\lambda_{2},\lambda_{3},
\lambda_{4},\lambda_{5},\lambda_{6})}(q_{b1},q_{b2},q_{c1},
q_{c2},k_{1},k_{2})&\equiv&B_{Fi}^{(k)}(q_{b1},q_{b2},q_{c1},
q_{c2},k_{1},k_{2})\nonumber\\
&=&\sum_{m=1}^{9}\sum_{j=1}^{4} f_{i,m,j}E_{m,j,k}\,.
\label{befun}
\end{eqnarray}
Here $i$ and $k$ denote the 36 Feynman diagrams (terms) and the 64
possible helicities of all the particles in the two factorized
processes. The correspondence of helicities on $k=1,\cdots,64$ is
given as described in Table~\ref{helicity}. The 36 functions
$i=1,\cdots,36$ are labelled as the 36 Feynman diagrams, $i.e.$
$i=1a,1b,\cdots,5d$. The coefficient $f_{i,m,j}$ corresponding to
QED-like diagrams can be directly read out easily, but as for the
diagrams involving three- and four- gluon vertices we decompose
them by applying the formulae obtained in the subsection B.

\subsection{The amplitude $B_{Fi}^{(k)}(q_{b1},q_{b2},q_{c1},
q_{c2},k_{1},k_{2})$}


\begin{table}
\begin{center}
\caption{The expansion coefficients $f_{i,m,j}$ for the functions
$B_{Fi}^{(k)}(q_{b1},q_{b2},q_{c1}, q_{c2},k_{1},k_{2})$ which are
grouped into the $cb$ subset directly or indirectly through a
proper decomposition (the coefficients $f_{i,m,j}$ are not listed
here if they are equal to zero in a whole row).}
\begin{tabular}{|c||c|c|c|c|c|}
\hline
 & \multicolumn{5}{|c|}{$j=1$}\\
\hline
 ~~~$m$~~~   & ~~\(1\)~~ & ~~\(2\)~~ & ~~\(3\)~~ & ~~\(4\)~~ & ~~~~~~~\(5\)~~~~~~~ \\
\hline\hline
~\(f_{3a,m,j}\)~ & 1 & 0 & 0 & 0 & 0  \\
\hline
\(f_{3b,m,j}\) & 0 & 1 & 0 & 0 & 0 \\
\hline
\(f_{3d,m,j}\) & 0 & 0 & 1 & 0 & 0 \\
\hline
\(f_{3e,m,j}\) & 0 & 0 & 0 & 1 & 0 \\
\hline
\(f_{4c,m,j}\)& -1 & 0 & 1 & 0 &-\(2(q_{c2}\cdot k_{1})\)  \\
\hline
\(f_{4d,m,j}\)  & 0 & -1 & 0 & 1 &\(2(q_{c1}\cdot k_{1})\)  \\
\hline
\(f_{4f,m,j}\) & 1 & -1 & 0 & 0 &\(2(q_{b1}\cdot k_{2})\)  \\
\hline
\(f_{4g,m,j}\) & 0 & 0 & 1 & -1 &-\(2(q_{b2}\cdot k_{2})\) \\
\hline
\(f_{5b,m,j}\) & 1 & -1 & -1 & 1 &~\(s_{b}+s_{c}-s_{2}\)~ \\
\hline
\end{tabular}
\label{tab1}
\end{center}
\end{table}

All the terms for the subprocess may be divided into four subsets
according to the manner how the gluons attach to the fermion
lines. For the $cc$ set, both gluons attach directly to the
$c$-quark line; for the $cb$ set, the gluon-1 attaches to the
$c$-quark line while the gluon-2 attaches to the $b$-quark line;
for the $bc$ set, the gluon-1 attaches to the $b$-quark line while
the gluon-2 attaches to the $c$-quark line; for the $bb$ set, both
gluons attach directly to the $b$-quark line in the relevant
Feynman diagrams. The diagrams involving three- and four- gluon
vertex (vertices) which are shown in Fig.\ref{obc} and
Fig.\ref{oo1} are decomposed by applying the decomposition
formulae in subsection B. Hence each of the terms corresponding to
the diagrams in general turns to four terms and then they may be put
into four subsets separately according to the resulting
characteristics of the decomposed term. The results are shown in
Table~\ref{tab1} and Table~\ref{tab2}. In Table~\ref{tab1} and
Table~\ref{tab2}, we list the exact results for the $cb$ and $cc$
sets, and the results for the $bc$ and $bb$ sets by performing
interchanges of the initial gluons and the final quark lines. In
Table~\ref{tab1} and Table~\ref{tab2} one may see, for example,
how the Feynman diagrams in Figs.\ref{obc}c, \ref{obc}d and
Fig.\ref{oo1}b are decomposed, and how each of the decomposed term
is divided into a $cb$ or a $cc$ set. We summarize the relations
among the functions
$B_{Fi}^{(k)}(q_{b1},q_{b2},q_{c1},q_{c2},k_{1},k_{2})$ and show
them in Table~\ref{tab1} and Table~\ref{tab2} respectively. In
Table~\ref{tab1}, the functions
$B_{Fi}^{(k)}(q_{b1},q_{b2},q_{c1}, q_{c2},k_{1},k_{2})$ are
listed either directly or through a proper decomposition in order
to relate them to the $cb$ subset through functions $E_{m,j,k}$
and suitable coefficients $f_{i,m,j}$. In Table~\ref{tab2}, the
functions $B_{Fi}^{(k)}(q_{b1},q_{b2},q_{c1}, q_{c2},k_{1},k_{2})$
are listed either directly or indirectly through a proper
decomposition relating to the $cc$ subset via suitable
coefficients $f_{i,m,j}$. In addition, in Table~\ref{tab2}, we
have decomposed the matrix element $M_{5d}$ of Fig.\ref{oo1}d, as
shown in Eq.(\ref{Moo4}), into three parts according to the three
different color factors:
$C_{5d1}=C_{2ij}+C_{3ij}-C_{1ij}-C_{4ij}$,
$C_{5d2}=C_{5ij}-C_{1ij}+C_{3ij}$,
$C_{5d3}=C_{5ij}-C_{2ij}+C_{4ij}$. The parameters used in
Table~\ref{tab1} and Table~\ref{tab2} are as follows:
$s_{b}=(q_{b1}+q_{b2})^{2}$, $s_{c}=(q_{c1}+q_{c2})^{2}$,
$s_{2}=(k_{2}-q_{b1}-q_{b2})^{2}$, $f_{1}=2(q_{c1}-q_{c2})\cdot
k_{2}+s_{b}$, $f_{2}=2(q_{c1}-q_{c2})\cdot k_{1}+s_{b}$,
$f_{3}=2k_{1}\cdot k_{2}-2q_{c2}\cdot k_{1}-2q_{c2}\cdot k_{2}$,
$f_{4}=2k_{1}\cdot k_{2}-2q_{c1}\cdot k_{1}-2q_{c1}\cdot k_{2}$.

\begin{table}
\begin{center}
\caption{The expansion coefficients $f_{i,m,j}$ for the functions
$B_{Fi}^{(k)}(q_{b1},q_{b2},q_{c1}, q_{c2},k_{1},k_{2})$ which are
grouped into the $cc$ subset directly or indirectly through a proper
decomposition (the coefficients $f_{i,m,j}$, which are equal to zero in a
whole row, are not listed here).}
\begin{tabular}{|c||c|c|c|c|c||c|c|c|c|c|}
\hline
 &\multicolumn{5}{|c||}{$j=1$}&\multicolumn{5}{|c|}{$j=2$}\\
\hline $m$ & ~~~\(6\)~~~ & ~~~\(7\)~~~ & ~~~\(8\)~~~ & ~~~\(9\)~~~
& ~~~\(5\)~~~&
~~~\(6\)~~~ & ~~~\(7\)~~~ & ~~~\(8\)~~~ & ~~~\(9\)~~~ &~~~ \(5\)~~~\\
\hline\hline
\(f_{1a,m,j}\) & 1 & 0 & 0 & 0 & 0 & 0& 0& 0& 0& 0\\
\hline
\(f_{1b,m,j}\) & 0 & 0 & 0 & 0 & 0 & 1& 0& 0& 0& 0\\
\hline
\(f_{1c,m,j}\) & 0 & 1 & 0 & 0 & 0 & 0& 0& 0& 0& 0\\
\hline
\(f_{1d,m,j}\) & 0 & 0 & 0 & 0 & 0 & 0& 1& 0& 0& 0\\
\hline
\(f_{1e,m,j}\) & 0 & 0 & 1 & 0 & 0 & 0& 0& 0& 0& 0\\
\hline
\(f_{1f,m,j}\) & 0 & 0 & 0 & 0 & 0 & 0& 0& 1& 0& 0\\
\hline \(f_{1g,m,j}\) & 1 & 0 & 0 &\(\frac{f_{3}}{2}\)
&0 & -1& 0& 0& -\(\frac{f_{3}}{2}\)& 0\\
\hline \(f_{1h,m,j}\) & 0 & 0 & 1 &
\(\frac{f_{4}}{2}\) & \(2f_{4}\) & 0& 0& -1& \(\frac{f_{4}}{2}\)& -\(2f_{4}\)\\
\hline \(f_{4a,m,j}\) & -1 & 1 & 0 & \(2q_{c2}\cdot k_{2}\) &
0& 0& 0& 0& 0& -\(2q_{c2}\cdot k_{2}\) \\
\hline \(f_{4b,m,j}\) & 0 & 0 & 0 &0 & \(4q_{c1}\cdot k_{2}\) & 0&
-1& 1& -\(2q_{c1}\cdot k_{2}\)& -$2q_{c1}\cdot k_{2}$ \\
\hline \(f_{4c,m,j}\) & 0 & 0 & 0 &0 &-\(2q_{c2}\cdot k_{1}\)& -1&
1&0& \(2q_{c2}\cdot k_{1}\)& 0 \\
\hline \(f_{4d,m,j}\) & 0 & -1 & 1 & -\(2q_{c1}\cdot k_{1}\) &
-\(2q_{c1}\cdot k_{1}\) &0&0&0&0& $4q_{c1}\cdot k_{2}$ \\
\hline \(f_{5a,m,j}\) & 1 & -\(1\) & 0 &\(2q_{c2}\cdot k_{2}\) &
-\(4q_{c1}\cdot k_{2}\) &0&-1&1&$2q_{c1}\cdot k_{2}$&\(f_{1}\)\\
\hline \(f_{5b,m,j}\) & 0 & -\(1\) & 1 & \(2q_{c1}\cdot k_{1}\) &
\(f_{2}\) &1&-1&0&\(2q_{c2}\cdot k_{1}\)& -$4q_{c1}\cdot k_{1}$ \\
\hline \(f_{5c,m,j}\) & -\(1\) & 0 & \(1\) &
-\(\frac{f_{3}-f_{4}}{2}\)& \(2f_{4}\)&1&0&-1&
$\frac{f_{3}-f_{4}}{2}$&-$2f_{4}$\\
\hline
\(f_{5d1,m,j}\) & 0 & 0 & 0 & 0 & \(1\)&0&0&0&0&$-1$ \\
\hline \(f_{5d2,m,j}\) & 0 & 0 & 0 & \(\frac{1}{2}\) & 0
&0&0&0&$\frac{1}{2}$&-1\\
\hline \(f_{5d3,m,j}\) & 0 & 0 & 0 & \(\frac{1}{2}\) & -\(1\)
&0&0&0&\(\frac{1}{2}\)&0\\
\hline
\end{tabular}
\label{tab2}
\end{center}
\end{table}

\subsection{The `basic functions' $E_{m,j,k}(q_{b1}, q_{b2}, q_{c1},
q_{c2}, k_{1}, k_{2})$}

To obtain the nine `basic functions' $E_{m,j,k}\;(m=1,\cdots,9)$,
let us first define the functions which correspond to various
kinds of quark lines (different $\gamma$ structures of the fermion
lines), where $q^{2}_{1}=q^{2}_{2}=m^{2}$ and
$q^{2}_{0}=k^{2}=k^{2}_{1}=k^{2}_{2}=0$:
\begin{eqnarray}
f_{0}(q_{1},q_{2},\lambda_{1},\lambda_{2})&=& \langle
q_{0\lambda_{1}}|(\slash\!\!\!q_{1}+m) \gamma_{\delta}
(\slash\!\!\!q_{2}-m)|q_{0\lambda_{2}}\rangle \,,\nonumber \\
f_{1}(q_{1},q_{2},k,\lambda_{1},\lambda_{2},\lambda_{3})&=&
\langle q_{0\lambda_{1}}|(\slash\!\!\!q_{1}+m)\gamma_{\delta}
(\slash\!\!\!k-\slash\!\!\!q_{2}+m)\slash\!\!\!\epsilon^{\lambda_{3}}
(k,q_{0})(\slash\!\!\!q_{2}-m)|q_{0\lambda_{2}} \rangle \,,\nonumber \\
f_{2}(q_{1},q_{2},k,\lambda_{1},\lambda_{2},\lambda_{3})&=&
\langle q_{0\lambda_{1}}|(\slash\!\!\!q_{1}+m)
\slash\!\!\!\epsilon^{\lambda_{3}} (k,q_{0})
(\slash\!\!\!q_{1}-\slash\!\!\!k+m)\gamma_{\delta}(\slash\!\!\!q_{2}-m)
|q_{0\lambda_{2}}\rangle \,,\nonumber \\
f_{3}(q_{1},q_{2},k,\lambda_{1},\lambda_{2},\lambda_{3})&=&
\langle q_{0\lambda_{1}}|(\slash\!\!\!q_{1}+m)
\slash\!\!\!\epsilon^{\lambda_{3}} (k,q_{0})
(\slash\!\!\!q_{2}-m)|q_{0\lambda_{2}}\rangle \,,\nonumber \\
f_{4}(q_{1},q_{2},k_{1},k_{2},\lambda_{1},\lambda_{2},\lambda_{3},\lambda_{4})&=&
\langle q_{0\lambda_{1}}|(\slash\!\!\!q_{1}+m)\gamma_{\delta}
(\slash\!\!\!k_{1}+\slash\!\!\!k_{2}-\slash\!\!\!q_{2}+m)
\slash\!\!\!\epsilon^{\lambda_{3}} (k_{1},q_{0})\cdot \nonumber\\
&&(\slash\!\!\!k_{2}-\slash\!\!\!q_{2}+m)
\slash\!\!\!\epsilon^{\lambda_{4}}(k_{2},q_{0})
(\slash\!\!\!q_{2}-m)|q_{0\lambda_{2}}\rangle \,,\nonumber \\
f_{5}(q_{1},q_{2},k_{1},k_{2},\lambda_{1},\lambda_{2},\lambda_{3},\lambda_{4})&=&
\langle q_{0\lambda_{1}}|(\slash\!\!\!q_{1}+m)
\slash\!\!\!\epsilon^{\lambda_{3}}(k_{1},q_{0})
(\slash\!\!\!q_{1}-\slash\!\!\!k_{1}+m) \gamma_{\delta}\cdot
\nonumber\\
&& (\slash\!\!\!k_{2}-\slash\!\!\!q_{2}+m)
\slash\!\!\!\epsilon^{\lambda_{4}}(k_{2},q_{0})
(\slash\!\!\!q_{2}-m)|q_{0\lambda_{2}}\rangle \,,\nonumber \\
f_{6}(q_{1},q_{2},k_{1},k_{2},\lambda_{1},\lambda_{2},
\lambda_{3},\lambda_{4})&=&
\langle q_{0\lambda_{1}}|(\slash\!\!\!q_{1}+m)
\slash\!\!\!\epsilon^{\lambda_{3}}(k_{1},q_{0})
(\slash\!\!\!q_{1}-\slash\!\!\!k_{1}+m)\cdot \nonumber\\
& & \slash\!\!\!\epsilon^{\lambda_{4}}(k_{2},q_{0})
(\slash\!\!\!q_{1}-\slash\!\!\!k_{1}-\slash\!\!\!k_{2}+m)\gamma_{\delta}
 (\slash\!\!\!q_{2}-m)|q_{0\lambda_{2}}\rangle\,, \nonumber \\
f_{7}(q_{1},q_{2},k_{1},k_{2},\lambda_{1},\lambda_{2},\lambda_{3},\lambda_{4})
&=& \langle q_{0\lambda_{1}}|(\slash\!\!\!q_{1}+m)\gamma_{\delta}
\slash\!\!\!\epsilon^{\lambda_{3}} (k_{1},q_{0})
\slash\!\!\!\epsilon^{\lambda_{4}} (k_{2},q_{0})
(\slash\!\!\!q_{2}-m)|q_{0\lambda_{2}}\rangle \,,\nonumber \\
f_{8}(q_{1},q_{2},k_{1},k_{2},\lambda_{1},\lambda_{2},\lambda_{3},\lambda_{4})
&=& \langle q_{0\lambda_{1}}|(\slash\!\!\!q_{1}+m)
\slash\!\!\!\epsilon^{\lambda_{3}} (k_{1},q_{0})
\slash\!\!\!\epsilon^{\lambda_{4}} (k_{2},q_{0})
\gamma_{\delta}(\slash\!\!\!q_{2}-m)|q_{0\lambda_{2}}\rangle\,,\nonumber\\
f_{9}(q_{1},q_{2},k_{1},k_{2},\lambda_{1},\lambda_{2},\lambda_{3},\lambda_{4})
&=& \langle q_{0\lambda_{1}}|(\slash\!\!\!q_{1}+m)
\slash\!\!\!\epsilon^{\lambda_{3}} (k_{1},q_{0}) \gamma_{\delta}
\slash\!\!\!\epsilon^{\lambda_{4}} (k_{2},q_{0})
(\slash\!\!\!q_{2}-m)| q_{0\lambda_{2}}\rangle\,.
\label{basicf}
\end{eqnarray}
There are several ways to deal with these fermion lines. In
Ref.~\cite{ht}, a systematic way for doing the massive case is
proposed. Here we take another and more `direct' approach. When
 the massive fermions have time-like momenta
$q_{i}$ ($i=1,2$) and $\slash\!\!\!q_{i}$ are directly connected
to $|q_{0\lambda_{i}}\rangle$ or $\langle q_{0\lambda_{i}}|$ as in
Eq.(\ref{basicf}), we may introduce the light-like momenta by
defining
\begin{equation}
q^{\prime}_{i}=q_{i}-\frac{q^{2}_{i}}{2q_{i}\cdot q_{0}}q_{0} \, .
\label{m-l}
\end{equation}
Then $\slash\!\!\!q_{i}$ for massive fermions can be replaced by
the massless ones, $\slash\!\!\!q\prime_{i}$, without any
consequences. This is due to the relations
\begin{equation}
\slash\!\!\!q_{0}|q_{0\lambda_{i}}\rangle=0
\label{q0spin1}
\end{equation} or
\begin{equation}
\langle q_{0\lambda_{i}}|\slash\!\!\! q_{0}=0\,.
\label{q0spin2}
\end{equation}
If the massive fermions with momenta $q_{i}\;(i=1,2)$ are not
directly connected to $|q_{0\lambda_{i}}\rangle$ or $\langle
q_{0\lambda_{i}}|$, but to another light-like spinor, say
$|q'_{0\lambda_{i}}\rangle$ or $\langle q'_{0\lambda_{i}}|$, then
we may do the same thing if the momentum $q_0$ and the spinor
$|q_{0\lambda_{i}}\rangle$ or $\langle q_{0\lambda_{i}}|$ are
replaced by $q'_0$ and $|q'_{0\lambda_{i}}\rangle$ or $\langle
q'_{0\lambda_{i}}|$, accordingly. In this way we can turn the
massive terms into massless terms, and then they can be dealt with
similarly as in the massless cases\cite{cal1,cal2,zdl,aes,ld}.

The nine basic functions may be expressed in terms of the above ten functions as
\begin{eqnarray}
E_{1,1,k}&=&
f_{1}(q_{c1},q_{c2},k_{1},\lambda_{3},\lambda_{4},\lambda_{5})
\cdot f_{2}(q_{b1},q_{b2},k_{2},\lambda_{1},\lambda_{2},
\lambda_{6})\,, \nonumber \\
E_{2,1,k}&=&
f_{2}(q_{c1},q_{c2},k_{1},\lambda_{3},\lambda_{4},\lambda_{5})
\cdot f_{2}(q_{b1},q_{b2},k_{2},\lambda_{1},\lambda_{2},
\lambda_{6})\,,  \nonumber\\
E_{3,1,k}&=&
f_{1}(q_{c1},q_{c2},k_{1},\lambda_{3},\lambda_{4},\lambda_{5})
\cdot f_{1}(q_{b1},q_{b2},k_{2},\lambda_{1},\lambda_{2},
\lambda_{6})\,,  \nonumber \\
E_{4,1,k}&=&
f_{2}(q_{c1},q_{c2},k_{1},\lambda_{3},\lambda_{4},\lambda_{5})
\cdot f_{1}(q_{b1},q_{b2},k_{2},\lambda_{1},\lambda_{2},
\lambda_{6})\,,  \nonumber \\
E_{5,1,k}&=&
f_{3}(q_{b1},q_{b2},k_{2},\lambda_{1},\lambda_{2},\lambda_{5})
\cdot
f_{3}(q_{c1},q_{c2},k_{1},\lambda_{3},\lambda_{4},
\lambda_{6})\,, \nonumber \\
E_{6,1,k}&=&f_{0}(q_{b1},q_{b2},\lambda_{1},\lambda_{2}) \cdot
f_{4}(q_{c1},q_{c2},k_{1},k_{2},\lambda_{3},\lambda_{4},
\lambda_{5},\lambda_{6})\,, \nonumber\\
E_{7,1,k}&=&f_{0}(q_{b1},q_{b2},\lambda_{1},\lambda_{2}) \cdot
f_{5}(q_{c1},q_{c2},k_{1},k_{2},\lambda_{3},
\lambda_{4},\lambda_{5},\lambda_{6})\,, \nonumber \\
E_{8,1,k}&=&f_{0}(q_{b1},q_{b2},\lambda_{1},\lambda_{2}) \cdot
f_{6}(q_{c1},q_{c2},k_{1},k_{2},\lambda_{3},
\lambda_{4},\lambda_{5},\lambda_{6})\,, \nonumber\\
E_{9,1,k}&=& f_{0}(q_{b1},q_{b2},\lambda_{1},\lambda_{2}) \cdot
f_{7}(q_{c1},q_{c2},k_{1},k_{2},\lambda_{3},\lambda_{4},
\lambda_{5},\lambda_{6})\,. \label{efunc}
\end{eqnarray}
Applying the exchange symmetries between the two gluons and among
the quarks, the following relations may be obtained:
\begin{eqnarray}
E_{1,3,k}&=&E_{4,2,k},\,\,E_{2,3,k}=E_{2,2,k},
\,\,E_{3,3,k}=E_{3,2,k}\,, \nonumber\\
E_{4,3,k}&=&E_{1,2,k},\,\,E_{5,3,k}=E_{5,2,k},\,\,
E_{1,4,k}=E_{4,1,k}\,, \nonumber\\
E_{2,4,k}&=&E_{2,1,k},\,\,E_{3,4,k}=E_{3,1,k},\,\,
E_{4,4,k}=E_{1,1,k}\,, \nonumber\\
E_{5,4,k}&=&E_{5,1,k},\,\,E_{9,4,k}=E_{9,1,k}+E_{9,2,k}-E_{9,3,k}\,.
\label{efunc1}
\end{eqnarray}
Furthermore, for the diagrams involving three gluon vertices, by
using a proper decomposition and the results in Tables.\ref{tab1}
and \ref{tab2}, $E_{6,j,k}$ may be replaced as
\begin{eqnarray}
E_{6,1,k}&=&E_{7,1,k}+2q_{c2}\cdot k_{2}E_{9,1,k}-
E_{3,2,k}+E_{1,2,k}\,, \nonumber\\
E_{6,2,k}&=&E_{7,2,k}+2q_{c2}\cdot k_{1}E_{9,2,k}-
E_{3,1,k}+E_{1,1,k}\,, \nonumber\\
E_{6,3,k}&=&E_{7,3,k}+2q_{b2}\cdot k_{2}E_{9,3,k}-
E_{3,4,k}+E_{1,4,k}\,, \nonumber\\
E_{6,4,k}&=&E_{7,4,k}+2q_{c2}\cdot k_{1}E_{9,4,k}-
E_{3,3,k}+E_{1,3,k}\,;
\label{efunc2}
\end{eqnarray}
and $E_{7,j,k}$ may be replaced as
\begin{eqnarray}
E_{7,1,k}&=&-E_{4,1,k}+E_{2,1,k}+E_{8,1,k}-
2q_{c1}\cdot k_{1}(2E_{5,1,k}-2E_{5,2,k}+E_{9,1,k})\,,  \nonumber\\
E_{7,2,k}&=&-E_{4,2,k}+E_{2,2,k}+E_{8,2,k}-
2q_{c1}\cdot k_{2}(2E_{5,2,k}-2E_{5,1,k}+E_{9,2,k})\,, \nonumber\\
E_{7,3,k}&=&-E_{4,3,k}+E_{2,3,k}+E_{8,3,k}-
2q_{b1}\cdot k_{1}(2E_{5,3,k}-2E_{5,4,k}+E_{9,3,k})\,, \nonumber\\
E_{7,4,k}&=&-E_{4,4,k}+E_{2,4,k}+E_{8,4,k}- 2q_{b1}\cdot
k_{2}(2E_{5,4,k}- 2E_{5,3,k}+E_{9,4,k})\,. \label{efunc3}
\end{eqnarray}

Thus the seven kinds of basic functions
$E_{m,j,k}\;(m=1,\cdots,5,8,9$, $j=1,\cdots,4$, $k=1,\cdots\,64)$
may be written in a very compact form. As an explicit
example, $E_{m,1,1}$
($m=1,\cdots,5,8,9$) is shown in one of the appendices.

\subsection{Rearrangement of the color factor for the amplitude}
As shown in section II, there are only five independent color
factors, and we may choose them as $C_{mij}\;(m=1,\cdots,5)$,
where $i, j \;(1,2,3)$ are the indices of the final $b$ and
$\bar{c}$ quarks' colors respectively. Thus the whole amplitude
may be rewritten as
\begin{eqnarray}
M^{(\lambda_{1}, \lambda_{4}, \lambda_{5},
\lambda_{6})}(q_{b1},q_{b2},q_{c1},q_{c2},k_{1},k_{2})&=&
\sum_{i=1}^{36}M_{i}^{(\lambda_{1}, \lambda_{4}, \lambda_{5},
\lambda_{6})}(q_{b1},q_{b2},q_{c1},q_{c2},k_{1},k_{2})\nonumber\\
&=&\sum_{m=1}^{5}C_{mij}M^{\prime(\lambda_{1},\lambda_{4},\lambda_{5},
\lambda_{6})}_{m}(q_{b1},q_{b2},q_{c1},q_{c2},k_{1},k_{2})\;,
\end{eqnarray}
where $C_{mij}\;(m=1,\cdots,5)$ are the five independent color
factors, defined in subsection A.
With Eq.(\ref{matrixd}),
$M^{\prime(\lambda_{1},\lambda_{4},\lambda_{5}, \lambda_{6})}_{m}
(q_{b1},q_{b2},q_{c1},q_{c2},k_{1},k_{2})$ (in short notation
$M^{\prime}_{m}$) can be further factorized as
\begin{equation}
M^{\prime}_{m}= \sum_{\lambda_{2},\lambda_{3}}
C_{mij}M^{\prime(\lambda_{1},\lambda_{2},\lambda_{3},\lambda_{4},\lambda_{5},
\lambda_{6})}_{Fm} (q_{b1},q_{b2},q_{c1},q_{c2},k_{1},k_{2})
D_{2}B_{B_c(B_c^*)}^{(\lambda_{2},\lambda_{3})}(q_{b2},q_{c1})\;,
\end{equation}
where $M^{\prime(\lambda_{1},\lambda_{2},\lambda_{3},
\lambda_{4},\lambda_{5}, \lambda_{6})}_{Fm}
(q_{b1},q_{b2},q_{c1},q_{c2},k_{1},k_{2})$ (in short notation
$M^{\prime}_{Fm}$) is the amplitude of the $2\rightarrow 4$ free
quark process $gg\to c+\bar{c}+b+\bar{b}$.

Owing to the fact that each of the terms in the amplitude
$M^{\prime}_{m}\,(M^{\prime}_{Fm})$ is related to one of the 36
Feynman diagrams, the explicit formulae for
$M^{\prime}_{m}(M^{\prime}_{Fm})\;(m=1,\cdots,5)$ may be written
down directly. With the nine kinds of basic functions
$E_{i,m,k}$ Eqs.(\ref{efunc}-\ref{efunc3}), $M^{\prime}_{Fm}$ may
be written as
\begin{eqnarray}
M^{\prime}_{F1}&=& \frac{D_{1}}{2}\Bigl(2 (X_{3a} + X_{4c} +
X_{5b}) E_{1,1,k} - 2 X_{4e} E_{2,4,k} -2 X_{4c} E_{3,1,k} + 2
X_{4e}
E_{4,4,k} \nonumber\\
& &-  X_{5d} (2 E_{5,1,k}- 4 E_{5,2,k} + E_{9,1,k} + E_{9,2,k}) +2
\{-(X_{1g} + X_{5c}) E_{6,1,k} \nonumber \\
& & +(X_{1b} + X_{1g} +X_{5c}) E_{6,2,k}+  X_{5c} (E_{8,1,k} -
E_{8,2,k}) - X_{2h} E_{8,3,k} +\nonumber\\
& &(X_{2f} + X_{2h}) E_{8,4,k} +2 X_{4e}E_{5,4,k}q_{b1}\cdot
k_{2}\}+ 4 X_{4c} E_{5,1,k} q_{c2}\cdot k_{1} -(X_{1g} +\nonumber\\
& & X_{5c}) (E_{9,1,k} - E_{9,2,k}) f_{3} + X_{5c}(4 E_{5,1,k} - 4
E_{5,2,k} +E_{9,1,k} -\nonumber\\
& &  E_{9,2,k}) f_{4} + X_{2h} (-4 E_{5,3,k} + 4
E_{5,4,k} - E_{9,3,k} + E_{9,4,k}) f_{6} -\nonumber \\
& & 2 X_{5b} \{E_{2,1,k} + E_{3,1,k}
- E_{4,1,k} - E_{5,1,k} (s_{c}-s_{1} + s_{b})\}\Bigr)\,,\\
M^{\prime}_{F2}&=&\frac{D_{1}}{2}\Bigl(2 (X_{3e} + X_{4a} +
X_{5a})
E_{1,2,k} - 2 X_{4g} E_{2,3,k} -2 X_{4a} E_{3,2,k} + \nonumber\\
& & 2 (X_{1a} + X_{1g} + X_{5c}) E_{6,1,k}  -(X_{1g}+X_{5c})
E_{6,2,k} +X_{5c} (E_{8,2,k}-E_{8,1,k}) + \nonumber\\
& &  (X_{2e} + X_{2h}) E_{8,3,k} - X_{2h} E_{8,4,k}) - X_{5d} (2
E_{5,2,k} + E_{9,1,k} +E_{9,2,k}) + \nonumber\\
& &2 X_{4g} (E_{4,3,k} + 2 E_{5,3,k} q_{b1}\cdot k_{1}) +4 X_{4a}
E_{5,2,k} q_{c2}\cdot k_{2} + (X_{1g} + X_{5c}) (E_{9,1,k} - \nonumber\\
& & E_{9,2,k})f_{3} + X_{5c}(4 E_{5,2,k} - E_{9,1,k} + E_{9,2,k})
f_{4} + 4 E_{5,1,k}\cdot (X_{5d}-  X_{5c} f_{4}) + \nonumber\\
& & X_{2h} (4 E_{5,3,k} - 4 E_{5,4,k} + E_{9,3,k}-E_{9,4,k})f_{6}-
2 X_{5a}\{E_{2,2,k}+E_{3,2,k} -\nonumber \\
& & E_{4,2,k}-E_{5,2,k} (-s_{2} + s_{b} + s_{c})\}\Bigr)\,,\\
M^{\prime}_{F3}&=&\frac{D_{1}}{2}\Bigl(-2(X_{4c} + X_{5b})
E_{1,1,k} +
2 X_{4e} E_{2,4,k} +2 (X_{3c} + X_{4c}) E_{3,1,k} +\nonumber\\
& & 2 X_{3d} E_{4,1,k} + 2 (X_{5c} E_{6,1,k} - X_{5c} E_{6,2,k}
+X_{2a} E_{6,4,k} +X_{2g} (-E_{6,3,k} +\nonumber\\
& & E_{6,4,k}) +X_{1d} E_{7,2,k} + X_{2d} E_{7,4,k} -(X_{1h} +
X_{5c}) E_{8,1,k} +\nonumber \\
&&(X_{1f} + X_{1h} + X_{5c}) E_{8,2,k}) + X_{5d} (2 E_{5,1,k} -
4 E_{5,2,k} + E_{9,1,k} +\nonumber\\
& &E_{9,2,k}) -2 X_{4e} (E_{4,4,k} + 2 E_{5,4,k} q_{b1}\cdot
k_{2})
-4 X_{4c} E_{5,1,k} q_{c2}\cdot k_{1}\nonumber\\
& & +X_{5c} (E_{9,1,k} - E_{9,2,k}) f_{3} -(X_{1h} + X_{5c}) (4
E_{5,1,k} - 4 E_{5,2,k} + \nonumber\\
& & E_{9,1,k} - E_{9,2,k}) f_{4} - X_{2g} (E_{9,3,k} - E_{9,4,k})
f_{5} + 2 \{ (X_{3b} + X_{5b})
E_{2,1,k} +\nonumber \\
&& X_{5b} [E_{3,1,k} - E_{4,1,k} - E_{5,1,k} (-s_{1} + s_{b} + s_{c})]\}\Bigr)\,,\\
M^{\prime}_{F4}&=&\frac{D_{1}}{2}\Bigl(-2 (X_{4a} + X_{5a})
E_{1,2,k}
+2 (X_{3g} + X_{4a}) E_{3,2,k} + 2 X_{3h} E_{4,2,k} -\nonumber\\
& &2 X_{5c} E_{6,1,k} + 2 (X_{5c} E_{6,2,k} + (X_{2a} + X_{2g})
E_{6,3,k} - X_{2g}E_{6,4,k} +\nonumber\\
& & X_{1c} E_{7,1,k} +X_{2c} E_{7,3,k} +
 (X_{1e} + X_{1h} + X_{5c}) E_{8,1,k} - (X_{1h} + X_{5c})
E_{8,2,k}) +\nonumber\\
& & X_{5d} (2 E_{5,2,k} + E_{9,1,k} + E_{9,2,k}) +2 X_{4g}
(E_{2,3,k} - E_{4,3,k} - \nonumber \\
&& 2 E_{5,3,k} q_{b1}\cdot k_{1}) -4 X_{4a} E_{5,2,k} q_{c2}\cdot
k_{2} -X_{5c} (E_{9,1,k} E_{9,2,k}) f_{3} - \nonumber\\
& &-(X_{1h} + X_{5c}) (4 E_{5,2,k} - E_{9,1,k} +
E_{9,2,k}) f_{4} +4 E_{5,1,k} (-X_{5d} + \nonumber\\
& & (X_{1h} + X_{5c}) f_{4}) +X_{2g} (E_{9,3,k} - E_{9,4,k})
f_{5} + 2 \{(X_{3f} + X_{5a}) E_{2,2,k} +\nonumber\\
& &X_{5a} [E_{3,2,k}
- E_{4,2,k} - E_{5,2,k}(-s_{2} + s_{b} + s_{c})]\}\Bigr)\,,\\
M^{\prime}_{F5}&=&D_{1}\Bigl(-(X_{5b} E_{1,1,k}) + (X_{4d} +
X_{5b}) E_{2,1,k} -(X_{3d} + X_{4d}) E_{4,1,k} - X_{3h} E_{4,2,k}
-\nonumber\\
& & X_{2a} (E_{6,3,k} + E_{6,4,k}) - X_{1e} E_{8,1,k} -X_{1f}
E_{8,2,k} - X_{5d}(E_{5,1,k} + E_{5,2,k} \nonumber\\
& & - E_{9,1,k} - E_{9,2,k})+ X_{4h} (-E_{1,3,k} + E_{3,3,k} -2
E_{5,3,k} q_{b2}\cdot k_{1}) + \nonumber\\
& &X_{4f} (-E_{1,4,k} + E_{3,4,k} -2 E_{5,4,k} q_{b2}\cdot k_{2})
-2 X_{4d} E_{5,1,k} q_{c1}\cdot k_{1} +\nonumber\\
& & X_{4b} (E_{2,2,k} - E_{4,2,k} - 2 E_{5,2,k} q_{c1}\cdot k_{2})
-X_{5b} (-E_{3,1,k} + E_{4,1,k} +\nonumber\\
& &E_{5,1,k} (-s_{1} + s_{b} + s_{c})) -X_{5a} \{E_{1,2,k} -
 E_{2,2,k} - E_{3,2,k} + \nonumber \\
& & E_{4,2,k} + E_{5,2,k}(-s_{2} + s_{b} + s_{c})\}\Bigr)\,,
\end{eqnarray}
where $X_{i}$ are the scalar factors of the propagators,
labelled by the corresponding Feynman diagram,
$s_{1}=(k_{1}-q_{b1}-q_{b2})^{2}$, $f_{5}=2k_{1}\cdot
k_{2}-2q_{b2}\cdot k_{1}-2q_{b2}\cdot k_{2}$ and
$f_{6}=2k_{1}\cdot k_{2}-2q_{b1}\cdot k_{1}-2q_{b1}\cdot k_{2}$.

\subsection{Programme check for the subprocess $gg\rightarrow
B_c(B_c^*) +b+\bar{c}$ and hadronic production of $B_c$}

With all the above preparations, it is straightforward to write
the programme for computing the cross section of the subprocess.
Applying the pQCD factorization theorem, various cross sections of
the $B_c(B_c^*)$-hadronic production may be obtained by
integrating the hadron structure functions over the cross section
of the subprocess $gg\to B_c(B_c^*)+b+\bar{c}$.

Various cross-sections of the subprocess $gg\rightarrow B_c(B_c^*)
+b+\bar{c}$ may be calculated, once the amplitude (with initial
state color factors averaged and final state color factors summed
up) has been computed. One only has to integrate over the proper
phase space according to the requirements. Since the final state
of the process is a massive three-body state, the phase space
integrations are carried out numerically. As in the more general
case, a Monte-Carlo simulation integration over the phase space is
a practical solution, when one is averaging over the helicities of
the initial gluons and spins of the quarks (and spins of $B_c^*$
if $gg\rightarrow B_c^*+b+\bar{c}$ is considered). To do the phase
space integration, we first use the routine RAMBOS~\cite{rkw} to
generate the requested phase space points (the energy-momentum
conservation is guaranteed and some additional constraints for
specific requests are matched). We then use the VEGAS~\cite{gpl}
program (with necessary revisions to suit the present problem) to
perform the integrations. The VEGAS program is useful for
obtaining accurate total cross-sections, smooth distributions for
the $p_\mathrm{T}$\footnote{$p_\mathrm{T}$ is the transverse
momentum with respect to the beam direction.} and the rapidity $Y$
of the $B_c$-meson, and so on. When running VEGAS, the most
important samples for the matrix element squared are taken first.
Then by taking an adequate number of points for the integration,
we obtain a final result, which is stable with respect to
increasing the number of points and compatible with the requested
statistical error.

To check the program, we calculate the total cross-section of the
subprocess, $gg\rightarrow B_c(B^{(*)}_c) +b+\bar{c}$, as well as
the $B_c$ $p_\mathrm{T}$ and $Y$ distributions for various subprocess
center-of-mass energies, using the same parameters ($e.g.$
$\alpha_s$, $m_b$, $m_c$, $m_{B_c}$ and $f_{B_c(B_c^*)}$) as in
Refs.\cite{prod1,prod2}. The results are compared with those in
Refs.\cite{prod1,prod2} and shown in Table~\ref{tab3} and
Table~\ref{tab4}. The $B_c$-$p_\mathrm{T}$ and $B_c$-$Y$
distributions for various subprocess center-of-mass energies are
compared with Ref.\cite{prod1} and shown in Fig.\ref{subpt1}. The
integrated cross sections versus the center-of-mass energy of the
subprocess for a fixed value of $\alpha_s=0.2$ are also shown in
Fig.\ref{totsubshat}. One may see the agreement between our
results and those in
Ref.\cite{prod1,prod2} from the tables and figures. Since the
present programme and that used in Ref.\cite{prod1,prod2} are
totally different, the agreement is a very solid check for
both of them.

\begin{table}
\caption{Comparison of total cross sections for $gg\rightarrow B_c
(B^{*}_c) +b+\bar{c}$ with the corresponding results of
Ref.\cite{prod1}. The input parameters are $m_b=$~4.9 GeV,
$m_c=$~1.5 GeV, $M_{B^{*}c}=m_{b}+m_{c}$, $f_{Bc}=$~0.480 GeV,
$\alpha_{s}=0.2$. The number in parenthesis shows the Monte Carlo
uncertainty in the last digit. The cross sections are expressed in
nb.}
\begin{center}
\begin{tabular}{|c||c|c|c|}
\hline\hline
$\sqrt{\bar{s}}$  & 20 GeV & 30 GeV  & 60 GeV  \\
\hline \(\sigma_{B_c}\) & $0.6579(5)\times 10^{-2}$ &
$0.9465(8)\times 10^{-2}$
& $0.7872(8)\times 10^{-2}$   \\
\hline \(\sigma_{B_c}\)\cite{prod1} & $0.661(7)\times 10^{-2}$ &
$0.949(8)\times 10^{-2}$ & $0.782(9)\times 10^{-2}$   \\
\hline\hline \(\sigma_{B^{*}_c}\) & $0.1606(1)\times 10^{-1}$ &
$0.2460(3)\times 10^{-1}$ & $0.2033(2)\times 10^{-1}$  \\
\hline \(\sigma_{B^{*}_c}\)\cite{prod1} & $0.160(2)\times 10^{-1}$
&
$0.244(3)\times 10^{-1}$ & $0.203(3)\times 10^{-1}$  \\
\hline\hline
\end{tabular}
\label{tab3}
\end{center}
\end{table}

\begin{table}
\caption{Comparison of total cross sections for $gg\rightarrow B_c
+b+\bar{c}$ with the corresponding results of Ref\cite{prod2}. The
input parameters are $m_b=3m_c$, $M_{B_c}=6.30GeV$,
$f_{Bc}=0.480GeV$, $\alpha_{s}=0.2$. The number in parenthesis
shows the Monte Carlo uncertainty in the last digit. The cross
sections are expressed in nb.}
\begin{center}
\begin{tabular}{|c||c|c|c|c|}
\hline\hline
$\sqrt{\bar{s}}$ & 20 GeV  &  30 GeV  & 60 GeV & 80 GeV  \\
\hline \(\sigma_{B_c}\) & $0.6853(5)\times 10^{-2}$ &
$0.9731(8)\times 10^{-2}$
& $0.7997(9)\times 10^{-2}$  & $0.6244(9)\times 10^{-2}$  \\
\hline \(\sigma_{B_c}\)\cite{prod2} & $0.686(2)\times 10^{-2}$ &
$0.971(4)\times
10^{-2}$ & $0.793(5)\times 10^{-2}$ & $0.623(5)\times 10^{-2}$  \\
\hline\hline
\end{tabular}
\label{tab4}
\end{center}
\end{table}

According to pQCD, the production cross section is formulated
\begin{equation}
d\sigma=\sum_{ij}\int dx_{1}\int
dx_{2}F^{i}_{H_{1}}(x_{1},\mu_{F})\times
F^{j}_{H_{2}}(x_{2},\mu_{F})d\hat{\sigma}_{ij\rightarrow
B_{c}X}(x_{1},x_{2},\mu_{F}), \label{pqcdf}
\end{equation}
where $F^{i}(x,\mu_{F})$ is this distribution function of the
parton $i$ in the hadron $H$, $d\hat{\sigma}_{ij\rightarrow
B_{c}X}(x_{1},x_{2},\mu_{F})$ is the cross section for the
relevant inclusive production ($i+j\rightarrow B_{c}+X$). With the
pQCD factorization formula Eq.(\ref{pqcdf}), various cross
sections for the hadronic production of the meson $B_c(B_c^*)$ can
be computed by integrating over two more dimensions of the
structure functions of the incoming hadrons.

The programme for the hadronic production of the
$B_c(B_c^*)$ mesons has also been checked by evaluating the hadronic
production of $B_c$ at Tevatron. The explicit example is to
produce the $B_c$ meson by using the next to leading order running
$\alpha_s$, the characteristic energy scale $Q^2=\bar{s}/4$
(NQ2=$1$) of the production and the CTEQ3M set of parton
distribution functions. The results for the distributions and the
cross sections agree with those in Ref.\cite{prod2}.

\section{The programme: BCVEGPY generator}

The programme BCVEGPY is a generator
for hadronic production of $B_c$ mesons in form of a Fortran package,
based on the
dominant subprocess $gg\to B_c(B_c^*)+b+\bar{c}$.
Concerning its implementation in PYTHIA, BCVEGPY is
written in the same format as in PYTHIA (including common block variables).
Thus it is easy to implement straightforward in
PYTHIA as an external process, and in this way all the functions
of PYTHIA can be utilized in connection with the use of BCVEGPY.

\subsection{Structure of the program}

The BCVEGPY Fortran package contains five files: bcvegpy.for,
genevnt.for, sqamp.for, foursets.for and pythia6208.for. The
routine bcvegpy.for is the main program of BCVEGPY which takes
care of the necessary input parameters for the event generation
and outputs a variety of results, such as the distribution of the transverse
momentum $p_\mathrm{T}$ and rapidity $Y$ of the produced $B_{c}$,
$etc.$. In the routine genevnt.for, there is a function
TOTFUN and five subroutines: EVNTINIT, UPINIT, UPEVNT, BCPYTHIA
and PHPOINT. The subroutines EVNTINIT and UPINIT are used for
initializing the parameters when running BCVEGPY in the PYTHIA
environment. The event generation starts by calling PYEVNT (a
PYTHIA routine) just after calling the subroutines EVNTINIT and
UPINIT. The routine UPEVNT (a PYTHIA user routine) is then called,
that allows the implementation of an external processes. The
functioning of the subroutine UPEVNT is determined by an input
flag, which is a number read in from the file input.dat. When the
input flag switches on, the generation of complete events may be
carried out. UPEVNT calls the subroutine BCPYTHIA and the full
information, $i.e.$ the status code, the mother code, the color flow
$etc.$ for the final state particles ($B_c(B_c^*)$
meson and two jets $b$ and $\bar{c}$). The allowed
phase-space for the production ($i.e.$ energy-momentum consevation)
controls the production completely. The phase-space is generated
by calling the subroutine PHPOINT. The function TOTFUN
computes the value of the integrand for the phase space
integration by calling the subroutine AMP2UP which is in
sqamp.for. The subroutine AMP2UP is written according to the
techniques described in the previous sections and its purpose is
to compute the square of the amplitude for the hard subprocess
$gg\rightarrow B_{c}(B_{c}^{*})+b+\bar{c}$. In files sqamp.for and
foursets.for, there are quite a lot subroutines and functions,
that are needed for calculating the square of the amplitude. All of them
will be explained in Appendix C. When running the package BCVEGPY,
the PYTHIA library must be linked; in particular, the generator is
designed to be interfaced to the PYTHIA version 6.2\cite{pythia}.
For reference, we include the file pythia6208.for in our package.
In order to increase the phase space integration accuracy,
one may apply the subroutine VEGAS (in sqamp.for) to optimize
the sampling of the phase space points for phase space integration
before calling PYTHIA.

\subsection{Use of the generator}

BCVEGPY can be used for generating a huge sample of $B_c(B_c^*)$
meson for hadronic collisions efficiently (based on the mechanism
with $gg\to B_c(B_c^*)+b+\bar{c}$ as the main subprocess).
Furthermore, if the generator is implemented in PYTHIA, a complete
event with two hadronized quark jets and a decayed
$B_c(B_c^*)$ meson can be simulated. If one would like to
simulate only the hadronic production of the $B_c(B_c^*)$ meson,
BCVEGPY alone is sufficient. Furthermore, users may choose
either the hadronic production or the subprocess $gg\to
B_c(B_c^*)+b+\bar{c}$ only, just by setting the value of the flag ISUBONLY
equal to $0$ to switch on the integration for the parton
distribution (structure) functions, or setting the flag equal to $1$ for the
subprocess only. Since the cross section of $B_c$ production in
hadron collisions is small compared to the $B^+, B^0,
B_s$ production, the efficiency for producing $B_c$ events
through fragmentation of a
$\bar{b}$ quark, as done in the standard PYTHIA, is too low (the
ratio of the signal to the background is too small). For
experimental feasibility studies of $B_c$ mesons, a very large
sample of $B_c$ events is needed. Therefore BCVEGPY, which is
powerful for generating $B_c$ events only with full information in
hadron-hadron collision, is very useful.

Users may communicate with or give instructions to the program
through an input file (input.dat). The output files include
1s0.dat (the $B_c$ total cross-section) or 3s1.dat (the $B_c^{*}$
total cross-section); pt.dat (the $B_c$-$p_\mathrm{T}$
distribution); shat.dat (the $B_c$-$\sqrt{\bar{s}}$ distribution);
rap.dat (the $B_c$-$Y$ rapidity distribution); pseta.dat (the
$B_c$-$\eta$ pseudorapidity distribution) and grade.dat (the
sampling importance function obtained by VEGAS). The input.dat
file allows users to setup the generation parameters and requests,
while the output files collect the relevant event information.

The sequential order and the format of the
parameters in the file input.dat should be
the same as in the Table~VI. The parameters
specified in the input file are:

\begin{table}
\begin{center}
\label{input} \caption{The parameter values in the
sequential order in the input.dat file.}
\begin{tabular}{|l|} \hline   PMBC
$\;\;\;$PMB$\;\;\;$ PMC$\;\;\;$ FBC\\ PTCUT $\;\;\;$ETACUT$\;\;\;$
ECM $\;\;\;$IBCSTATE$\;\;\;$
IGENERATE$\;\;\;$ IVEGASOPEN\\
NUMBER $\;\;\;$ NITMX \\
NQ2 $\;\;\;$NPDFU$\;\;\;$ NEV \\
ISHOWER$\;\;\;$
MSTP(51)\\
IDWTUP$\;\;\;$ MSTU(111)$\;\;\;$ PARU(111)\\
ISUBONLY$\;\;\;$ SUBENERGY$\;\;\;$ IGRADE\\
\hline
\end{tabular}
\end{center}
\end{table}

\noindent $\bullet $ PMBC, PMB, PMC=: masses of the $B_c$ meson,
$b$ quark and $c$ quark respectively (in units GeV);

\noindent $\bullet $ FBC=: decay constant for the $B_c$ meson (in
units GeV);

\noindent $\bullet $ PTCUT=:  $p_\mathrm{T}$ cut
for the $B_c (B_c^*)$ meson (in units GeV, value can be
freely selected by users);

\noindent $\bullet $ ETACUT=:  $Y$ cut for the $B_c
(B_c^*)$ meson (value can be freely selected by the users);

\noindent $\bullet $ ECM=: total energy for the hadron collision
(in units GeV);

\noindent $\bullet $ IBCSTATE=: state of the $B_c (B_c^*)$ meson:
IBCSTATE=$1$ for $B_c[1^{1}S_{0}]$ and IBCSTATE=$2$ for
$B_c^*[1^{3}S_{1}]$;

\noindent $\bullet $ IGENERATE=: whether to generate complete
events. IGENERATE=$0$, when users wish the simulation to
stop after the generation of the `final state' containing the $B_c$ meson, $b$-jet and
$\bar{c}$-jet of the subprocess $gg\to B_c+b+\bar{c}$;
IGENERATE=$1$, when users wish that complete events including the $B_c$
production are to be generated. In the latter case, IDWTUP=$1$;

\noindent $\bullet $ IVEGASOPEN=: whether switch on/off the VEGAS
subroutine: IVEGASOPEN=$1$ for using VEGAS; IVEGASOPEN=$0$ for not
using VEGAS;

\noindent $\bullet $ NUMBER=: total number of times for calling the
integrand (VEGAS parameter, see VEGAS manual). The parameter is
needed only when IVEGASOPEN=$1$ in each iteration;

\noindent $\bullet $ NITMX=: upper limit for the number of
iterations (VEGAS parameter, see VEGAS manual). The parameter
is needed only when IVEGASOPEN=$1$;

\noindent $\bullet $ NQ2=: choice of $Q^2$, the type of the
characteristic energy scale squared in the production (in units GeV$^2$). Here seven
choices are available: i). NQ2=$1$: $Q^2=\bar{s}/4$ ($\bar{s}$ is
the squared center-of-mass energy of the subprocess);
ii). NQ2=$2$: $Q^2=\bar{s}$; iii). NQ2=$3$:
$Q^2=p_{TB_c}^2+m_{B_c}^2$; iv). NQ2=$4$:
$Q^2=(\sqrt{p_{TB_c}^2+m_{B_c}^2}+\sqrt{p_{Tb}^2+m_{b}^2}
+\sqrt{p_{Tc}^2+m_{c}^2})^2$; v). NQ2=$5$:
$Q^2=(\sqrt{p_{TB_c}^2+m_{B_c}^2}+\sqrt{p_{Tb}^2+
m_{b}^2}+\sqrt{p_{Tc}^2+m_{c}^2})^2/9$; vi).
NQ2=$6$: $Q^2=p_{Tb}^2+m_{b}^2$ for the $\alpha_s$
in parton distribution functions and in the coupling to the
$b$-quark line, and $Q^2=4m_c^2$ for the
$\alpha_s$ in the coupling to the $\bar{c}$-quark line; vii).
NQ2=$7$: $Q^2=p_{Tb}^2+m_{b}^2$;

\noindent $\bullet $ NPDFU=: choice of the collision type of
hadrons. The assignments can be found in PYTHIA manual, $e.g.$
NPDFU=$1$ for $p-\bar{p}$ and NPDFU=$2$ for $p-p$;

\noindent $\bullet $ NEV=: number of the events for the hadronic
production $h+h\to B_c+\cdots$ ($h$ means a hadron) to be
generated;

\noindent $\bullet $ ISHOWER=: whether to switch on/off the showers,
including initial and final states, multiple interactions,
hadronization; $e.g.$ ISHOWER=$1$ for `on' and ISHOWER=$0$ for `off'
(see PYTHIA manual);

\noindent $\bullet $ MSTP(51)=: choice of the proton
parton-distribution set; $e.g.$ MSTP(51)=$2$ for CTEQ3M;
MSTP(51)=$7$ for CTEQ5L; MSTP(51)=$8$ for CTEQ5M $etc.$ (PYTHIA
parameter, see PYTHIA manual);

\noindent $\bullet $ IDWTUP=: master switch dictating how the
event weights and the cross-sections should be interpreted (PYTHIA
parameter, see PYTHIA manual); $e.g.$ when IDWTUP=$1$, parton-level
events have a weight at the input to PYTHIA. Events are then either
accepted or rejected, so that fully generated events at the output
have a common weight; when IDWTUP=$3$, parton-level events have a
unit weight at the input to PYTHIA $i.e.$ they are always accepted;

\noindent $\bullet $ MSTU(111)=: order of $\alpha_s$ in the
evaluation in PYALPS (a PYTHIA routine for calculating $\alpha_s$,
see PYTHIA manual); $e.g.$ MSTU(111)=$1$ for leading order (LO);
MSTU(111)=$2$ for next leading order (NLO);

\noindent $\bullet $ PARU(111)=: constant value of $\alpha_s$ (see
PYTHIA manual), which is used only when MSTU(111)=$0$;

\noindent $\bullet $ ISUBONLY=: whether to
keep the information only of the sub-process $gg\rightarrow
B_c(B^{*}_c) +b+\bar{c}$; ISUBONLY=$0$ for
the full hadronic production, $i.e.$ the structure functions
are connected; ISUBONLY=$1$ for the subprocess only.

\noindent $\bullet $ SUBENERGY=: the energy (in units GeV) of the
sub-process $gg\rightarrow B_c(B_c^{*}) +b+\bar{c}$. It is
needed only when ISUBONLY=$1$;

\noindent $\bullet $ IGRADE=: whether to use the grade generated
by previous running VEGAS when IVEGASOPEN=$0$; IGRADE=$1$ means to
use; IGRADE=$0$ means not to use.

In the package, two subroutines PHASE-GEN and VEGAS are needed
when integrating over the phase space:

\vspace{2mm} \fbox{\bf The subroutine PHASE-GEN(YY,ET,WT)}
\vspace{2mm}

The subroutine is included in the file sqamp.for and is called by
another subroutine PHPOINT in genevnt.for. The purpose is to
evaluate the allowed phase-space points for the sub-process
$gg\rightarrow B_c(B^{*}_c) +b+\bar{c}$  when the center-of-mass
energy of the sub-process is fixed, and to return a non-zero weight for
each allowed phase space point.
The energy-momentum conservation is integrated out in order to reduce
the dimension of the phase space integration by four. Since the
subprocess is a three body final state,
the nine-dimensional phase space integration of the process
turns into a five-dimensional one with a proper Jacobi determinant,
when PHASE-GEN has been applied.

The variables in the routine are:

$\bullet$ YY(5)=: a five-dimensional random number with a range
from $0$ to $1$ for each dimension, corresponding to the five
independent integration variables for the phase-space;

$\bullet$ ET=: center-of-mass energy for the subprocess (in units
GeV); ET can be chosen freely when ISUBONLY=$1$, otherwise when
ISUBONLY=$0$, ET is determined by PHPOINT;

$\bullet$ WT=: the returned weight for each generated phase-space
point.

\vspace{2mm} \fbox{\bf The subroutine
VEGAS(FXN,NDIM,NCALL,ITMX,NPRN)} \vspace{2mm}

$\bullet$ FXN=: the integrand calling the function TOTFUN in
genevnt.for;

$\bullet$ NDIM=: number of integration dimensions for the
generator; NDIM=$5$ when ISUBONLY=$1$; NDIM=$7$ when ISUBONLY=$0$;

$\bullet$ NCALL=: maximum total number of the times to call the
integrand in each iteration set by the user;

$\bullet$ ITMX=: maximum number of allowed iterations set by the
user;

$\bullet$ NPRN=: print out level; (see VEGAS manual) $e.g.$
NPRN=$2$, when printing out only the cross section values and errors.

Note that the units of all the output data is well explained in
the programme.

\subsection{Generator checks}
The whole Fortran package is checked by examining the gauge
invariance of the amplitude. The matrix element vanishes
when the polarization vector of an initial gluon is substituted by
the momentum vector of this gluon.

For further checking, we have performed several test runs. By
setting the parameter ISUBONLY=$1$ in the input.dat file, we
obtain the transverse momentum $p_\mathrm{T}$ and rapidity $Y$
distributions for the produced $B_c$, and the total cross-section
for the sub-process $gg\rightarrow B_c(B^{*}_c) +b+\bar{c}$. The
results, which are shown in the previous section, coincide well
with several groups' results\cite{prod2}.

\begin{table}
\caption{Generation parameters used in the sample generation.}
\begin{center}
\begin{tabular}{l}
\hline
.............. INITIAL PARAMETERS ................\\
Bc IN $1^{1}S_0$\\
GENERATE EVNTS 30000000 FOR TEVATRON ENERGY(GEV) $0.20E+04$\\
\noindent M\_\{Bc\}=$6.400$$\;\;$M\_\{B\}=
$4.900$$\;\;$M\_\{C\}=$1.500$
$\;\;$f\_\{Bc\}=$0.4800$ \\
Q2 TYPE= 1 $\;\;\;$ ALPHAS ORDER= 2\\
PARTON DISTRIBUTION FUNCTION: CTEQ 3M \\
USING PYTHIA MODEL FOR IDWTUP= 3\\
PTCUT =0.000$\;\;\;$GeV\\
NO RAPIDITY CUT\\
USING VEGAS: NUMBER IN EACH ITERATION= 300000  ITERATION= 27\\
.............. END OF INITIALIZATION ..............\\
\hline
\end{tabular}
\label{output}
\end{center}
\end{table}

When running the programme, the initialization is shown as a
screen snap-shot in Table~\ref{output}. Some output data are shown
by Figs.\ref{totpt1},\ref{totrap1}.

\section{Conclusions}

A $B_c$ meson generator BCVEGPY for
hadronic collisions,
based on the dominant sub-process $gg\rightarrow
B_c(B_c^*) +b+\bar{c}$,
has been developed and well-tested.
The generator has been interfaced with PYTHIA, which takes care of
producing the full event and filling the standard PYTHIA event
common block. In view of the prospects for $B_c$ physics at
Tevatron and at LHC, the generator offers a valuable platform for
further experimental studies.

\vspace{20mm}
{\bf\Large Acknowledgement:} The author (C.-H. Chang) would like
to thank Lund University for warm hospitality during his visit.
The authors would like to thank Dr.
Y.-Q. Chen and Mr. Y.-B. Sun for useful discussions, and
Prof. T. Sj\"ostrand for valuable advice and comments. The work was
supported in part by Nature Science Foundation of China (NSFC).

\appendix
\section{The helicity functions for the amplitude}

In this Appendix, we show an example how to calculate the
functions $E_{m,j,k}$.

For evaluating the inner product \(\langle p\cdot q \rangle\), we
introduce the notations \(k_{\pm}\), \(k_{\perp}\) for a
light-like momentum \(k^{\mu}\) and use the Weyl representation for
\(\gamma\)- matrices:
\begin{equation}
\gamma^{0}=\left( \begin{array}{cc} 0 & {\bf 1}\\ {\bf 1} & 0
\end{array}\right)\,,\,\,
\gamma^{i}=\left( \begin{array}{cc} 0 & -{\bf \sigma}^{i}\\ {\bf
\sigma}^{i} & 0 \end{array}\right)\,,\,\,(i=1,2,3)
\end{equation}
with Pauli matrices
\begin{equation}
\sigma^{1}=\left( \begin{array}{cc} 0 & {1}\\ {1} & 0
\end{array}\right)\,,\,\,\,
\sigma^{2}=\left( \begin{array}{cc} 0 & {-i}\\ {i} & 0
\end{array}\right)\,,\,\,\,
\sigma^{3}=\left( \begin{array}{cc} 1 & {0}\\ {0} & -1
\end{array}\right)\,,
\end{equation}
and
\begin{equation}
k_{\pm}=k_{0}\pm k_{z},\,\, k_{\perp}=k_{x}+i k_{y}=|k_{\perp}
|e^{i\varphi_{k}} =\sqrt{k_{+}k_{-}} e^{i\varphi_{k}}\,,
\end{equation}
where \({\bf 1}\) is the \(2 \times 2\) unit matrix and \({\bf
\sigma}^{i}\) (i=1,2,3) are the Pauli matrices. Note that here we
always have $k_+\geq 0, k_-\geq 0$, because $k_+k_-\equiv
k_0^2-k_z^2=k_x^2+k_y^2\geq 0$ due to
$k^2=k_0^2-k_z^2-k_x^2-k_y^2=0$. By suitable choice of the phase
we introduce the Weyl spinors
\begin{equation}
|k_{+}\rangle=\left( \begin{array}{c} \sqrt{k_{+}}\\
\sqrt{k_{-}}e^{i\varphi_{k}} \\ 0\\0 \end{array}\right),\,\,
|k_{-}\rangle=\left( \begin{array}{c} 0\\0\\\sqrt{k_{-}}e^{-i\varphi_{k}}\\
-\sqrt{k_{+}} \end{array}\right)\,,
\end{equation}
and
\begin{eqnarray}
\langle k_{1}\cdot k_{2} \rangle&=&\langle k_{1-}|k_{2+} \rangle
=\sqrt{k_{1-}k_{2+}}e^{i\varphi_{1}}-\sqrt{k_{1+}k_{2-}}e^{i\varphi_{2}}
\nonumber \\
&=& k_{1\perp}\sqrt{\frac{k_{2+}}{k_{1+}}}-
k_{2\perp}\sqrt{\frac{k_{1+}}{k_{2+}}}\,,
\end{eqnarray}
which appear to be explicitly antisymmetric. For the spinor
product \(\langle k_{1+}|\slash\!\!\!k_{3}|k_{2+}\rangle\), where
\(k^{2}_{i}=0\) (\(i=1,2,3\)),
\begin{eqnarray}
\langle k_{1+}|\slash\!\!\!k_{3}|k_{2+}\rangle &=& \langle
k_{1+}|k_{3-}\rangle\langle k_{3-}|k_{2+}\rangle \nonumber \\
&=&\frac{1}{\sqrt{k_{1+}k_{2+}}} (k_{1+}k_{2+}k_{3-}-k_{1+}
k_{2\perp}k_{3\perp}^{*}-k_{1\perp}^{*}k_{2+}k_{3\perp}+
k^{*}_{1\perp}k_{2\perp}k_{3+})\,,
\end{eqnarray}
and for the spinor product involving the polarization vector
$\epsilon(s_{z})$ of $B^{*}_{c}$,
\begin{eqnarray}
\langle q_{0+}| \slash\!\!\!\epsilon(s_{z})\,\,
\slash\!\!\!\!P|q_{0-} \rangle &=&
P^{\prime}_{+}\epsilon(s_{z})_{-}q^{*}_{0\perp}-(P^{\prime
}_{\perp})^{*} q_{0+}\epsilon(s_{z})_{-}-P^{\prime}_{+}
\epsilon(s_{z})_{\perp} \frac{q^{2}_{0\perp}}{q_{0+}}+
\epsilon(s_{z})_{\perp}(P^{\prime }_{\perp})^{*}
q^{*}_{0\perp}-\nonumber\\
& &P^{\prime}_{\perp} \epsilon(s_{z})^{*}_{\perp}q^{*}_{0\perp}+
q_{0+}\epsilon^{*}_{\perp}P^{\prime}_{-}+
\epsilon(s_{z})_{+}P^{\prime}_{\perp}\frac{q^{2}_{0\perp}}
{q_{0+}}-\epsilon(s_{z})_{+}P^{\prime}_{-}q^{*}_{0\perp}\,,
\end{eqnarray}
where $P^{\prime}=P-\frac{P^{2}}{2P\cdot q_{0}}q_{0}$.

Since the spinor products and the double inner spinor products are
used frequently in the program, we take $y_{i}\;(i=1,\cdots,42)$
to denote all the possible spinor products appearing in the
computation, $e.g.$ $y_{1}=\langle
k_{1+}|\slash\!\!\!q^{\prime}_{c2}|q_{0+}\rangle$ and $x_1, x_2$
to denote the double inner spinor products:
\begin{equation}
x_{1}= \langle q_{0}\cdot k_{1} \rangle \langle q_{0}\cdot k_{2}
\rangle,\,\,\,\;\; x_{2}=\langle q_{0}\cdot k_{1} \rangle^{*}
\langle q_{0}\cdot k_{2} \rangle^{*}\,.
\end{equation}
To simplify the results, we have introduced the light-like
momenta $q^{\prime}_{b1}$, $q^{\prime}_{b2}$, $q^{\prime}_{c1}$
and $q^{\prime}_{c2}$ in terms of time-like momenta $q_{b1}$,
$q_{b2}$, $q_{c1}$ and $q_{c2}$ respectively:
\begin{eqnarray}
q^{\prime}_{b1}&=&q_{b1}-\frac{q_{b1}^{2}}{2q_{b1}\cdot
q_{0}}q_{0}\,,\;\;
q^{\prime}_{b2}=q_{b2}-\frac{q_{b2}^{2}}{2q_{b2}\cdot
q_{0}}q_{0}\,, \nonumber \\
q^{\prime}_{c1}&=&q_{c1}-\frac{q_{c1}^{2}}{2q_{c1}\cdot
q_{0}}q_{0}\,,\;\;
q^{\prime}_{c2}=q_{c2}-\frac{q_{c2}^{2}}{2q_{c2}\cdot
q_{0}}q_{0}\,.
\end{eqnarray}

When the functions $E_{m,1,k}\,\,(m=1, \cdots, 5, 8, 9;\,\,k=1,
\cdots, 64)$ are given, $E_{m,j,k}\;(j=2, \cdots, 4)$ can be
obtained by interchanging the initial gluon momenta and the final
quark flavors. As shown below, due to the properties of the
helicities, the functions may be simplified, and may even become
zero for some helicities. Here we list $E_{m,1,1}\; (m=1, \cdots,
5, 8, 9)$ for an explicit example,
\begin{eqnarray}
E_{1,1,1}&=&
\frac{4{y_1}}{x_{1}}\Bigl({{m_b}}^2(({y_{11}}-{y_{12}}){y_{19}}{y_{27}}+
{y_{10}}({y_{34}}{y_{34}^{c}} - {y_5}{y_{5}^{c}}))\nonumber
\\ &+& {y_{10}}(
{y_{20}}({y_2}{y_{5}^{c}}-{y_{24}^{c}}{y_{34}^{c}})+{{m_c}}^2
{y_{35}}{y_{35}^{c}})\Bigr)\,,\nonumber\\
E_{2,1,1}&=&
\frac{-4}{x_{1}}\Bigl({{m_b}}^2{y_9}({y_{11}}{y_{19}}{y_{27}}+
{y_{10}}{y_{34}}{y_{34}^{c}})\nonumber\\
&+&{y_{10}}\{-({y_9}{y_{20}}{y_{24}^{c}}
{y_{34}^{c}})+{{m_c}}^2({y_2}{y_{29}}{y_{31}} +{y_9}{y_{35}}
{y_{35}^{c}})\}\Bigr)\,,\nonumber \\
E_{3,1,1}&=&
\frac{4{y_1}{y_{16}}}{x_{1}}\Bigl(({{m_c}}^2{y_{17}}-{y_{11}}{y_{32}})
{y_{17}^{c}}+{{m_b}}^2({y_5}{y_{5}^{c}}-{y_{34}}{y_{34}^{c}})\nonumber\\
&+&{y_{20}}\{(y_{7}-{y_2}){y_{5}^{c}}+{y_{24}^{c}}{y_{34}^{c}}\}-
{{m_c}}^2{y_{35}}{y_{35}^{c}}\Bigr)\,,\nonumber \\
E_{4,1,1}&=&
\frac{4{y_{16}}}{x_{1}}({y_9}\Bigl({y_{11}}{y_{32}}{y_{17}^{c}}+({{m_b}}^2
{y_{34}}-{y_{20}}{y_{24}^{c}}){y_{34}^{c}})\nonumber
\\&+&{{m_c}}^2\{({y_2}-{y_7}){y_{29}}{y_{31}}+{y_9}({y_{35}}{y_{35}^{c}}
-{y_{17}}{y_{17}^{c}})\}\Bigr)\,,\nonumber \\
E_{5,1,1}&=& \frac{2{y_1}{y_{16}}{y_{27}} {y_{29}}}{x_{1}}\,,\nonumber\\
E_{8,1,1}&=& \frac{4}{x_{1}}\Bigl({y_9}({y_{12}}-{y_{18}})
({{m_b}}^2{y_{34}}- {y_{20}} {y_{24}^{c}}){y_{34}^{c}}+
2{q_{c1}}\cdot{k_1} \langle q_{0}\cdot q_{b1}^{\prime} \rangle
\langle k_{2}\cdot k_{1} \rangle^{*}{y_{27}} {y_{36}}{y_{35}^{c}}
+ \nonumber\\
& & {{m_c}}^2\{({y_{2}} (y_{12}-{y_{18}})- {y_7}{y_{16}}){y_{29}}
{y_{31}}+ {y_9}(({y_{12}}-{y_{18}}){y_{35}}{y_{35}^{c}}-{y_{16}}
{y_{19}}{y_{29}})\}\Bigr) \\
E_{9,1,1}&=& 0\,,
\end{eqnarray}
where the index \(c\) means charge conjugation of the spinor,
\(|p_{-}\rangle=|p_{+}\rangle ^{c}\).

\section{Polarization vector for $B^*_{c}[1^{3}S_{1}]$ meson}

The expressions of the polarization vectors depend on the gauge
choice. Here we choose a cartesian basis for the polarization vectors:
\begin{eqnarray}
\epsilon^{z}(P)&=&\frac{1}{\sqrt{P_{0}^{2}-P_{z}^{2}}}
(P_{z}\,,0\,,0\,,P_{0})\,,\\
\epsilon^{x}(P)&=&\frac{|P_{z}|}{P_{z}M}\Biggl(\frac{P_{x}P_{0}}
{\sqrt{P_{0}^{2}-P_{y}^{2}-P_{z}^{2}}}\,,
\sqrt{P_{0}^{2}-P_{y}^{2}-P_{z}^{2}}\,, \frac{P_{x}P_{y}}
{\sqrt{P_{0}^{2}-P_{y}^{2}- P_{z}^{2}}}\,,\nonumber \\
&&\frac{P_{x}P_{z}} {\sqrt{P_{0}^{2}
-P_{y}^{2}-P_{z}^{2}}}\Biggr)\,,\\
\epsilon^{y}(P)&=&\left(\frac{P_{y}P_{0}}{\sqrt{P^{2}_{0}-P^{2}_{z}}
\sqrt{P_{0}^{2}-P_{y}^{2}- P_{z}^{2}}}, 0,
\frac{\sqrt{P^{2}_{0}-P^{2}_{z}}}{\sqrt{P_{0}^{2}-P_{y}^{2}-
P_{z}^{2}}},\frac{P_{y}P_{z}}{\sqrt{P^{2}_{0}-P^{2}_{z}}
\sqrt{P_{0}^{2}-P_{y}^{2}- P_{z}^{2}}}\right)\,,
\end{eqnarray}
which satisfy the conditions
\begin{equation}
\epsilon^{i}\cdot P=0\,,\,\;\;\; \epsilon^{i}
\cdot\epsilon^{j}=-\delta^{ij}\,,\,\,\;\;\;(i,j=x,y,z)\,.
\end{equation}

\section{Routines and functions for the helicity amplitude}

In this Appendix, subroutines and functions for calculating the
helicity amplitudes for the sub-process $gg\rightarrow
B_c(B^{*}_c) +b+\bar{c}$ are explained.

\noindent {\bf SUBROUTINE BUNDHELICITY}

\noindent Purpose: to compute the helicity amplitude of the bound
state part, $\bar{b}+c\rightarrow B_c (B_c^{*})$, where $B_{c}$ is
the lowest state of $1^{1}S_{0}$ and $B_c^*$ is the lowest one of
$1^{3}S_{1}$, with the expressions for the polarization as
presented in appendix B.

\noindent Integer IBCSTATE=: state of the double heavy meson,
IBCSTATE=$1$ for $B_c[1^{1}S_{0}]$; IBCSTATE=$2$ for
$B_c^*[1^{3}S_{1}]$.

\noindent Real*8 BUNDAMP(4)=: four helicity amplitudes of the
bound state part $c+\bar{b}\to B_c (B_c^*)$;  BUNDAMP(4)=$0$ for
$c+\bar{b}\to B_c$; BUNDAMP(4)=$1$ for $c+\bar{b}\to B_c$.

\noindent Real*8 POLAR(4,3)=: four Lorentz components of the three
polarization vectors $\epsilon^{k}\,,(k=x,y,z)$ for the $B_{c}^*$
state.

\noindent {\bf SUBROUTINE FREEHELICITY}

\noindent Purpose: to compute the helicity amplitude of the
process $gg\rightarrow \bar{b}+c+b+\bar{c}$. The functions
$M^{\prime}_{Fm}\;(m=1, \cdots, 5)$, $E_{m,j,k}\;(m=1,\cdots, 9;
j=1, \cdots, 4; k=1, \cdots, 64)$ are defined in the body of the
paper.

\noindent Real*8 PMOMUP(5,8)=: the momenta PMOMUP(5,j)
($j=1,\cdots,8$) in the process: PMOMUP(5,1) for the gluon-1,
PMOMUP(5,2) for the gluon-2, PMOMUP(5,3) for the $B_{c} (B_c^*)$,
PMOMUP(5,4) for the $b$, PMOMUP(5,5) for the $\bar{c}$,
PMOMUP(5,6) for the $\bar{b}$, PMOMUP(5,7) for the $c$,
PMOMUP(5,8) for the $q_{0}$ (the light-like reference momentum).

\noindent Real*8 PMOMZERO(5,8)=: eight light-like momenta obtained
from PMOMUP(5,8) according to Eq.(\ref{m-l}) respectively.

\noindent Real*8 COLMAT(5,64)=:
$M^{\prime(\lambda_1,\lambda_2,\lambda_3,\lambda_4,\lambda_5,\lambda_6)}_{Fm}
(\lambda_i=\pm,\;m=1,\cdots,5)$.

\noindent Integer
IDP,\,IDQ0,\,IDB1,\,IDB2,\,IDC1,\,IDC2,\,IDK1,\,IDK2=: symbols
(codes) for the particles in the processes: IDP for $B_c (B_c^*)$,
IDQ0 for the reference massless fermion, IDB1 for $b$-quark, IDB2
for $\bar{b}$-quark, IDC1 for $c$-quark, IDC2 for $\bar{c}$-quark,
IDK1 for gluon-$1$, IDK2 for gluon-$2$.

\noindent Complex*16 YUP(42)=: 42 possible non-zero spinor
products ($y_i, i=1, 2, \cdots, 42$).

\noindent Complex*16 XUP(2)=: two possible double-spinor inner
products ($x_1, x_2$).

\noindent Real*8 PROPUP(14,4)=: 14 possible denominators from the
propagators appearing in the four basic groups of the Feynman
diagrams ($cb,bc,cc,bb$).

\noindent Complex*16 BFUN(9,4,64)=: $E_{m,j,k}\; (m=1, 2, \cdots,
9; j=1, \cdots, 4; k=1, 2, \cdots, 64)$.
\newline

\noindent {\bf SUBROUTINE AMP2UP}

\noindent Purpose: to compute the square of the helicity amplitude
of the process $gg\rightarrow B_c(B_c^{*})+b+\bar{c}$ by
connecting the amplitude of the bound state part to that of the
free quark part. The helicities of the intermediate quarks $c,
\bar{b}$ are summed up, the whole amplitude is squared, and then
all the 16 possible squared helicity amplitudes (if particle
polarizations in the final state are not measured) are summed up.
Here we also consider the three kinds of color flows explicitly.

\noindent Real*8 AMP2CF(3) =: three results of the square of the
whole amplitude corresponding to the three kinds of color flows
for the sub-process $gg\rightarrow B_c(B^{*}_c)+b+\bar{c}$.

\noindent Real*8 FINCOL(5,16) =: 16 possible helicity amplitudes
for the process, where the helicities of the intermediate quarks
have been summed up.

\noindent {\bf SUBROUTINE FIRST}

\noindent Purpose: to compute the nine basic basic functions
$E_{m,1,k} (m=1, \cdots, 9; k=1, 2, \cdots, 64)$. The
correspondence of the nine functions to the Feynman diagrams is
shown in Table~\ref{tab1} and Table~\ref{tab2}. In the present
case: IDK1=$1$, IDK2=$2$, IDP=$3$, IDQ0=$8$, IDB1=$4$, IDB2=$6$,
IDC1=$7$, IDC2=$5$.

\noindent {\bf SUBROUTINE SECOND}

\noindent Purpose: to compute the nine basic functions, $E_{m,2,k}
(m=1, \cdots, 9; k=1, \cdots, 64)$ obtained from $E_{m,1,k}$ by
gluon exchange. Here IDK1=$2$, IDK2=$1$, IDP=$3$, IDQ0=$8$,
IDB1=$4$, IDB2=$6$, IDC1=$7$, IDC2=$5$.

\noindent {\bf SUBROUTINE THIRD}

\noindent Purpose: to compute the nine basic basic functions,
$E_{m,3,k} (m=1, \cdots, 9; k=1, \cdots, 64)$ obtained from
$E_{m,1,k}$ by $b, c$ quark exchange and $\bar{b}, \bar{c}$
antiquark exchange. Here IDK1=$1$, IDK2=$2$, IDP=$3$, IDQ0=$8$,
IDB1=$7$, IDB2=$5$, IDC1=$4$, IDC2=$6$.

\noindent {\bf SUBROUTINE FOURTH}

\noindent Purpose: to compute the nine basic basic functions,
$E_{m,4,k} (m=1, \cdots, 9; k=1, \cdots, 64)$ obtained from
$E_{m,3,k}$ by gluon exchange. Here IDK1=$2$, IDK2=$1$, IDP=$3$,
IDQ0=$8$, IDB1=$7$, IDB2=$5$, IDC1=$4$, IDC2=$6$.

\noindent {\bf FUNCTION DOTUP(I,J)}

\noindent Purpose: to compute the dot-product of two momenta.

\noindent Integer I,J=: codes of the two particles.

\noindent {\bf FUNCTION INPUP(IP,JP)}

\noindent Purpose: to compute the inner product of two spinors
with light-like momenta based on the formula presented in APPENDIX
A.

\noindent Integer IP, JP=:  codes of the two particles.

\noindent {\bf FUNCTION SPPUP(IP,KP,JP)}

\noindent Purpose(s): to calculate the spinor product with the
formula presented in Appendix A.

\noindent Integer IP,JP,KP=:  codes of the three particles.\newline

\noindent {\bf FUNCTION POLSPPUP(I)}

\noindent Purpose: to compute the spinor product involving the
$B^{*}_c$ meson polarization vectors with the formula presented in
Appendix B.

\noindent Integer I=: one of the three polarization vectors for
$B^{*}_c$.

\clearpage

\begin{figure}
\includegraphics[width=0.43\textwidth]{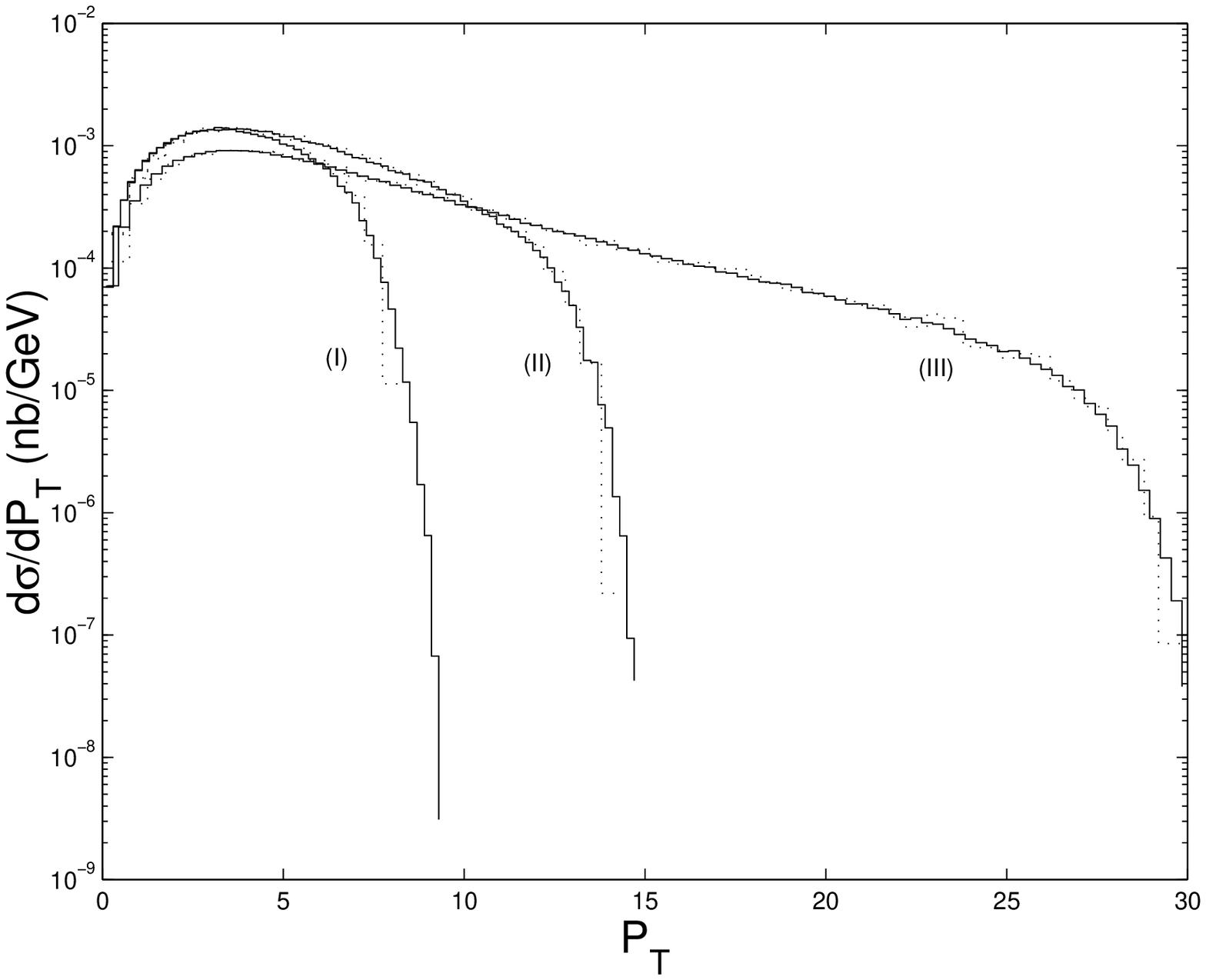}%
\includegraphics[width=0.43\textwidth]{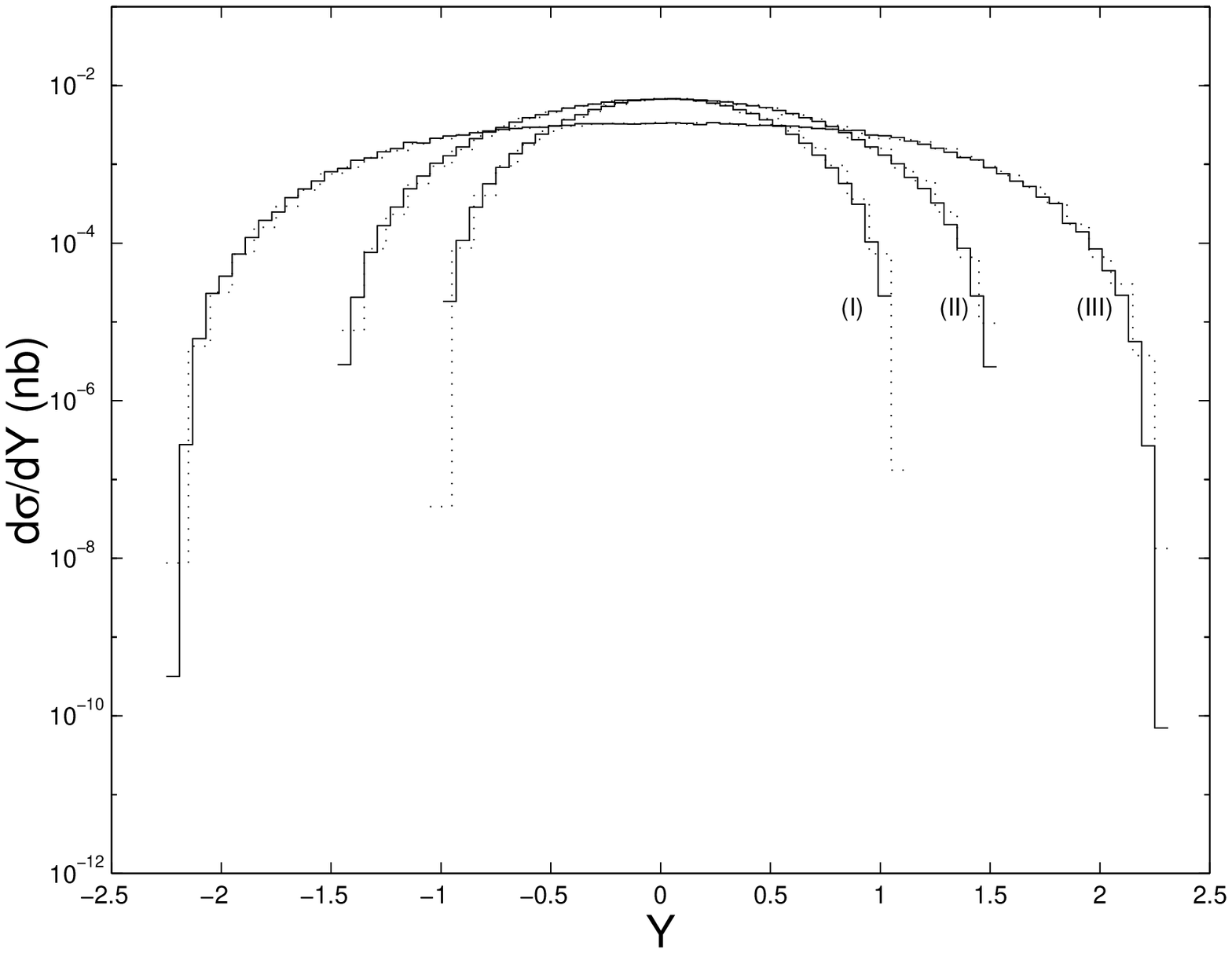}
\vspace{-4mm} \caption{$B_c$-$p_\mathrm{T}$ and
$B_c$-$Y$(rapidity) distributions for the subprocess
$gg\rightarrow B_c+b+\bar{c}$ with different center-of-mass
energies $\sqrt{\bar{s}}=$20 GeV(I), 30 GeV(II) and 60 GeV(III).
The solid line shows the present results and the dotted line shows
the results obtained in Ref.\cite{prod1}.} \label{subpt1}
\end{figure}

\begin{figure}
\centering
\includegraphics[width=0.43\textwidth]{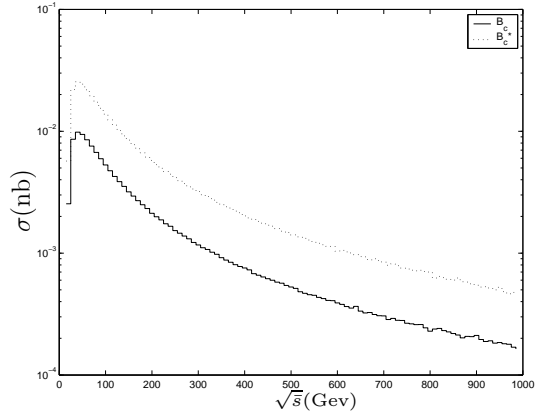}
\caption{Integrated cross sections for the subprocess $gg\rightarrow
B_c(B^{*}_{c})+b+\bar{c}$, with $\alpha_s=0.2$, $m_b=4.9$~GeV and
$m_c=1.5$~GeV. The solid (dotted) line corresponds to the
$B_c\;(B^{*}_{c})$ production.} \label{totsubshat}
\end{figure}

\begin{figure}
\centering
\includegraphics[width=0.43\textwidth]{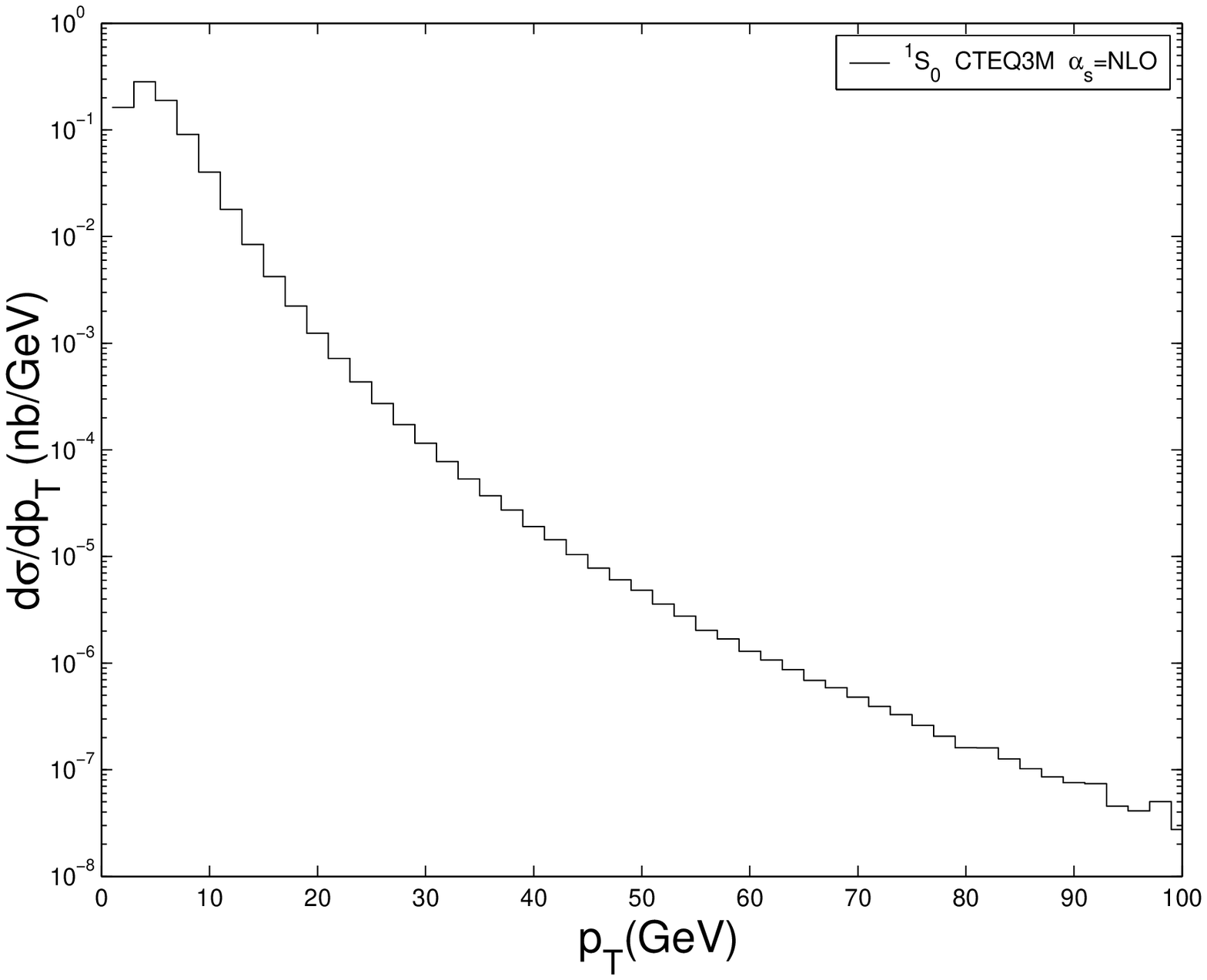}%
\includegraphics[width=0.43\textwidth]{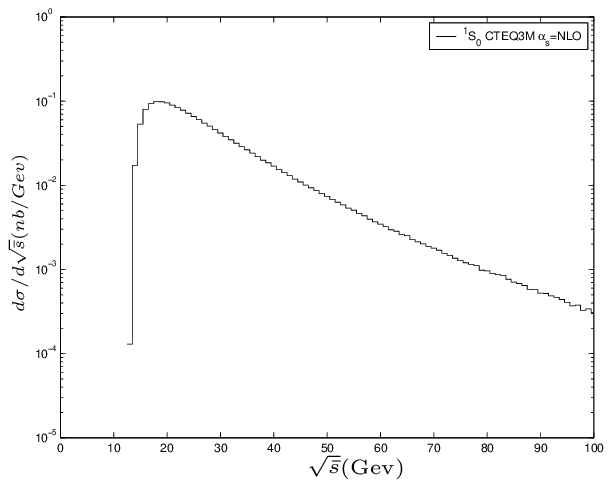}%
\caption{$B_c$-$p_\mathrm{T}$ and $\sqrt{\bar{s}}$ distributions
for the CTEQ3M parton distribution function by using $\alpha_s$ in
the next-to-leading order (NLO) and adopting the characteristic
energy scale squared $Q^2=\bar{s}/4$, where $\bar{s}$ is the
squared center-of-mass energy of the subprocess.}
\label{totpt1}
\end{figure}

\begin{figure}
\centering
\includegraphics[width=0.43\textwidth]{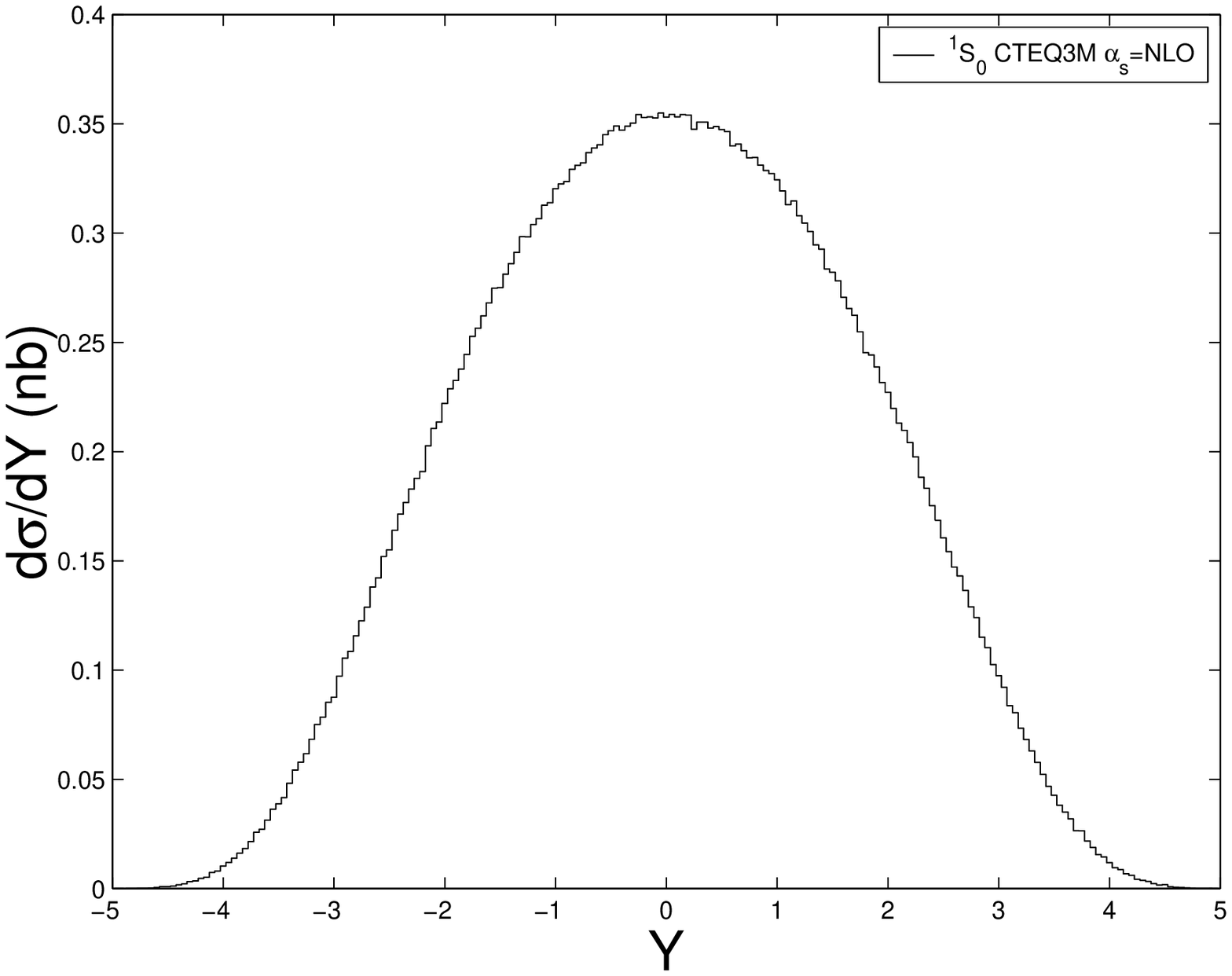}%
\includegraphics[width=0.43\textwidth]{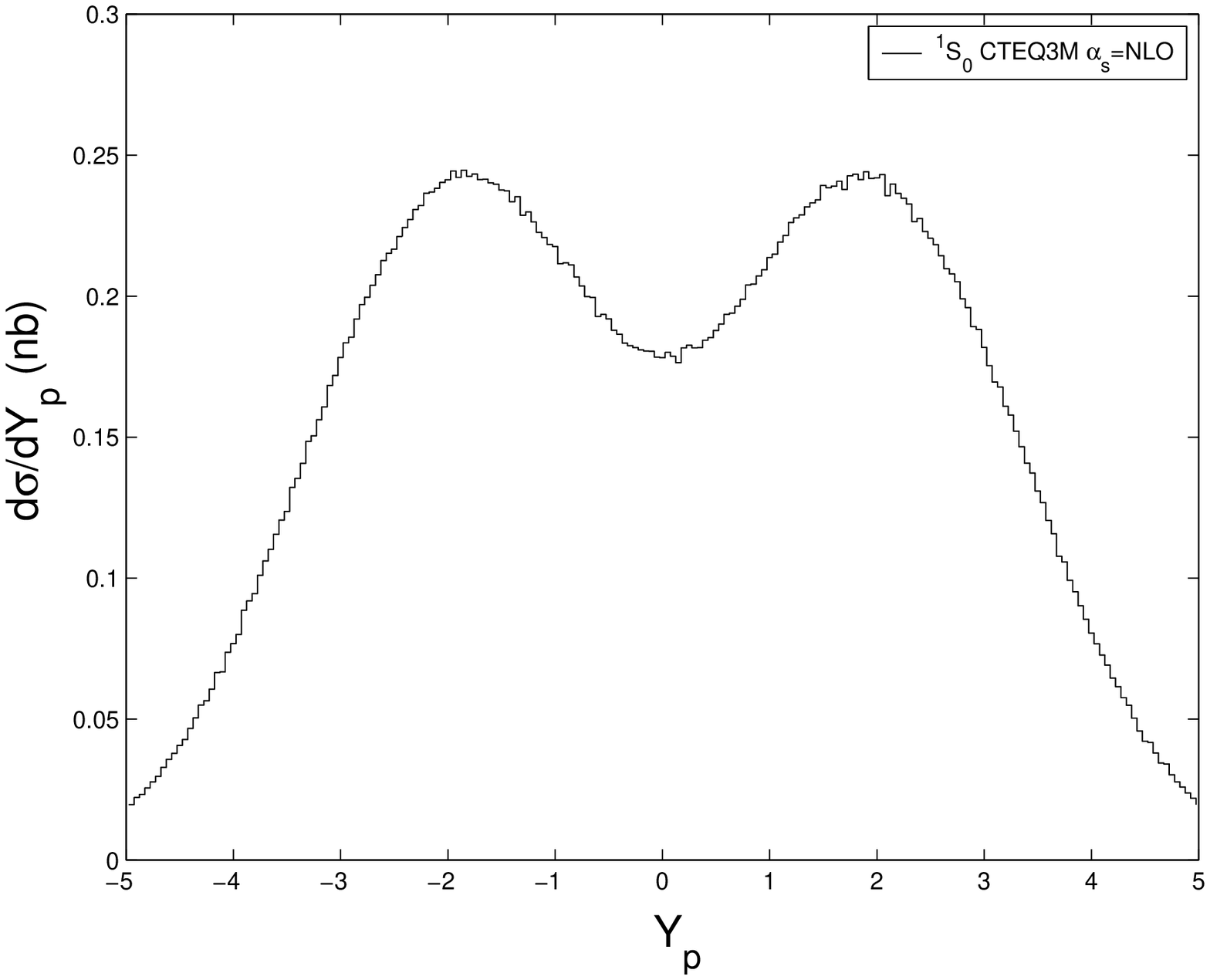}%
\caption{$B_c$-$Y$ (rapidity) and $B_c$-$Y_{p}$ (pseudorapidity)
distributions for the CTEQ3M parton distribution function, by
using $\alpha_s$ in the next-to-leading order (NLO) and adopting
the characteristic energy scale squared $Q^2=\bar{s}/4$, where
$\bar{s}$ is the center-of-mass energy of the subprocess.}
\label{totrap1}
\end{figure}

\end{document}